\documentclass[numbers]{article}
\usepackage{jcappub}

\usepackage{jcappub}
\usepackage{lmodern}
\usepackage{float}
\usepackage{graphicx}
\usepackage{amsfonts}
\usepackage{amsmath}
\usepackage{mathtools}
\usepackage{hyperref}
\usepackage{xcolor}
\usepackage{subcaption}
\usepackage{cellspace}
\hypersetup{colorlinks = true, urlcolor = blue, linkcolor = blue, frenchlinks = true, pdfborder = {0 0 0}}
\setcitestyle{square}

\usepackage{romannum}
\usepackage{soul}

\begin{document}
\title{Aspects of Everpresent $\Lambda$ (\Romannum{2}): Cosmological Tests of Current Models}
\author{Santanu Das, Arad Nasiri, and Yasaman K. Yazdi}
\affiliation{Theoretical Physics Group, Blackett Laboratory, Imperial College London, SW7 2AZ, UK}
\emailAdd{santanu.das@imperial.ac.uk, a.nasiri21@imperial.ac.uk, ykouchek@imperial.ac.uk}

\abstract{This paper investigates Everpresent $\Lambda$, a  stochastic dark energy model motivated by causal set theory and unimodular gravity, and confronts it with two key observational data sets, Supernova Ia (SN Ia) and Cosmic Microwave Background (CMB) data.  A key feature of this model is that $\Lambda$ fluctuates over time and on average the magnitude of its fluctuations is of the order of the dominant energy density  (be it radiation or matter) for the given epoch.  In particular, we focus on a phenomenological implementation of Everpresent $\Lambda$ known as Model 1. The random fluctuations in Everpresent $\Lambda$ realizations are generated using seed numbers, and we find that for a small fraction of seeds Model 1 is capable of producing realizations that fit SN Ia data better than $\Lambda$CDM.  We further investigate what features distinguish these realizations from the more general behaviour, and find that the ``good'' realizations have relatively small fluctuations at low redshifts ($z<1.5$), which do not closely track the matter density.  We find that Model 1 struggles to improve on $\Lambda$CDM at describing the CMB data.  However, by suppressing the values of $\Lambda$ near the last scattering surface, as suggested in \cite{Zwane_2018}, we find a large improvement in the best fit of the model, though still with a $\chi^2$ value much larger than that of $\Lambda$CDM.  We also study the allowed variation of the dark energy density by the CMB constraints in a more model-independent manner, and find that some variation (especially prior to recombination) is possible and in fact can lead to improvement over $\Lambda$CDM and reduce the Hubble tension, in line with some early dark energy proposals.  However, for the kinds of variations considered, the favoured  fluctuations are smaller in magnitude than is typical in current Everpresent $\Lambda$ models. 
} 
\maketitle
\section{Introduction}

After the surprising discovery that the expansion of the Universe is accelerating \cite{SupernovaCosmologyProject:1998vns, SupernovaSearchTeam:1998fmf}, one of the leading candidates for dark energy has been the cosmological constant. Along with dark matter, it has been combined into the very successful $\Lambda$CDM model for describing the evolution of the Universe. 
It is not only effective at the background level but also robust at linear perturbations, giving accurate predictions for the matter power spectrum and for the micro-Kelvin level fluctuations in the photon temperature observed in the cosmic microwave background (CMB).

The cosmological constant is the simplest dark energy candidate; however, several issues such as discrepancies in $H_0$ inferred from CMB measurements assuming the $\Lambda$CDM model and its direct measured value from supernovae (dubbed the Hubble tension)~\citep{Di_Valentino_2021,riess2018new,Riess_2019, riess2020expansion,wong2020h0licow,di2021combined,riess2021cosmic,riess2022comprehensive}, discrepancies in $\sigma_8$ of density perturbations at the scale of $8h^{-1}$Mpc inferred from late time measurement like the cluster number counts or weak lensing \citep{secco2022dark,li2023kids} and inferred from the CMB alone~\citep{aghanim2020planck}, discrepancies in the amount of lensing in the CMB angular power
spectrum~\cite{troxel2018cosmological,asgari2021kids}, in the curvature $\Omega_k$ and other specific parameters~\citep{Addison_2016,Kitching_2016,Couchot_2017,lusso2019tension,verde2013planck,addison2018elucidating,evslin2017isolating,guandalin2023theoretical,luongo2022larger,horstmann2022inference,zhao2021tomographic,bengaly2018probing}
have led to the development of alternative models of dark energy. Apart from this, the ad hoc value of the cosmological constant is another theoretical motivation for searching for other dark energy models. 


In recent years, cosmological model building has taken on a new flavour, in part following some perplexing observational findings. 
In addition to the Hubble tension, baryon acoustic oscillation (BAO) data~\cite{BOSS:2013igd,BOSS:2014hwf} appears to allow for a negative cosmological constant~\cite{Wang:2018fng, Aubourg:2014yra,colgain2023putting, malekjani2023negative, colgain2023mcmc} for the relevant epoch. See also the latest DESI results \cite{DESI:2024mwx} which seem to favor an evolving dark energy over a cosmological constant. As a result, models with a variable cosmological “constant” are being considered more seriously. Some of these examples include~\cite{Sorkin:2007bd,deCesare:2016dnp,Cree:2018mcx}.

In this paper we will be considering one of the earliest  fluctuating cosmological constant models, known as Everpresent $\Lambda$. Everpresent $\Lambda$ dates back to 1987 and originates from a heuristic argument by Sorkin~\cite{originallambda,sorkin1991spacetime,sorkin1994role} based on unimodular gravity and causal set theory. This model predicted that the value of the cosmological constant should fluctuate, but with a standard deviation that at
any time is of the order of the total energy density in the Universe at that time. This prediction was verified for the current epoch with the supernova observations made in 1998. It was seen that, at present, the contribution of dark energy is of the order of the matter density. Subsequently, more practical phenomenological models have been proposed for producing full expansion histories that have the character of Everpresent $\Lambda$. The first such cosmological model based on Sorkin’s argument was proposed in~\cite{Ahmed_2004}. This model has further been studied in~\cite{Ahmed_2013,Zwane_2018}. A second model was also proposed in~\cite{Zwane_2018}, which is computationally more efficient, but has much of the characteristics of the first phenomenological model. 

In an earlier paper~\cite{das2023aspects1}, we considered the theoretical motivations and statistical properties of Everpresent $\Lambda$ and its phenomenological implementation. In the present paper, we will focus on the observational signatures and evidence for and against these models. As we show in the subsequent sections, the model successfully fits SN Ia for a fraction of its realizations, while at least in its present form, it struggles to explain the CMB data. The reason is partly due to the ISW effect. As we will see, its contribution in Everpresent $\Lambda$ is very different from its contribution in $\Lambda$CDM as it is non-vanishing in the matter dominated epoch. The SW effect will also turn out to be modified non-trivially in Everpresent $\Lambda$.

The primary aims of this paper are to determine to what extent the existing implementations of Everpresent $\Lambda$ are able to explain cosmological data, and whether they can be modified to improve the agreement.  While we use the $\Lambda$CDM fit to data as a reference point that we compare the Everpresent $\Lambda$ fits to, it is not necessarily our goal to quantitatively improve on it at this stage. As the Everpresent $\Lambda$ models are still works in progress, we wish to find the qualitative features of those realizations that fit the data well or relatively better.

This paper is organized as follows. In the next section, we briefly introduce the two phenomenological implementations of Everpresent $\Lambda$ we have mentioned. Then, in Section~\ref{sec:cosmological_observations} we briefly describe the kinds of cosmological observations which can be used to constrain dark energy, before we compare the predictions of current models of Everpresent $\Lambda$ with supernova observations in Section~\ref{sec:supernova}. In Section~\ref{sec:cmb_comparison}, we compare the models against CMB data. In Section~\ref{sec:cmb_constraints}, we provide theoretical analyses of the constraints on $\Lambda$ from the epochs before, during, and after recombination. We then numerically study arbitrary fluctuations of the dark energy, allowing for variations as large as the ambient density, and constrain it in each of the three epochs separately, showing that only small deviations in dark energy from $\Lambda$CDM are allowed. Finally, we discuss our findings in Section~\ref{sec:discussion_conclusion}. There are also two appendices: Appendix~\ref{appendixA} with a technical discussion of MCMC implementations for stochastic models and Appendix~\ref{appendixB} with a derivation of the width of the visibility function during recombination, used in Section~\ref{sec:cmb_constraints}.

\section{An Overview of Everpresent $\Lambda$}

The theoretical details of Everpresent $\Lambda$  were discussed in depth in~\cite{das2023aspects1}. In addition, two phenomenological models of it, known as Models 1 and 2, were reviewed and the statistics and properties of Model 1 were studied in detail. In this section we provide a brief overview of the equations involved in generating an expansion history within these models of Everpresent $\Lambda$. In the following sections, we will present our results from comparing the outputs of these two models (primarily Model 1) with cosmological observations.
\subsection{Model 1}
Everpresent $\Lambda$ is a consequence of spacetime discreteness and the quantum mechanical conjugacy between spacetime four-volume $V$ and the cosmological constant $\Lambda$. The uncertainty in the value of the cosmological constant is then proportional to the inverse square root of the four-volume. In \cite{Ahmed_2004} a phenomenological model (known as Model 1) for Everpresent $\Lambda$ was proposed in which the four-volume of the past lightcone of an observer at each cosmic time is used to evolve and set the scale of fluctuations of $\Lambda$. Therefore, $\Lambda$ will fluctuate around some mean (assumed to be zero), with the size of the fluctuations decaying over time as the observer's past lightcone grows.

At a given time $t$, if the four-volume of the past light cone of an observer is $V_t$, then $\Lambda_t$ should be a random number drawn from a Gaussian distribution with mean $0$ and standard deviation $\sim1/\sqrt{V_t}$. The equations for calculating the expansion history and the dark energy density can be written as

\begin{eqnarray}
V_{t} &=&\frac{4 \pi}{3} \int_0^t d t^{\prime} a\left(t^{\prime}\right)^3\left(\int_{t^{\prime}}^{t} d t^{\prime \prime} \frac{1}{a\left(t^{\prime \prime}\right)}\right)^3 \,,\label{vt}\\
    \Lambda_{t}&=&\frac{\Lambda_{t-dt}V_{t-dt}+8\pi\alpha\, \xi_t\sqrt{V_{t}-V_{t-dt}}}{V_t}
    \,, \label{lambdat}\\
    {\rho_\Lambda}_t &=& \frac{\Lambda_t}{8\pi G} \,.
\end{eqnarray}

\noindent Here $a(t)$ is the scale factor at cosmic time $t$, $\xi_t$ is a random number drawn from a standard normal distribution, and ${\rho_\Lambda}_t$ is the dark energy density. We adopt units in which the speed of light is set to unity, $c=1$, throughout this work. The parameter $\alpha$ controls the magnitude of the fluctuations. We can use this dark energy density to integrate the Friedmann equation given by 
\begin{equation} \label{friedmann}
    \frac{\dot{a}^2}{a^2}=\frac{8 \pi G }{3}\Bigg(\rho_t+{\rho_\Lambda}_t\Bigg) = \frac{8 \pi G \rho^0_{cr}}{3}\Bigg(\Omega^0_m a^{-3} + \Omega^0_r a^{-4} +\frac{{\rho_\Lambda}_t}{\rho^0_{cr}}\Bigg)\,.
\end{equation}

\noindent $\rho^0_{cr}=3H_0^2/(8\pi G)$ is the critical density of the Universe at present, where $H_0$ is the Hubble parameter at present, taken to be approximately 70 km/s/Mpc, and not to be confused with the non-fixed  $H_0$ values that appear later in this paper which are the outcomes of simulations. $\Omega^0_m$, and $\Omega^0_r$ are the present time density parameters for all the matter components (baryons and CDM) and the radiation components (photons and massless neutrinos), respectively.  In other words, we take the present day matter and radiation densities, and scale them appropriately with $a$ to find their densities at time $t$. At any given time $t$, we can use \eqref{friedmann} to calculate the change in the scale factor after a time $dt$.  We then calculate the new volume by substituting $a$ into \eqref{vt} and repeat the steps from \eqref{vt} to \eqref{friedmann} to recursively get the full expansion history of the Universe. 

Sometimes, we will get random numbers that make \eqref{friedmann} become negative before the histories reach the present day. This renders \eqref{friedmann} inadequate to describe the ensuing dynamics and we abandon such simulations. The correct dynamics to replace \eqref{friedmann} for these simulations is not known at present. When this happens, we refer to it as an $H^2<0$ crash \citep{das2023aspects1}. This situation is reminiscent of the tunneling proposal in quantum cosmology, where the absence of classical solutions and existence of only complex saddle points for the gravitational path integral indicate quantum behavior. In a Lorentzian path integral approach, such contributions are suppressed compared to classical evolution \citep{lehners2023review}. For now, we simply discard these non-classical realizations of the model. A proper quantum mechanical treatment of these cases is beyond the scope of this paper, and is left for future work.

Another issue in this approach is that the value of the dark energy density at time $t=0$ is infinite (or undefined, if $\Lambda_0=0$), as the volume of the past lightcone at $t=0$ is zero. However, it is important to remember that the current model is only applicable within the domain of validity of semi-classical gravity. The initial state of the universe falls within the quantum gravity regime, and we do not expect Model 1 to adequately describe that era. Therefore, instead of taking the initial condition to be at the Big Bang singularity, we set it during the radiation-dominated era by specifying an initial non-zero four-volume $V_0$ (calculated using a radiation-dominated universe without dark energy). A discussion about initial conditions can also be found in~\cite{das2023aspects1}.   

\subsection{Model 2}
Due to the volume integrals in Model 1, solving for the background equations is computationally intensive. Another approach is to identify the key aspects of the dark energy history in model 1, and attempt to construct a phenomenological model that can instantly create a dark energy history obeying those features. One such model (known as Model 2) was proposed in \cite{Zwane_2018}. The authors highlighted the essential features of Model 1 as the stochastic nature of dark energy and the auto-correlation time equalling the Hubble time. Moreover, $\Omega_\Lambda$ must be allowed to change sign. Since $\Omega_\Lambda\leq1$, it cannot be the outcome of a Gaussian random variable. However, if one constrains $-1\leq\Omega_\Lambda\leq1$ and introduces $\hat{\Omega}_{\Lambda}(\lambda)$ such that 

\begin{eqnarray}
\Omega_{\Lambda}(\lambda)=\tanh \left[\hat{\Omega}_{\Lambda}(\lambda)\right]\,,
\end{eqnarray}

\noindent where $\lambda \equiv \log(a)$, then $\hat{\Omega}_{\Lambda}(\lambda)$ can be modeled as a continuous Gaussian random variable with a mean of $0$. The authors chose a Gaussian ansatz for the covariance matrix:

\begin{eqnarray}
\left\langle\hat{\Omega}_{\Lambda}\left(\lambda_1\right) \hat{\Omega}_{\Lambda}\left(\lambda_2\right)\right\rangle=\tilde{\alpha}^2 e^{-\frac{\mu\left(\lambda_1-\lambda_2\right)^2}{2}}\,.
\end{eqnarray}

\noindent $\tilde{\alpha}$ and $\mu$ are free parameters of the model and the $\tanh$ function is chosen to restrict $\Omega_{\Lambda}$ to the range $(-1\;,1)$. If $\mu=\mathcal{O}(1)$, this model roughly reproduces the basic features of Model 1 simulations while reducing the computational cost and complexity. For further details on this model, refer to \cite{das2023aspects1}.

\section{Cosmological Observations}\label{sec:cosmological_observations}

Observational validation is essential not only to test the predictions of any theoretical model but also to discern between different models. Different observational data are available, and many upcoming experiments will provide additional data for distinguishing between different expansion histories. The cosmological observations used to test the expansion history can be divided into two primary categories: background observations and perturbation measurements.

The most crucial data set for testing the background expansion history of the Universe comes from type Ia Supernova (SN Ia). All  SN Ia have approximately the same luminosity; hence, they can be used as standard candles~\cite{Riess_2019,scolnic2018complete,scolnic2022pantheon+,riess2022comprehensive} and can constrain the expansion history up to redshift $z \approx 2$. Apart from supernovae, there are also many other measurements of the Hubble parameter. The Hubble parameter is estimated directly using the differential age methodology, i.e. by measuring the age between pairs of passive evolving galaxies with similar metallicity and separated by a small redshift interval~\cite{Jimenez_2002}. Other measurements of the Hubble parameter come from e.g.  observations of the tip of the red giant branch (TRGB)~\cite{M_ller_2018,freedman2023progress,anand2022comparing} and Cepheid amplitudes~\cite{Riess_2020,Sabour2014}. These late-time background expansion history measurements constrain $H_0$ to lie in the range $71 \-- 75 \text{km/s/MPc}$. 

Another kind of expansion tracer comes from cosmological perturbations. The cosmic microwave background (CMB) is the most critical probe in this context \cite{rosenberg2022cmb,seraille2024constraining,aghanim2020planck}. Another important type of observation is the baryon acoustic oscillations (BAO). The most recent compilation of BAO data can be found in~\cite{Nunes_2020, PhysRevD.93.023530,Jailson2016,de_Carvalho_2018}. Several intensity mapping experiments are also upcoming, which promise to provide a 3D map of the Universe at different redshifts, e.g.   HERA~\citep{DeBoer_2017, Carilli_2020}, CHIME~\citep{Amiri_2022,Mandana2022}, SKA~\cite{barry2022ska}, MeerKAT~\citep{Cunnington2022}, LOFAR~\citep{van_Haarlem_2013}, HIRAX~\citep{Newburgh_2016}, FAST~\cite{li2023fast}, GBT~\cite{grasha2020evolution,wolz2022h}, PAON4~\cite{ansari2020design}, Tianlai~\citep{Zhang_2016}. Therefore, in the near future, we will be able to resolve the expansion histories and hence the dark energy as a function of time. 

There are additional data sets that can also constrain dark energy. Strong gravitational lens systems (SLS) offer a unique opportunity to study the $\Omega_m - w$ plane because their confidence regions are almost orthogonal to those of standard rulers such as BAO and CMB. Different methods have been used to constrain cosmological parameters with strong gravitational lensing \cite{Chae_2002,Biesiada_2010,Cao_2012,Cao_2015}. Second order perturbations such as weak lensing and the kSZ effect may also shed light on early dark energy~\cite{Waizmann_2008, DeDeo2005, abbott2018dark, haridasu2022scrutinizing, ye2023shape}. 

In this paper, we focus on Supernova and CMB data.  

\section{Fitting Model 1 to Supernova Data} \label{sec:supernova}

The cosmological distance modulus is an important observable in astronomy. It measures the difference between the apparent magnitude and the absolute magnitude of an object and is an indicator of distance in astronomy. The distance modulus is defined by
\begin{equation} \label{distancemodulus}
\mu=m_B-M_B=5 \log _{10} d_L+25    ,
\end{equation}
where $d_L$ is the luminosity distance of an object (in MPc) and can be calculated as 

\begin{equation}
d_L = (1+z)\int_0^z \frac{d z^{\prime}}{H(z^{\prime})}
= \frac{1}{a}\int_a^1 \frac{da'}{a'^2 H(a')}\,.
\end{equation}

\noindent SN Ia are widely used as standard candles for cosmological distance measurements, and they can provide a strong constraint on the Hubble parameter. We use the Pantheon+SH0ES sample of the apparent magnitude and redshift of 1550 SN Ia events with redshift $z\geq0.01$ together with 42 calibrator events \cite{riess2022comprehensive}. As the expansion history for Everpresent $\Lambda$ differs from the standard $\Lambda$CDM model, it affects the distance modulus $\mu$. 

Since the fluctuations in $\Lambda$  are of the order of the inverse square root of the spacetime volume, which in an FLRW type universe is of the order of the Hubble parameter squared, the fluctuations will always be of the order of the dominant energy density.  A priori, one might expect that since the fluctuations of $\Lambda$ for small redshifts track the matter density, Everpresent $\Lambda$ is more like CDM rather than $\Lambda$CDM, and should not improve on $\Lambda$CDM in describing the supernova data. Although this is statistically correct, we shall surprisingly see that there exist realizations of Model 1 that fit the observations better than $\Lambda$CDM.  We will then investigate the characteristics of these realizations.

\subsection{MCMC Analysis}

Everpresent $\Lambda$ is a stochastic model of the Universe. Given a pair of values for the parameters $(\Omega_m^0h^2,\alpha)$, and a series of Gaussian random numbers, Model 1 of Everpresent $\Lambda$ provides a unique expansion history of the Universe (note that the dimensionless Hubble parameter $h$ represents $h= H_0 / (100 \mathrm{km/s/Mpc})$ and is not a function of redshift). The series of random numbers is characterized by a seed $s$. That is, given an integer number $s$ as a seed parameter, we uniquely generate a series $\xi_1,\xi_2,...$ of samples from the standard normal distribution, and use them in discrete time steps to evolve $\Lambda$ according to \eqref{lambdat}. Therefore, given  $(\Omega_m^0h^2,\alpha,s)$, we can get a unique expansion history. We also add $M_B$, the absolute magnitude of SN Ia, as a nuisance parameter in the MCMC. The constraint on $M_B$ arises from measurements of the distance modulus using Cepheids located in the same host galaxies as the 42 calibrator Type Ia supernovae, as conducted by the SH0ES team.

To explore the viability of the model, we choose 90,000 different seeds, and for each of them, we run a Markov chain Monte Carlo (MCMC) process independently and determine how often we obtain a likelihood as good as $\Lambda \text{CDM}$ for the supernova data.\footnote{In Appendix \ref{appendixA} we describe our method for doing an MCMC analysis to determine the posterior distribution of the model parameters.}  
In all of these runs, we use flat priors on $\Omega^0_m h^2$, $\alpha$, and $M_B$. We allow the parameters to vary in a large range, namely,  $\Omega^0_m h^2\in [0.01,1]$, $\alpha\in[0.001, 0.025]$, and $M_B\in[-30.0,-10.0]$. The range for $\alpha$ was chosen based on the behaviour observed in \cite{das2023aspects1}, so that the chains would not hit the boundaries of the ranges. In that paper values of $\alpha$ around 0.012 produced the most runs with $H_0$ comparable to the observed range (when dividing the history of the Universe into 10,000 timesteps).  The majority of runs encounter an $H^2<0$ crash when $\alpha>0.015$, and a value of $H_0$ that is too small when $\alpha<0.005$.  Therefore the best fit $\alpha$ should be obtainable within the given range, without hitting the boundaries.  The desired value of $\Omega^0_m h^2$ rounded to 0.14 is well within the given range and we shall see that the distribution obtained in this parameter does not overlap with the boundaries of its range either.

\begin{figure}
    \centering
        \begin{subfigure}[b]{0.45\textwidth}
        \centering
        \includegraphics[width=\textwidth]{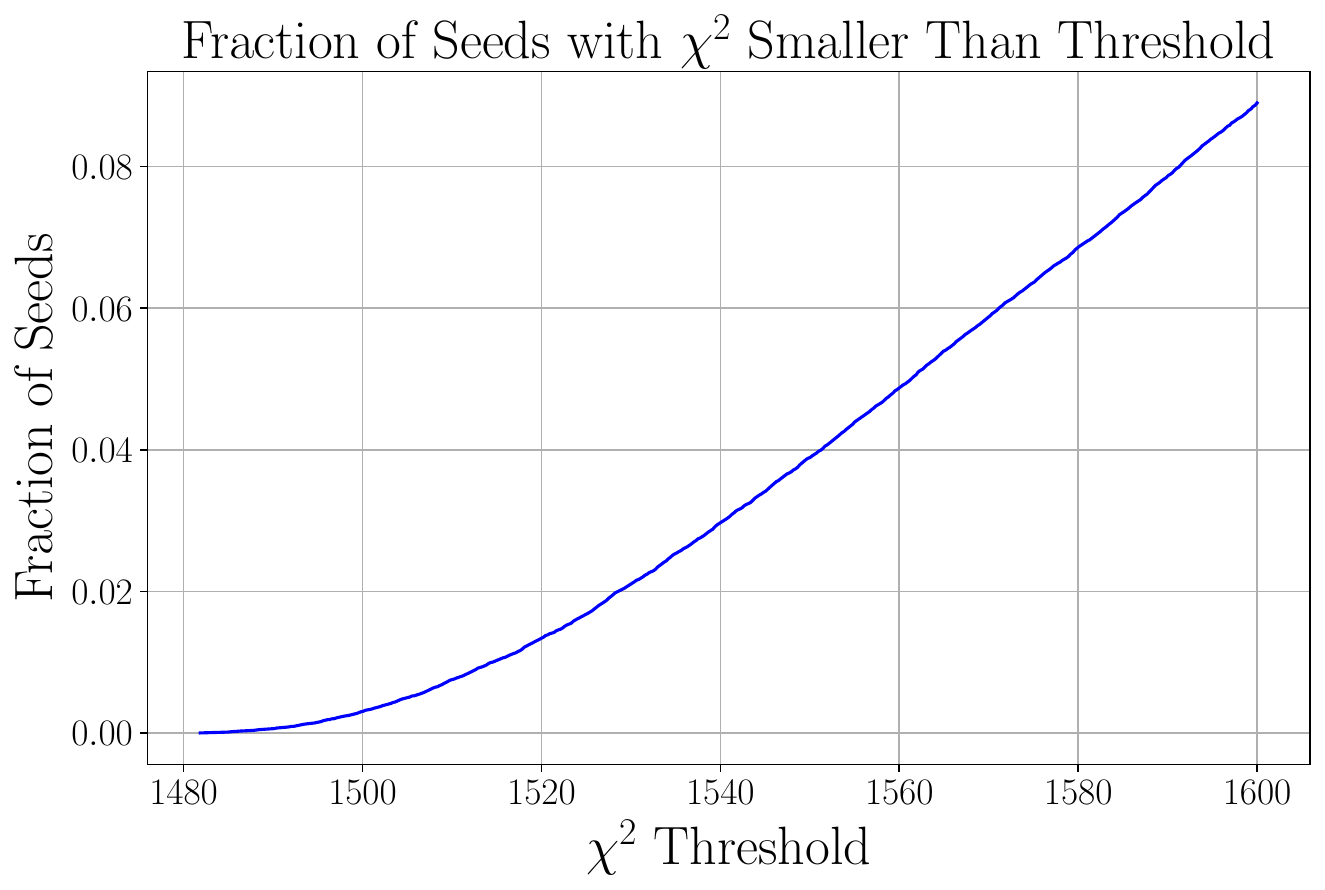}
        \caption{}
        \label{fracvschi2}
    \end{subfigure}
    \hfill
    \begin{subfigure}[b]{0.45\textwidth}
        \centering
        \includegraphics[width=\textwidth]{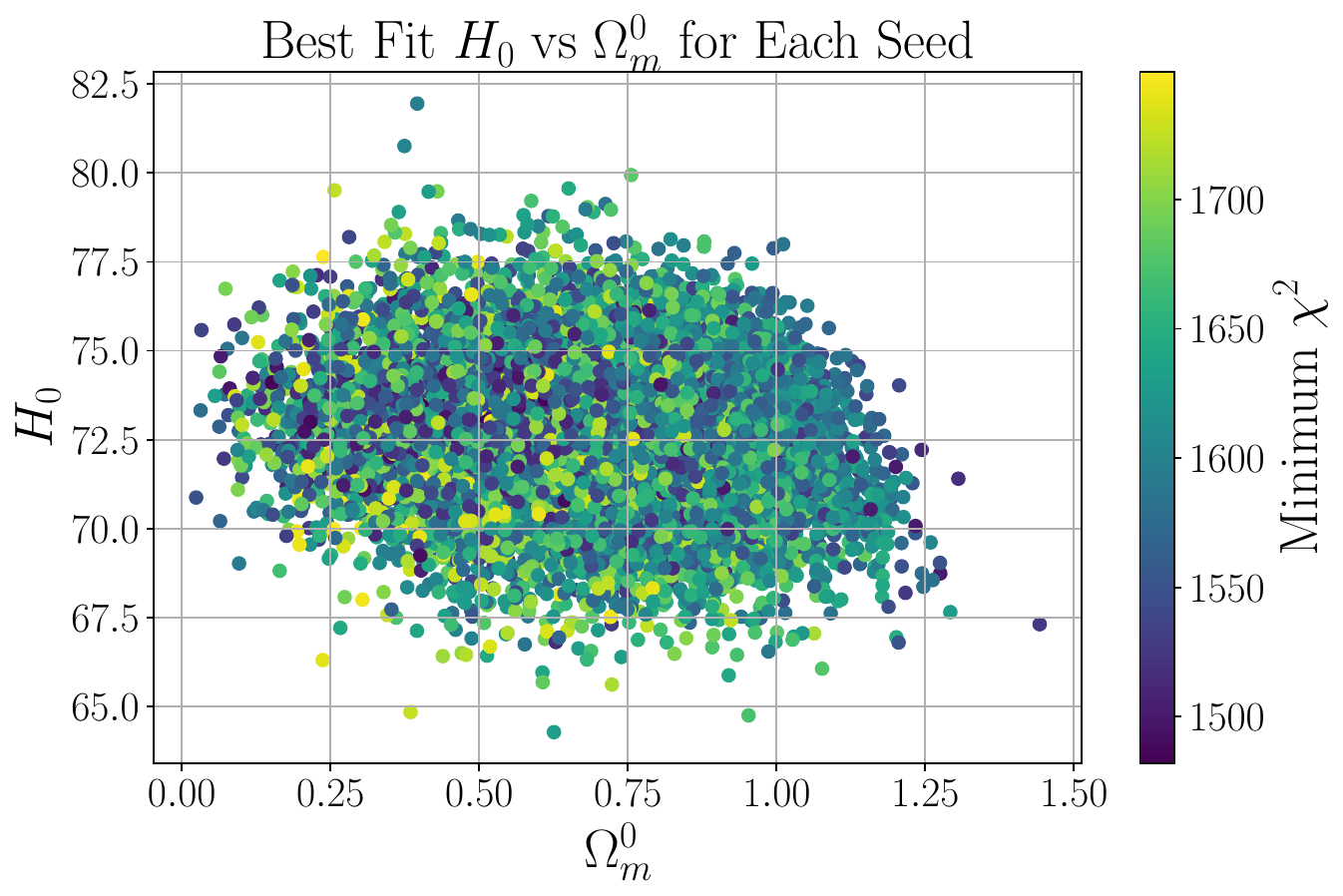}
        \caption{}
        \label{scatterH0vsOmegam_many}
    \end{subfigure}
        \vfill
    \begin{subfigure}[b]{1\textwidth}
        \centering
        \includegraphics[width=\textwidth]{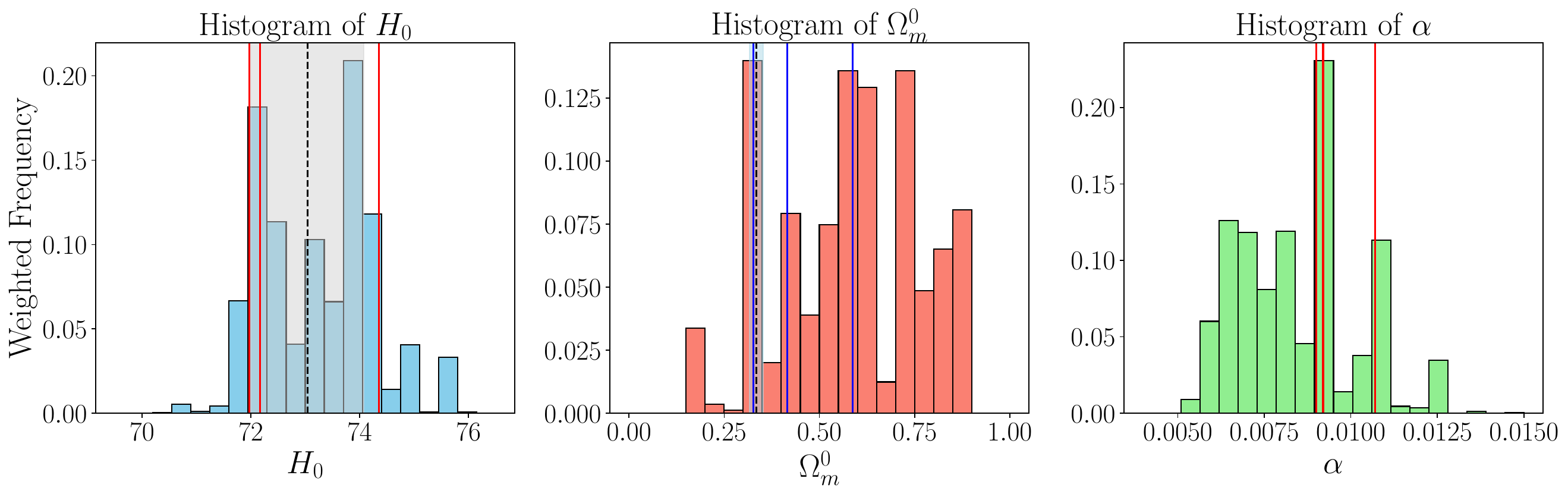}  
        \caption{}
        \label{bestfithistograms}
    \end{subfigure}
 \caption{(a): The fraction of seeds that have a best-fit $\chi^2$ smaller than some threshold on the $x$ axis. For example, a fraction of $0.003$ of seeds have a best-fit $\chi^2<1500$. (b): The scatter plot of best-fit ($\Omega_m^0$,$H_0$) for 16767 seeds with a best-fit $\chi^2$ smaller than 1750. The plot has a mild Pearson correlation coefficient of $-0.13$, showing that the seeds that can reduce $H_0$ and have a reasonable $\Omega_m^0\sim0.3$ are rare. Also the figure shows clearly that the likelihood surface is not smooth if we change the seed. (c): The histograms represent the posterior averages of the parameters $H_0$, $\Omega_m^0$, and $\alpha$ from different seeds. For each seed, averages are taken after the burn-in period, and each seed is weighted by its posterior $P(s_i|D)$. The three vertical colored lines show the average parameters of the three best seeds. The dashed vertical lines and the shaded regions represent the mean and 1$\sigma$ confidence interval as reported by the SH0ES model-independent measurement for $H_0$ \citep{riess2022comprehensive} and the flat $\Lambda$CDM fit to Pantheon+ for $\Omega_m^0$ \citep{brout2022pantheon+}.}
    \label{sneIadists}
\end{figure}

We try to find seeds that  provide fits to the data close to or better than the standard model does.  $\Lambda$CDM gives $\chi^2\approx 1485.3$ for the Pantheon+SH0ES data.  We therefore consider only seeds with a $\chi^2$ close to that, and find that out of the 90,000 seeds, 55 give a $\chi^2$  in the range $[1481.0,\,1490.0]$. 16 of these seeds give a better likelihood than $\Lambda$CDM. This is summarized in Fig. \ref{fracvschi2}; for each value $x$ on the horizontal axis, the plot shows the fraction of the seeds that have a minimum $\chi^2$ smaller than $x$.

Figures \ref{scatterH0vsOmegam_many} and \ref{bestfithistograms} show the summary statistics of the best-fit and average parameters of different seeds. Fig. \ref{scatterH0vsOmegam_many} shows the scatter of $(\Omega_m^0,H_0)$, where the best-fit point of each seed is colored according to its $\chi^2$. The same parameters can give a very good or bad fit depending on the choice of the seed; hence, the likelihood surface is extremely discontinuous. This further justifies running each MCMC chain for a fixed seed number. However, the variance of this scatter plot should not be interpreted as the variance of the parameters themselves. To properly bound the parameters, it is essential to weight each seed's average parameters with the corresponding likelihood. This approach is known as Bayesian Model Averaging \citep{hoeting1999bayesian,paradiso2024convenient,paradiso2024evaluating}. It can be employed for summarizing the posterior behavior of any stochastic model. For example, the Bayesian average and variance of $H_0$ for Model 1 are given by
\begin{equation}
    \langle H_0\rangle = \sum_{i=1}^n H_{0i} P(s_i | D),\ \ \text{var}(H_0)=\sum_{i=1}^n \Big(\text{var}(H_0|s_i,D)+H_{0i}^2\Big) P(s_i | D) - \langle H_0\rangle^2
\end{equation}
where the sum is over all the considered seeds, $H_{0i}=\mathbb{E}(H_0|s_i,D)$ is the posterior average of $H_0$ for seed $s_i$, $\text{var}(H_0|s_i,D)$ is the variance of $H_0$ for seed $s_i$, $D$ represents the data set, and 
\begin{equation}
    P(s_i | D) = \frac{P(D | s_i) P(s_i)}{\sum_{j} P(D | s_j) P(s_j)}
\end{equation}
is the posterior probability of the seed $s_i$. $P(D | s_i)$ is the marginalized likelihood of the data for a given seed, and is estimated by the harmonic mean of the likelihoods in a given MCMC chain~\citep{paradiso2024convenient}. Note that $P(s_i)$ is the probability of a particular seed, which we assume to be uniform across all considered seeds. Given that the value of \(\Omega_m^0\) has been robustly measured using various independent methods such as cluster abundance and weak lensing experiments (e.g.,~\cite{Bahcall1992, Costanzi2019, 2020Abdullah}), there is little flexibility for modifying \(\Omega_m^0\). However, since Model 1 allows for a broad range of \(\Omega_m^0\) values, \(P(s_i)\) could theoretically be adjusted to prefer seeds that produce an \(\Omega_m^0\) consistent with these established measurements; nevertheless, we do not pursue this adjustment further. The result of the Bayesian Model Averaging is
\begin{equation}
\begin{aligned}
 H_0 &= 73.27 \pm 1.27, \\
 \Omega_m^0 &= 0.58 \pm 0.19, \\
 \alpha &= 0.0085 \pm 0.0017.
\end{aligned}
\end{equation}
The corresponding weighted 1-d distributions are plotted in Fig. \ref{bestfithistograms}. Note that, due to the relatively small number of seeds with a good $\chi^2$, the distributions appear less smooth. For smoother plots, it is necessary to utilize a pool of at least a million seeds. 

Our $H_0$ is consistent with the model-independent result by SH0ES \citep{riess2022comprehensive}:
\begin{eqnarray}
    H_0^{\text{SH0ES}}= 73.04 \pm 1.04.
\end{eqnarray}
The model independence is achieved by Taylor expanding $H$ around its present value and constraining the expansion coefficients. Strictly speaking, this method does not accommodate stochastic models where $H$ cannot be assumed to be smooth, including Everpresent $\Lambda$. Nevertheless, we find a similar constraint to that of SH0ES, although with a larger error bar, owing to the fact that the fluctuations in $H$ are relatively small at low redshift, see the right panel in Fig. \ref{distHagain}.

Unlike deterministic models, $\Omega_m^0$ is pretty much unconstrained for Everpresent $\Lambda$. To better understand why $\Omega_m^0$ does not get constrained by the Pantheon+ data, let us look at the expression for $\chi^2$, ignoring the likelihood for calibrator events and also the covariance between different events for the sake of argument: 
\begin{eqnarray}
&\chi^2 &= \sum_i \left(\frac{1}{\Delta m^2_i}\right)\left[5 \log _{10} \left((1+z_i)\int_0^{z_i} \frac{d z^{\prime}}{H\left(z^{\prime}\right)}
\right)+ 25 +M_B - m_i \right]^2  \\ &=& \sum_i \left(\frac{1}{\Delta m^2_i}\right) \left[5 \log _{10} \left(\int_0^{z_i} \frac{d z^{\prime}}{\sqrt{\Omega^0_m (1+z)^3 + (1-\Omega^0_m) g(z)}}
\right) + 5 \log _{10} (1+z_i) - 5 \log _{10} \left(H_0\right)+ 25 +M_B- m_i \right]^2 \nonumber
\end{eqnarray}

\noindent Here, the summation is over all the supernova data points, and $m_i$ is the apparent magnitude of the i-th supernova. $g(z)$ is some function by which the dark energy density is varying in redshift for the particular seed. $g(z)$ is completely different for different seeds. The supernova data roughly follows a straight line in the $\mu$ vs $\log(1+z)$ plot. $g(z)\rightarrow 1$ as $z \rightarrow 0$, and hence the integrand in the second line does not depend on $\Omega_m^0$. Hence, the $y$-intercept of the line $\mu$ vs $\log(1+z)$ for small redshifts ($z\ll1$) solely depends on $H_0$. Therefore, the overall intercept of the plot fixes $H_0$. $\Omega_m^0$, on the other hand, controls the deviation from the straight line at high redshifts ($z>1$) through the first term. 
The best fit $\Omega^0_m$ should then be found from $\frac{\partial \chi^2}{\partial\Omega_m^0} = 0$.  As each seed gives a completely different $g(z)$, the value of $\Omega_m^0$ also changes significantly from one seed to another, as seen in our results.

\begin{figure}

    \centering

         \includegraphics[width=0.8\textwidth]{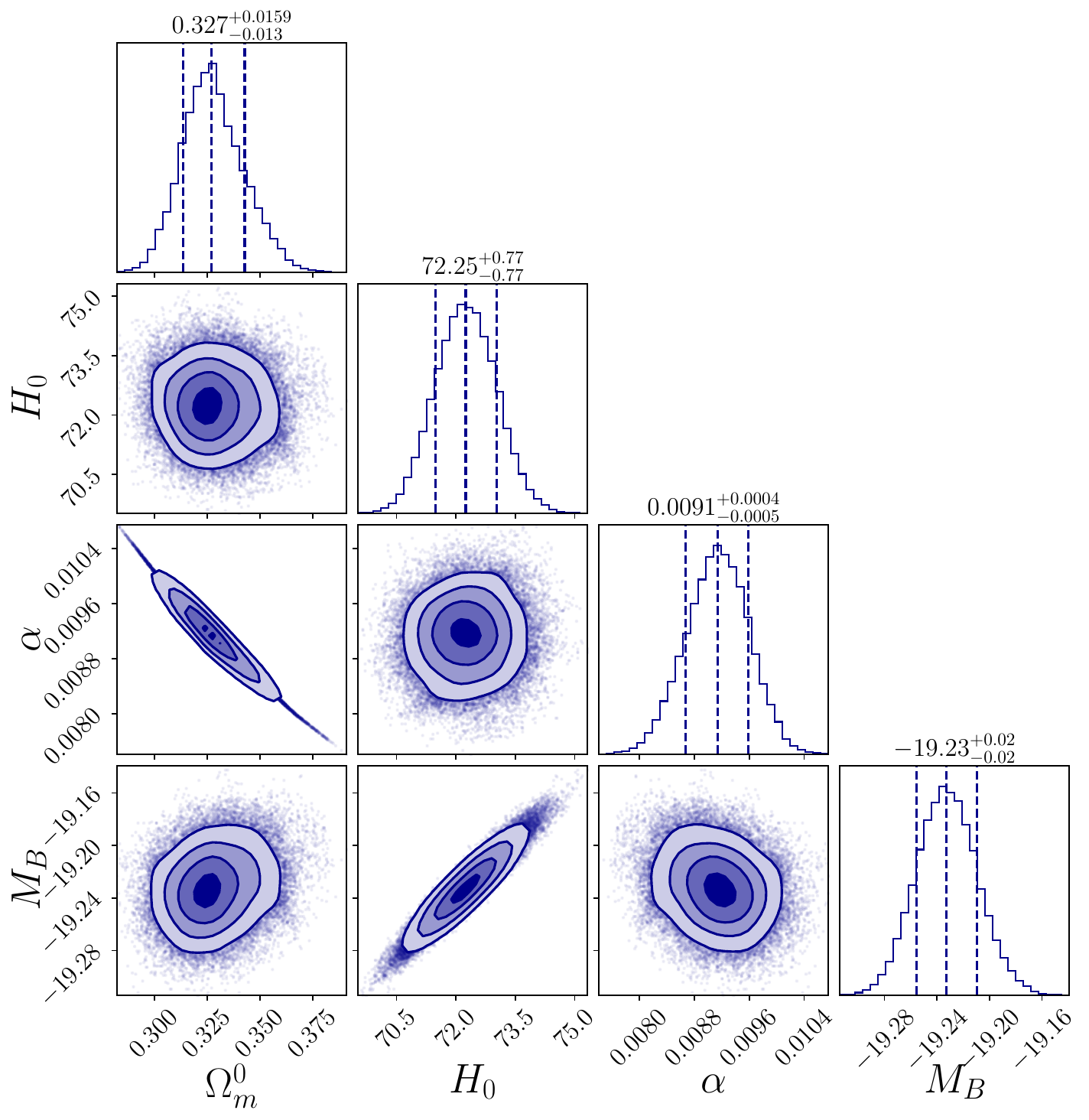}

    \caption{The corner plot for the best seed, with $0.5\sigma$, $1\sigma$, $1.5\sigma$, and $2\sigma$ contour lines. The mean and $1\sigma$ error of each parameter is above its 1-d distribution. The MCMC parameters are $(\Omega_m^0h^2,\alpha,M_B)$. $H_0$ and $\Omega_m^0$ are derived parameters. Note that the perfect correlation between $\alpha$ and $\Omega^0_m$ is an artifact of having fixed the seed number (see Section 4.4 of \cite{das2023aspects1}).}
    \label{sneIacorner}
\end{figure}

In Fig.~\ref{sneIacorner}, we show the results from the best seed that has a best-fit $\chi^2=1481.9$. Its best-fit parameters are
\begin{equation}
\begin{aligned}
 H_0 &= 72.18, \\
 \Omega_m^0 &= 0.33, \\
 \alpha &= 0.0092.
\end{aligned}
\end{equation}
The figure shows the 2-dimensional contour plots and 1-d likelihood distributions for $\Omega^0_m$, $H_0$, $\alpha$, and $M_B$. $H_0$ is a derived parameter for the analysis. The plots show that $\Omega^0_m$ and $H_0$ are almost uncorrelated. As $\alpha$ is a function of $\Omega^0_m$, they are inversely related (for further discussions see Section 4.4 of \cite{das2023aspects1}). 

\subsection{Diagnosing Good Seeds by a Single Parameter}

Our analysis shows that it is possible to get expansion histories from Model 1 that fit the supernova data better than the standard $\Lambda$CDM. However, the probability of getting such seeds is small (16 out of 90000, or about 0.017\%).  That it was able to outperform $\Lambda$CDM at all, shows that there is some merit to Everpresent $\Lambda$. However, the small number of these runs raises the question of the practical ability of Model 1 to generate data that can be compared with observations. For a deterministic and non-random model, it is straightforward to assess its success with the data as it comes with a fixed $\chi^2$ value. However, more care is needed to judge the results from a stochastic model such as Everpresent $\Lambda$. An important question in this regard is: what are the physical features of those realizations that give the best fits to the supernova data, and why are their seed numbers rare.

\begin{figure}
    \centering
    \begin{subfigure}[b]{\textwidth}
         \centering
         \includegraphics[width=.9\textwidth]{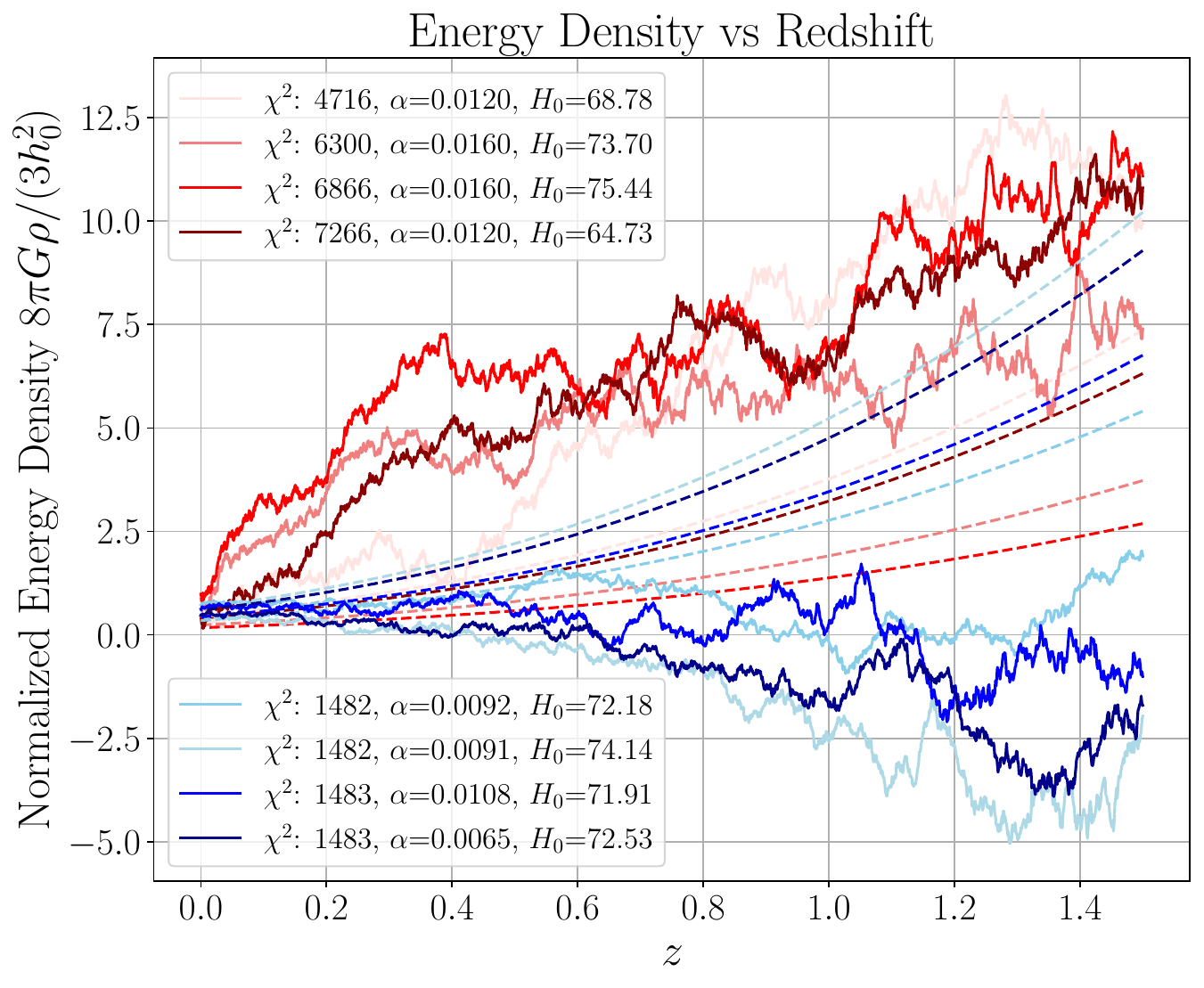}
     \end{subfigure}
    \caption{$\Lambda/(3h_0^2)$ versus redshift is plotted for the 4 realizations of Model 1 with smallest $\chi^2$ for Pantheon+SH0ES data set, with blue solid curves. The dashed curves are the corresponding $8\pi G\rho_m/(3h_0^2)$. Note that $h_0=70.2$km/s/MPc is a fixed constant we use for normalization. For comparison, the same quantities are plotted in red for 4 realizations that give very high $\chi^2$ for the supernova data. In the 4 ``good'' realizations, $\Lambda$ tends to cross the corresponding matter density fewer times, and stays smaller than the matter density for a larger redshift interval.}
    \label{rhoxsn1a}
\end{figure}

In Fig.~\ref{rhoxsn1a}, 4 of the seeds with the lowest $\chi^2$ (“good" seeds) and 4 with a relatively high $\chi^2$ (“bad" seeds) are shown for comparison. The solid fluctuating lines are the energy density in $\Lambda$ (normalized using a fixed constant $h_0=70.2$km/s/MPc, not equal to their Hubble constant), and the dashed curves are the corresponding matter density. Visual inspection provides a distinction between these two extremes: In the range $z<1.5$, a good seed generally has a smaller $\Lambda$ than a bad seed. Moreover, bad seeds are more closely tracking the matter density, while for good seeds $\Lambda$ fluctuates as if it is independent from matter density in the range $z<1.5$. A more quantitative distinction can be made using the so-called $Om(z)$ diagnostic.

The $Om(z)$ diagnostic was first introduced in \cite{sahni2008two} as a discriminator between different dark energy models. It is defined as 
\begin{equation}
Om(z)=\frac{\left(\frac{H(z)}{H_0}\right)^2-1 }{(1+z)^3-1}.
\end{equation}

\noindent For $\Lambda$CDM, $Om(z)=\Omega_m^0$, and for pure CDM, $Om(z)=1$. It has the property that at small redshifts $z<1$, it magnifies any deviations from the standard dark energy equation of state, $w=-1$. For example, for a quintessence dark energy (i.e. $-1<w<0$), as a function of increasing redshift $z$, $Om$ is decreasing and positively curved, while for a phantom dark energy (i.e. $w<-1$) it is increasing and negatively curved. In both cases, $Om$ converges to $\Omega_m^0$ for $z\gg1$.

\begin{figure}
    \centering
        \begin{subfigure}[b]{0.48\textwidth}
        \centering
        \includegraphics[width=\textwidth]{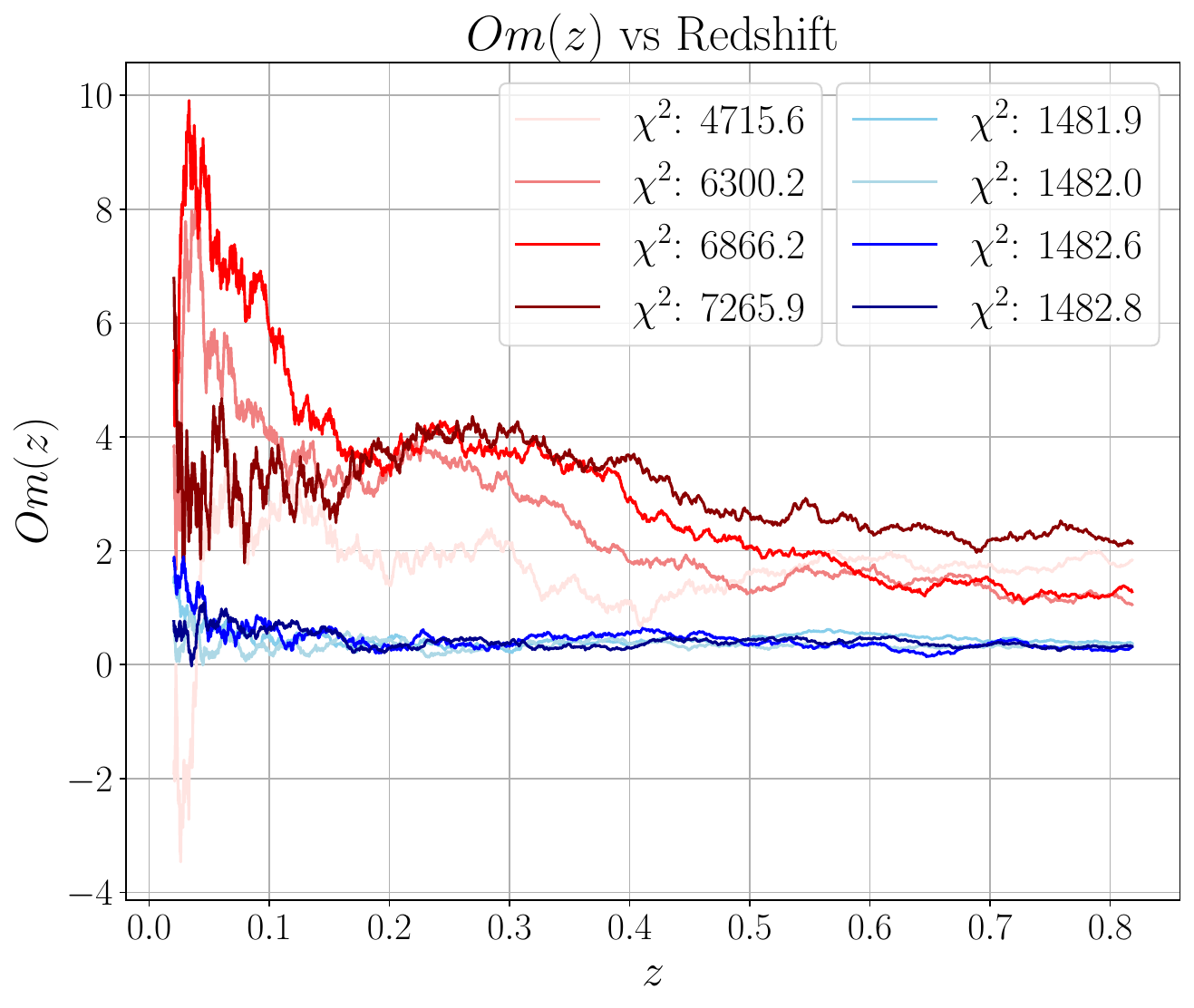}
        \caption{}
    \end{subfigure}
    \hfill
    \begin{subfigure}[b]{0.48\textwidth}
        \centering
        \includegraphics[width=\textwidth]{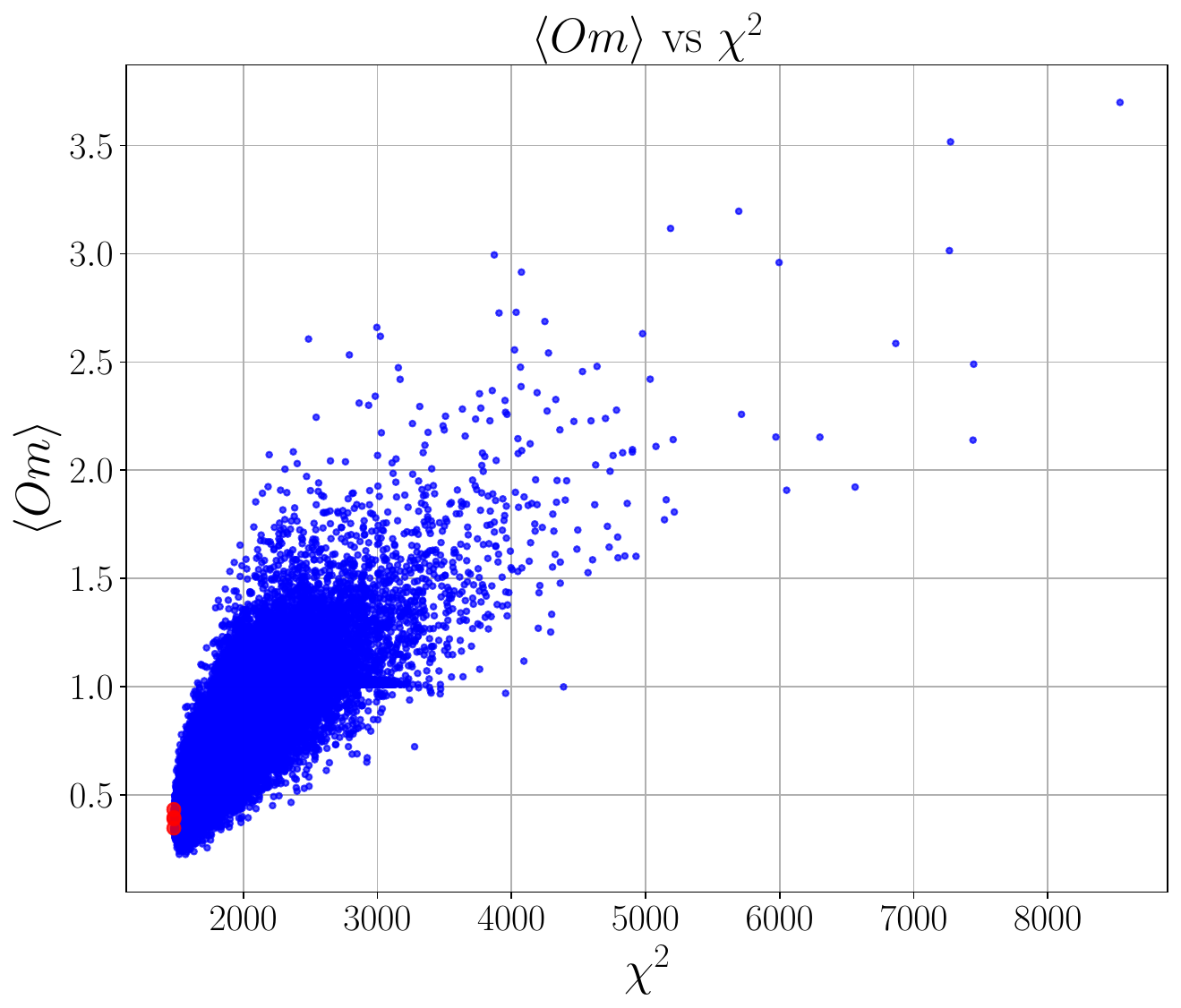}
        \caption{}
    \end{subfigure}
    \caption{Left plot: $Om(z)$ versus redshift for the same 8 realizations as in Fig.~\ref{rhoxsn1a}. The distinction between good and bad seeds is clear here. After large oscillations for $z<0.1$, the seeds with a good fit to the supernova data settle and fluctuate around some horizontal line, while the seeds with a bad fit have a bump-like feature around $z\sim0.3$. Right plot: For all 90000 seeds, we compute $\langle Om\rangle=\frac{1}{0.7}\int_{.1}^{.8}Om(z)dz$. A scatter plot of $\langle Om\rangle$ versus the $\chi^2$ is shown. There is a strong correlation between these two quantities. The Pearson correlation coefficient of the plot is 0.85. The 4 best seeds are marked with red dots.}
    \label{Om}
\end{figure}

In the left plot of Fig. \ref{Om}, we have plotted the $Om$ function versus $z$ for 4 of the best and worst seeds. The $Om$ diagnostic makes their distinction even more apparent. Bad seeds have a bump around $z\sim0.3$ and then decrease with redshift. This is like a combination of quintessence and matter-like behaviour. For good seeds, however, $Om$ fluctuates around some horizontal line and has much smaller values. The plot on the right shows the scatter plot of the average of $Om$ in the interval $z\in [0.1,0.8]$ versus the best fit $\chi^2$ for all 90,000 seeds. There is a strong correlation between these two quantities. The good seeds tend to have a smaller $\langle Om\rangle$. In fact, we find that for the 70 seeds with $\chi^2<1060$, $\langle Om\rangle=0.27\pm0.03$. Therefore, not only is the $Om$ behaviour of a good seed reminiscent of constant $\Lambda$, it also fluctuates around the $Om$ value of $\Lambda$CDM and not around $\Omega_m^0$ of the same run (recall that the value $\Omega_m^0$ of each seed, according to Fig. \ref{sneIadists}, is not concentrated around 0.3).

Regarding the practicality of the model, this result suggests that to obtain good seeds with a higher probability, one should focus on a sub-space of seeds that give smaller values of $\langle Om\rangle$ in the low-redshift range $z\lesssim 0.8$. It further indicates that to find good fits to data for a stochastic model such as Everpresent $\Lambda$, an understanding of the physical constraints that have to be satisfied is required in order to strategically limit one's search to seed numbers that are more likely to satisfy those constraints.  For the supernova data, the single number $\langle Om\rangle$ has turned out be a good constraint parameter. 

Next, given that only a small fraction of the seeds produce better fits to the data than $\Lambda$CDM, we would like to further understand if their behaviour is markedly different from the typical behaviour of Everpresent $\Lambda$.

\subsection{Understanding Our Results With Mock Data}

To confirm the exceptionality of the good seeds, we perform a mock data analysis. Our procedure for generating mock supernova data is as follows: for a realization of Model 1, we find the value of the distance modulus from \eqref{distancemodulus} at 1580 redshifts greater than $0.01$, corresponding to the non-calibrator SN Ia events in the Pantheon+ dataset. Using their full covariance matrix, we add Gaussian random noise to the corresponding distance moduli. In this way we obtain supernova data that could occur in a universe described by a particular Everpresent $\Lambda$ realization. 

As an interesting first example, consider the following case: Using an arbitrary seed and fixing the parameters $(\alpha,H_0,\Omega^0_m)=(0.01, 70.4, 0.28)$, we generated mock data. After running the MCMC for this mock data set, the best fit for Model 1 yielded the parameters $(\alpha,H_0,\Omega^0_m)=(0.0047, 71.0, 0.81)$. This again confirms the observation from the previous subsection: The supernova data does not constrain $\Omega^0_m$ for Model 1.  The model can fit the data, whether real or mock, with almost any value of $\Omega^0_m$. In other words, the model does not make a prediction for $\Omega_m^0$, even for a data set generated by itself.

We are then interested in the following fraction for this mock data set, $D$:
\begin{equation}
    f_D\coloneqq \frac{\#\{seed=s|\text{Model 1 with $seed=s$ gives a better fit to $D$ than }\Lambda\text{CDM}\}}{\text{Total number of seeds}}.
\end{equation}
We previously saw that if we take $D$ to be the Pantheon+SH0ES data set, then $f_D=16/90000$. The question then is what the fraction $f_D$ is if we take $D$ to be a mock data set generated by Model 1. Furthermore, we are interested to know if the mock data generated by an arbitrary seed has different characteristics from the mock data generated by a good seed. 

\begin{figure}
        \centering
        \begin{subfigure}[b]{0.5\textwidth}
        \centering
        \includegraphics[width=1.05\textwidth]{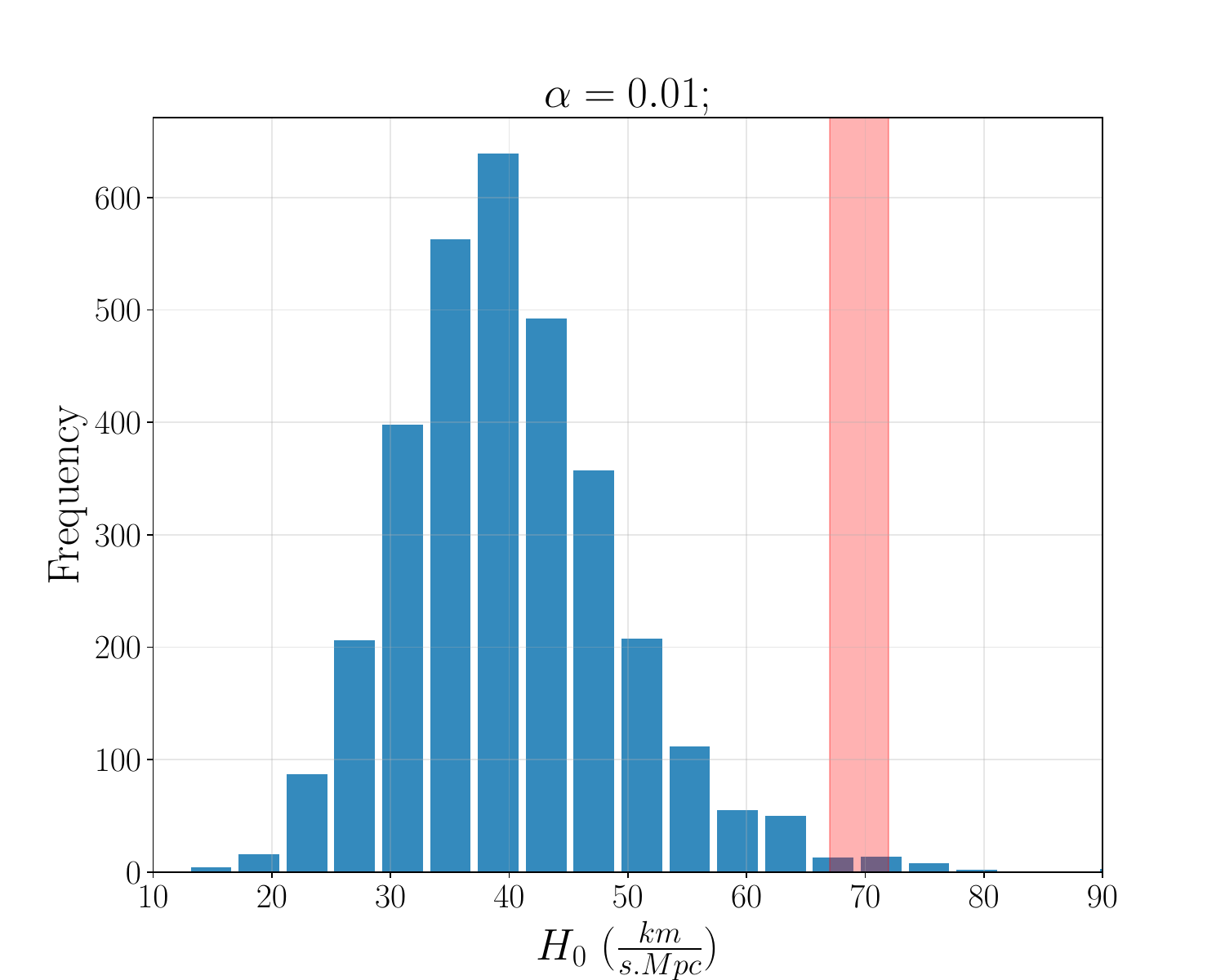}
        \caption{}
    \end{subfigure}
    \hfill
    \begin{subfigure}[b]{0.48\textwidth}
        \centering
        \includegraphics[width=\textwidth]{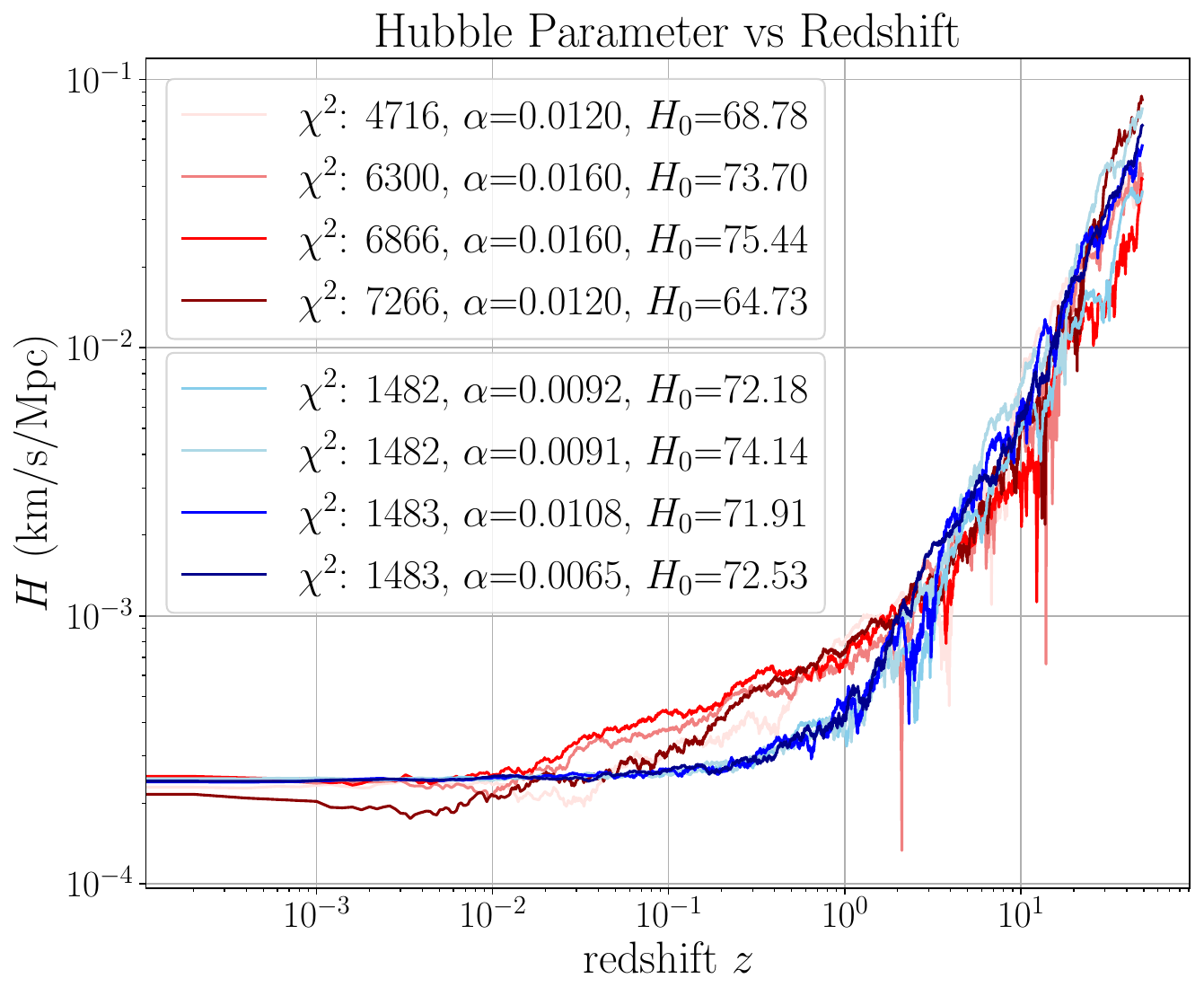}
        \caption{}
    \end{subfigure}
    \caption{Left plot: The distribution of $H_0$ for Model 1 with $\Omega_m^0h^2=0.14$, $\alpha=0.01$ (taken from \cite{das2023aspects1}). The average and standard deviation is given by $40.3\pm9.3$. Notice the difference between this distribution and the posterior distribution in Fig. \ref{sneIadists}. The range $[67,72]$ for $H_0$ is shaded in red. 
    Right plot: The Hubble parameter $H$ as a function of redshift for four of the best and four of the worst seeds. Good seeds show much smaller fluctuations at $z\lesssim1.0$.
    }
    \label{distHagain}
\end{figure}

First, let us investigate the question for an arbitrary seed. We try to be realistic with the choice of the seed $s$ for generating the mock data. In the left plot of Fig.~\ref{distHagain}, we show the distribution of the present day Hubble parameter for Everpresent $\Lambda$ realizations with parameters $\Omega^0_mh^2=0.14$ and $\alpha=0.01$. Most of the expansion histories give a value of the Hubble parameter close to $40$ km/s/MPc. Only a few realizations give Hubble parameters close to $72$ km/s/MPc, i.e. close to the present cosmology. To better mimic the real data, we pick a set of 10 seed numbers from the tail of this distribution, as the value of $H_0\sim72$km/s/MPc lies in the tail. The expansion history for each of these seeds give values of $H_0$ in the range $\sim[69,72]$ km/s/MPc. There is no further constraint on the choice of seed.
We then do a similar analysis as before. We take 10,000 arbitrary seeds and then try to find the best fit $\chi^2$ for each of these seeds with each of the mock data sets. For the 10 mock data sets, we find that $f_D$ lies in the interval $[0.026,0.205]$, with an average of 0.076. Thus it is around three orders of magnitude larger than the $f_D$ for the real data. 

However, if we instead use one of the good seeds to generate the mock data, $f_D$ turns out to be 1/10000, which is compatible with the real data. This analysis further confirms that the seed numbers that are in the lowest corner of the right plot in Fig.~\ref{Om} are distinguished, and they more closely resemble the late time expansion history of our universe. 
This result shows that there are seeds that produce Everpresent $\Lambda$ histories in Model 1 that can be more and less typical, even within the subset that have reasonable values of $H_0$.  It further indicates that Everpresent $\Lambda$ is easily capable of producing a momentary value of $H_0$ that fits present observations, but that it is more difficult to produce a history that has the right character to fit all data.

\section{Fitting Model 1 to CMB Data}\label{sec:cmb_comparison}

Supernova data and other observations such as the TRGB and Cepheid amplitudes, give a direct measurement of the Hubble parameter. However, they can constrain  dark energy only at low redshifts. To constrain the expansion history at larger redshifts, we can utilize cosmological perturbation data from the CMB and BAO. In this section, we discuss our analysis of Model 1 (and briefly Model 2) of Everpresent $\Lambda$ with CMB data.

\subsection{Methodology}

To investigate Model 1, we use a similar technique as used in our supernova data analysis in the previous section. We take 100,000 seeds, where each seed gives a series of random numbers that can be used to generate a unique expansion history given $\Omega^0_m$ and $h$. For each seed, we try to find the best fit cosmological parameters through an MCMC analysis. However, unlike in the case of the supernova analysis, here we need to calculate cosmological perturbations (e.g. for metric, density, velocity, and also higher multipole components for the matter and radiation). There are several publicly available Boltzmann packages that can be used for calculating the CMB power spectrum, e.g., CMBAns~\cite{CMBAns1910,Das:2013gta,Das_2014}, and CAMB~\cite{lewis2011camb}. We use Model 1 to get the dark energy density $\rho_\Lambda$ as a function of the scale factor, $a$, and then feed it into the Boltzmann package to get the CMB power spectrum. In our model, dark energy is not a fluid, so we turn off the dark energy perturbations in the Boltzmann code. As a consistency check, we modified both CAMB and CMBAns, and confirmed that both gave the same results. However, we use CAMB for our analysis as CMBAns is not yet equipped with parallel computation capability. 

 We use SCoPE along with the Planck likelihood code (high-$\ell$, low-$\ell$, lensing likelihood)\footnote{\url{https://wiki.cosmos.esa.int/planck-legacy-archive/index.php/CMB_spectrum_&_Likelihood_Code}} for our MCMC analysis~\cite{das2014scope}. SCoPE is the Monte Carlo sampler that draws the samples from the given prior distribution, which is then used to calculate the dark energy density as a function of scale factor. These parameters and the calculated dark energy density are then passed on to the modified CAMB code to calculate the cosmological power spectrum.  The modified CAMB package we use, instead of taking the dark energy equation of state as a parameter, takes the dark energy density as a function of scale factor and calculates the power spectra. The Planck likelihood package obtains the likelihood for that power spectrum. SCoPE then uses this likelihood to generate the next sample. The standard $6$ parameters, i.e. $\Omega^0_m h^2$, $\Omega^0_b h^2$, $h$, $\kappa$, $n_s$, and $A_s$, are used for the MCMC run. Here one should note that Model 1 also has another parameter $\alpha$ (see \eqref{lambdat}). However, according to our discussion in Section \ref{sec:supernova},  for a given seed, $\alpha$ and $\Omega^0_m$ are related. As long as we use the standard $6$ parameters, $\alpha$ is not a free parameter, since we can infer it using the input parameters $\Omega^0_m h^2$ and $h$.

Calculating the expansion histories for Model 1 of Everpresent $\Lambda$ is time-consuming. Therefore, it is not feasible to run the MCMC process for 100,000 seeds separately. Even running the MCMC for a single seed is a lengthy process, if we want to calculate the expansion history for every choice of parameters. Therefore, instead of calculating the expansion history repeatedly, we can take advantage of some properties of Model 1 to speed things up.
The nature of the expansion history for a given seed does not depend on $\Omega^0_m h^2$. Instead, for a given seed (i.e. a series of random numbers), model outputs such as $\Lambda$, $H^2$, and $1/\sqrt{V}$ scale linearly with $\Omega^0_m h^2$. In other words, for a given series of random numbers, $\Lambda$ and  $H^2$ at any redshift are linearly proportional to the value of $\Omega^0_m h^2$.  The full calculation is discussed in detail in Section 4.4 of \cite{das2023aspects1}.  So we can make the process more efficient with the following simplifications:
First, we calculate the expansion history for a fixed $\Omega^0_m h^2$, say 0.14. The histories with different values of $\Omega^0_m h^2$ can be determined by the scaling property mentioned above.
Secondly, the expansion history ($H^2 (a)$) is a slowly varying function of $\alpha$. Therefore, we calculate the expansion histories for different $\alpha$'s at a fixed interval and store them. These values are then interpolated for calculating the expansion history for any other $\alpha$ and then scaled for other $\Omega^0_m h^2$.  Here it should be noted that in our calculations, we started the Everpresent $\Lambda$ simulations from about $z = 10^5$. Therefore, there is a significant redshift range which is radiation dominated. The scaling relation holds, approximately, for a completely matter dominated or completely radiation dominated era. It does not hold during matter and radiation equality. The radiation density ($\Omega^0_r$) of the Universe is fixed from the current CMB temperature, and it does not change when we change $\Omega^0_m$. Therefore, for any $\Omega^0_m$ we can calculate the matter-radiation equality as $a_{eq}=\Omega^0_r/\Omega^0_m$ and scale the dark energy separately during the matter and radiation dominated era. As mentioned, the scaling relation of the dark energy using the above method does not work during the matter and radiation equality. However, this region is small in conformal time and does not change the power spectrum $C_\ell$, as we verified numerically.

Therefore, for any given seed, we make the fiducial choice $\Omega^0_m h^2=0.14$. We then generate expansion histories for 20 different $\alpha$'s starting from $\alpha = 0.001$ to $0.02$ at a regular interval of $\Delta\alpha=0.001$. The smallest $\alpha$ for which we get an $H^2<0$ crash is denoted by $\alpha^\text{max}$ for that seed. As $\alpha$ and $\Omega_m^0$ are inversely related, it corresponds to some ${\Omega^0_m}^\text{min}$. For any value of $\alpha$ ($\Omega^0_m$) greater (less) than $\alpha^\text{max}$ (${\Omega^0_m}^\text{min}$), the expansion history for that particular seed results in an $H^2<0$ crash (see Section 4.4 in \cite{das2023aspects1}). 
Since we only want to consider seeds that are capable of producing realistic outputs, we impose the following criteria to accept or reject a seed. Firstly, we only consider the seeds that give a present day Hubble parameter greater than 60 km/sec/Mpc for at least one choice of $\alpha$. Secondly, for large $\Omega^0_m$, the simulated CMB power spectra will not match the observational data. Therefore, for our analysis, we set an upper bound of 0.45 on ${\Omega^0_m}^\text{min}$. That is, we reject all seeds for which ${\Omega^0_m}^\text{min}> 0.45$. This also translates to a lower bound on $\alpha^\text{max}$ which is desirable since higher values of $\alpha$ have a better chance of giving the correct value of $H_0$. 
After excluding all seeds that did not meet these requirements, we were left with 3106 seeds out of our original 100,000. We can run the MCMC on this number of seeds in a reasonable time. For $\Lambda$CDM, the best fit $\chi^2$  for the CMB data is about 2339.0 for the given likelihood. Therefore as a further restriction, if after the burn-in period the $\chi^2$ for a seed is more than $3000.00$, we reject the seed. 

\subsection{Results}

\begin{figure}
    \centering
    \includegraphics[width=.98\textwidth,trim = 25 180 5 190,clip]{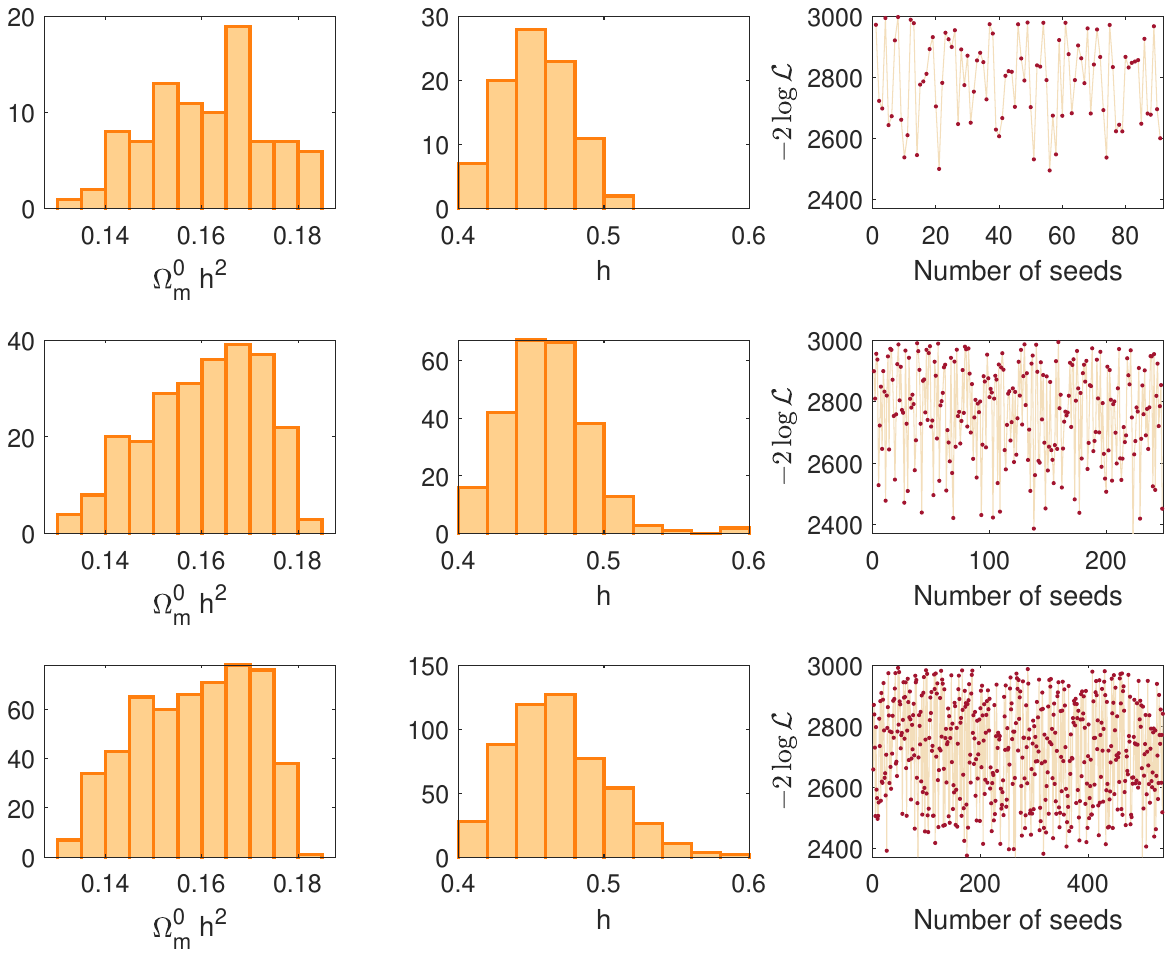}
    \caption{The results of our CMB analysis with Planck likelihood from 100,000 seeds. The first row is the output from Model 1; the second and the third row are from Model 1 with LSS suppression with $\Delta  z=100, \;200$, respectively. The first two columns are the histograms of the best fit  $\Omega^0_m h^2$ and $h$ for the seeds that give $\chi^2<3000.0$. The histograms show that $\Omega^0_m h^2$ is often higher than the standard $\Lambda$CDM value of $~0.14$, and most of the best fit $h$ are within the range $0.4\--0.5$, and it rarely goes to $0.6$.}
    \label{fig:my_model1}
\end{figure}

After reducing our initial seeds with the conditions outlined above, our MCMC analysis shows that there are 91 seeds that give a $\chi^2$ less than 3000.0.
The best fit $\chi^2$ is 2495.56, which is $\sim 157$ larger than the best fit of 
$\Lambda$CDM. In all of these cases, the best fit Hubble parameter is between 40 and 51 km/sec/Mpc. The value of $\Omega^0_m h^2$ in the MCMC chains lingers around 0.16 to 0.2 in most of the cases (82 out of 91 chains), occasionally going below 0.16 (the observed value is roughly 0.14). This tends to make $\Omega^0_m$ large and $\Omega^0_\Lambda$  small. In other words, it seems that the MCMC analysis favours small amounts of dark energy, since in most cases $\Omega^0_\Lambda \le 0.2$. However, despite this behaviour the best fit values are in some cases more realistic.

In the first row of Fig.~\ref{fig:my_model1}, we have shown some of the results of our analysis. The first and the second columns show the distribution of the best fit values of $\Omega_m^0 h^2$ and $h$ for the 91 seeds. The distribution of the best fit values of $\Omega_m^0 h^2$ have more values close to 0.14 than one might expect.  While the chains for these seeds spend more time in the range 0.16 to 0.20, they nevertheless give best fit values of $\Omega_m^0 h^2$ near 0.14.  However, it should be noted that the likelihood distributions for these seeds are not Gaussian. They are flatter, with multiple peaks. The last column shows the best fit $\chi^2$ of the final 91 seeds. Except for one seed, all others have a $\chi^2$ greater than 2500.  

\begin{figure}
    \centering
    \includegraphics[width=.8\textwidth,trim = 0 0 0 0,clip]{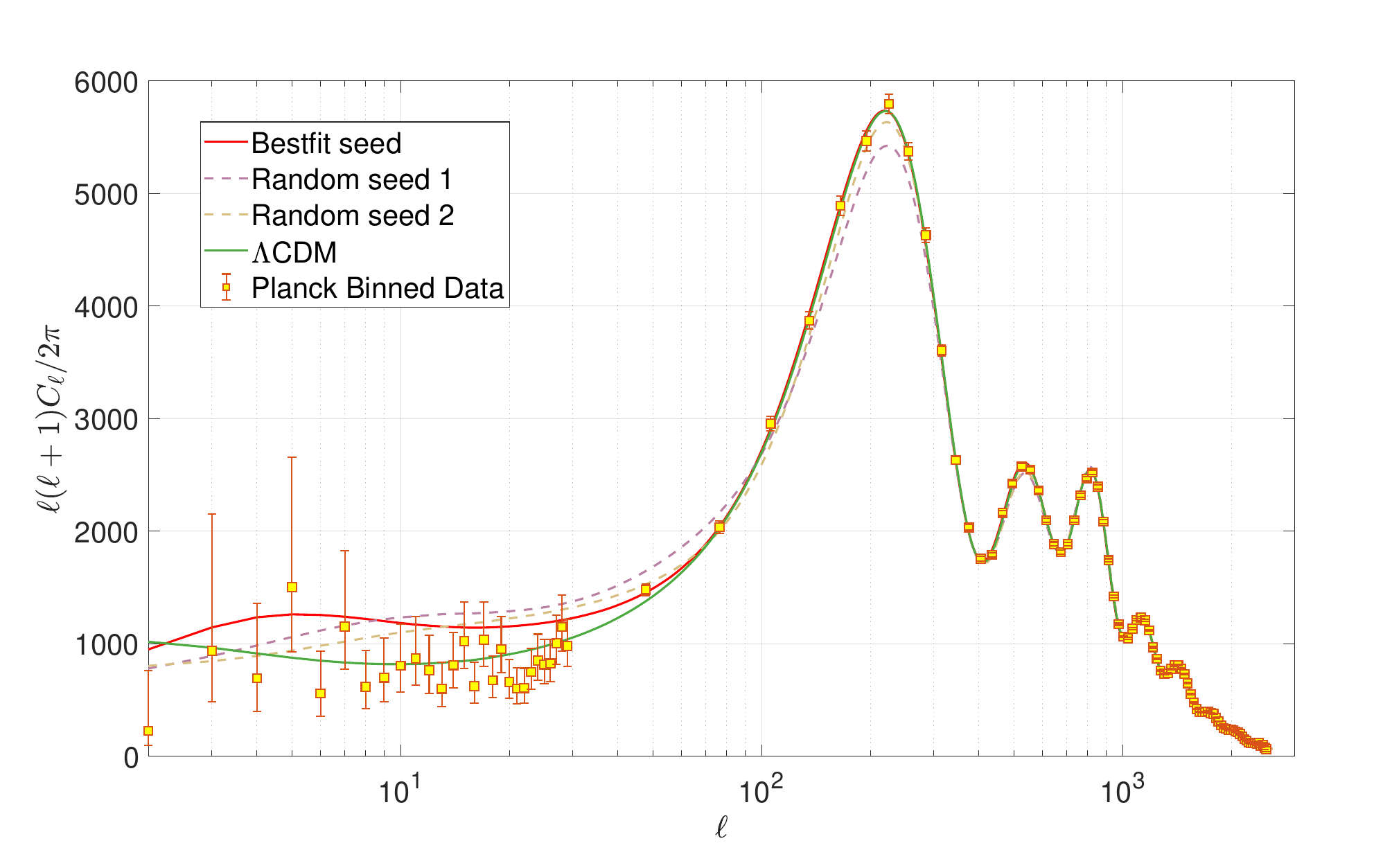}
    \caption{We show the CMB power spectrum data for different models. The green $C_\ell$ curve is the $\Lambda$CDM power spectrum. We show the best fit power spectrum for Model 1 out of 100,000 seeds in red. 
    Two other dashed lines show the best fit power spectrum for two other randomly chosen seeds from the $91$ seeds for which $\chi^2 < 3000.0$. The binned data from the Planck satellite is also displayed in the plot. For values of $\ell$ ranging from 2 to 29, the data are shown individually. After that, the data is presented in bins based on the value of $\ell$.}
    
    \label{fig:Cl_BestFit_Model_1}
\end{figure}

In Fig.~\ref{fig:Cl_BestFit_Model_1}, we compare the power spectrum from $\Lambda$CDM with the best fit values for our Model 1 analysis. We can see that  at low multipoles $C_\ell$ differs significantly. While there are departures from the best fit $\Lambda$CDM spectrum at high multipoles, especially near the peaks, it closely follows the shape. We have also shown the power spectrum for two other randomly selected seeds from the 91 seeds. They show a similar deviation from the $\Lambda$CDM trend. The Planck data is taken from~\footnote{\url{https://pla.esac.esa.int/\#cosmology}}.

Motivated by the claim in \cite{Zwane_2018} that with sufficiently small dark energy at the last scattering surface (LSS), the Everpresent $\Lambda$ models can explain the expansion history of the Universe for some seeds, we repeat our analysis with an artificial suppression of dark energy at the LSS. We implement this suppression for our  3106 seeds by multiplying the dark energy density by $1-\exp\left(-\frac{1}{2}\left(\frac{z-1100}{\Delta z}\right)^2\right)$ for $\Delta z = 100$ and $200$. This type of suppression increases the number of seeds that give a $\chi^2<3000$. For $\Delta z = 100$ and 200, we get 248 and 539 seeds respectively that yield a $\chi^2$ less than 3000. The best fit parameters and $\chi^2$ for both the analyses are given in Table~\ref{tab:table1}. While the $\chi^2$ remains greater than the $\Lambda$CDM value, it is interesting to note the large decrease ($\sim 136$) in the best fit $\chi^2$ the LSS suppression resulted in. 
The histograms for $\Omega^0_m h^2$ and $h$, and $\chi^2$ for the best fit for each seed are shown in the second and the third row of Fig.~\ref{fig:my_model1}. 
We see that increasing $\Delta z$ in the LSS supression, has the effect of giving more results with the observed value of 0.14 for $\Omega^0_m h^2$, as well as stretching the tail of $h$ to higher values, so that it comes closer to the accepted value.  We therefore partially confirm the claims of \cite{Zwane_2018}, although we still do not yet get better $\chi^2$ values than $\Lambda$CDM.
Besides the improvements in the fits to the CMB data, the analysis shows that $\Omega^0_\Lambda$ is small for most seeds (although the seeds with the best overall $\chi^2$ do have reasonable values of $\Omega^0_\Lambda$), i.e. the MCMC process tries to remove the dark energy component.  In summary, our analysis shows that our implementation of Model 1 of Everpresent $\Lambda$ taken together with the set of constraints and assumptions that we used to run the MCMC process, does not explain the CMB data at a level comparable to the standard model.

We also carried out tests using Model 2 of Everpresent $\Lambda$, with 3000 seeds. For each of the seeds, the value of $\tilde{\alpha}$ goes below $0.1$ or alternately $\Omega_m^0$ becomes higher than $0.9$. The Hubble parameter is close to $40$, which is too small. As $\tilde{\alpha}$ tends to a very small value, the value of $\mu$ becomes irrelevant. Therefore, $\mu$ does not converge to a fixed value for any of the seeds, leading the chains to not converge. 

In \cite{Zwane_2018} Model 2 was implemented using a modified equation of state parameter in CAMB. Their equation of state parameter $w(z)$ was restricted to a narrow negative range,  corresponding to a narrow positive energy density range within the allowed variation range of the model \cite{privatecomm}. This is a major point of departure from our treatment, in which we allow the dark energy density to fluctuate both in sign and in magnitude within the full range that the model allows it to. Therefore, our results should not be compared to those of this previous work. 
\begin{table}[]
\centering
\begin{tabular}{|p{0.20\columnwidth}|p{0.23\columnwidth}|p{0.23\columnwidth}|p{0.23\columnwidth}|}
\hline \rule{0pt}{0.4cm} &Model 1 & Model 1 with LSS suppression ($\Delta z = 100$) & Model 1 with LSS suppression ($\Delta z = 200$)\\[.5cm]
\hline \rule{0pt}{0.4cm} $\#seed_{-2 \log \mathcal{L}<3000.0}$ & $91$ & $248$  & $539$ \\[0.1cm]
\hline \rule{0pt}{0.4cm} $-2 \log \mathcal{L}$ & $2495.6$ &  $2360.5$ &  $2358.5$\\[0.1cm] 
\hline
\hline  \multicolumn{1}{|c|}{\rule{0pt}{0.4cm} } & \multicolumn{3}{c|}{ Best fit parameters  } \\[0.1cm]
\hline \rule{0pt}{0.4cm} $\Omega^0_m h^2$ & $0.1408$ & $0.1409$  & $0.1430$ \\[0.1cm]
\hline \rule{0pt}{0.4cm} $\Omega^0_b h^2$ & $0.0223$ &  $0.0222$ &  $0.0221$\\[0.1cm] 
\hline \rule{0pt}{0.4cm} $H_0$ & $43.27$ & $43.20$  & $42.68$ \\[0.1cm]
\hline \rule{0pt}{0.4cm} $\kappa$ & $0.0866$ &  $0.1319$ &  $0.1285$\\[0.1cm] 
\hline \rule{0pt}{0.4cm} $n_s$ & $0.9622$ & $0.9777$  & $0.9632$ \\[0.1cm]

\hline \rule{0pt}{0.4cm} $A_s$ & $3.0892$ &  $3.1918$ &  $3.1875$\\[0.1cm] 
\hline \rule{0pt}{0.4cm} $\alpha$ (derived) & $0.0040457$ &  $0.0036563$ &  $0.0025567$\\[0.1cm] 
\hline
\end{tabular}
    \caption{We tabulated the results of the CMB simulations with Planck likelihood from 100,000 seeds. We show the results for Model 1 and Model 1 with LSS suppression. For comparison, the $-2\log\mathcal{L}$ for $\Lambda$CDM is $\sim 2339.0$.}
    \label{tab:table1}
\end{table}

\section{Constraining Dark Energy Using CMB}\label{sec:cmb_constraints}

A key property of Everpresent $\Lambda$ is that the dark energy density is of the order the background energy density throughout the expansion history of the Universe. Models 1 and 2 capture this behaviour in a practical manner. However, both of these models are phenomenological, as we require a more developed theory of quantum gravity to inform the correct dynamics for Everpresent $\Lambda$. Furthermore, as discussed in the previous section, the present models do not fit the CMB data. Therefore, in this section, we try to understand how the variation in dark energy in different eras affect the CMB power spectrum. These studies can give an indication of the constraints on a fluctuating dark energy density. It can also inform how future models may need to be modified.

For our analysis, we only use the scalar power spectra. No tensor modes are considered. Ignoring CMB lensing and other higher-order effects, the CMB scalar power spectrum can be written as~\cite{Seljak_1996,CMBAns1910}

\begin{equation} \label{scalarpower}
C_\ell^{X X}=(4 \pi)^2 \int k^2 \mathrm{~d} k P(k)\left[\Delta_{X \ell}(k)\right]^2\,, \quad C_\ell^{T E}=(4 \pi)^2 \int k^2 \mathrm{~d} k P(k) \Delta_{T \ell} \Delta_{E \ell}\,,
\end{equation}

\noindent where $X\in (T,E)$. $P(k)$ is the primordial power spectrum from inflation and does not depend on the expansion history of the Universe. $\Delta_{X \ell}(k)$ are the brightness fluctuation functions for the temperature and the $E$-mode polarization power spectra and are given by  

\begin{equation}
\Delta_{T \ell}(k)=\int_0^{\tau_0} \mathrm{~d} \tau S_T(k, \tau) j_\ell(k(\tau_0-\tau)), \quad \Delta_{E \ell}(k)=\sqrt{\frac{(\ell+2) !}{(\ell-2) !}} \int_0^{\tau_0} \mathrm{~d} \tau S_P(k, \tau) j_\ell(k(\tau_0-\tau))\,.
\label{brightfunc}
\end{equation}

\noindent $j_\ell(x)$ are the spherical Bessel functions. $S_T(k, \tau)$ and $S_P(k, \tau)$ are known as the temperature and polarization source terms. These are the terms dependent on the cosmological perturbations and the expansion history of the Universe at different epochs. The source terms for the temperature and the $E$-mode polarization power spectra are given by  

\begin{eqnarray}
&S_T(k, \tau)=-g\left(\Delta_{T 0}+2 \dot{\vartheta}+\frac{\dot{\theta}_b}{k^2}+\frac{\Pi}{4}+\frac{3 \ddot{\Pi}}{4 k^2}\right)+e^{-\kappa}(\dot{\eta}+\ddot{\vartheta})+\dot{g}\left(\frac{\theta_b}{k^2}+\vartheta+\frac{3 \dot{\Pi}}{4 k^2}\right)+\frac{3 \ddot{g} \Pi}{4 k^2} \label{Tsource}\\
&S_P(k, \tau)=\frac{3 g \Pi(\tau, k)}{4 k^2\left(\tau_0-\tau\right)^2} \label{Esource}.
\end{eqnarray}

\noindent Overdots $\dot{(\;)}$ represent derivatives with respect to the conformal time $\tau$.  The temperature source term consists of 4  terms. The first term is the Sachs Wolfe (SW) term.  The second term is the integrated Sachs Wolfe (ISW) term.  The third is the velocity term. The last is the contribution from anisotropic stress, which is negligible compared to the other three terms (at least for standard $\Lambda$CDM). Here we can see that except for the ISW term, all other terms depend on the visibility function $g$, where $g=-\dot{\kappa} \exp (-\kappa)$. $\kappa=\int_\tau^{\tau_0} a\, n_e \,\sigma_T \,\mathrm{~d}\tau$, is the optical depth at $\tau$.  

All equations are written in the synchronous gauge. $\vartheta$ is given by $\vartheta=(\dot{h}+6 \dot{\eta}) / 2 k^2$, where $h$ and $\eta$ are the metric perturbation variables ($h$ here not to be confused with Hubble parameter). $\Pi$ is the anisotropic stress (plus polarization) term and is given by 

\begin{equation}
\Pi=\Delta_{T 2}+\Delta_{P 2}+\Delta_{P 0}\,.
\end{equation}
$\Delta_{P,T\, \ell}$ for $\ell\ge 0$ are the moments of the temperature and polarization perturbation variables (brightness fluctuation functions). 

The rest of the variables are perturbation variables for various components of the Universe. We take the Universe to consist of 5 components -- Cold dark matter, baryons (which includes leptons except for neutrinos), neutrinos, photons, and dark energy. Dark matter and dark energy do not couple to anything except through gravity. Neutrinos were coupled with baryons through the weak interaction before neutrino decoupling. However, our domain of interest for CMB calculations is much later than neutrino decoupling, and therefore neutrinos freestream through the medium. $\theta_b = \vec{k}.\vec{v_b}$, where $\vec{v_b}$ is the baryon velocity perturbation. 

Before the LSS, the optical depth is very high, causing $\exp (-\kappa)$ to decay deep inside the LSS. After the LSS  $\dot{\kappa}$ decays away due to the lack of free electrons. Therefore, $g=\Dot{\kappa}\exp(-\kappa)$ is only nonzero during last scattering and reionization, although the value during reionization is less than a 100th of the value at the LSS~\cite{CMBAns1910}. The contributions from the Sachs Wolfe and velocity term are only important before and during the recombination and marginally during reionization. The ISW, on the other hand, is important throughout the expansion history of the Universe since LSS.

From \eqref{Tsource} and \eqref{Esource} we can see that the expansion history of the Universe can be broken into three parts depending on how it affects the cosmological perturbations \-- 1. During recombination, 2. after recombination, and 3. before recombination. 
In the following three subsections, we first discuss what changes we can expect qualitatively in the CMB power spectrum if we increase or decrease the dark energy content in different sections of the Universe. We then numerically constrain the variation allowed in the expansion history before and after the last scattering surface to keep the $C_\ell$ unchanged. While other analyses have attempted to examine the permitted variations in the expansion history of the late-time universe~\cite{Bernal2016}, to the best of our knowledge, no previous studies have conducted such an extensive, model-independent analysis of the allowed expansion history throughout the entire history of the universe.

\subsection{Recombination}

In \eqref{Tsource} and \eqref{Esource}, we can see that all the terms except the ISW term are multiplied either by the visibility function $g$, or one of its derivatives $\dot{g}$ and $\ddot{g}$. The visibility function mainly depends on the background expansion history during recombination. However, the perturbation parameters inside the brackets of the SW, velocity, and ISW terms depend on the perturbations and the nature of dark energy even long before the LSS. Therefore, for a particular $k$ mode, the changes (due to a non-negligible dark energy component during recombination) in the perturbations inside the brackets, in general, cannot get cancelled by some change in $g$ or its derivatives. The variation of $g$ only depends on the background expansion and is not correlated with the modification in the perturbation equations. 

Suppose we add a non-negligible dark energy component during the recombination era. We would like to understand the effect on $g$ due to such a change. Recall that $g$ is given by 

\begin{equation}
    g=-\dot{\kappa}\exp(-\kappa) =\frac{d}{d\tau}\left( e^{-\kappa}\right)\qquad\implies\qquad  \int g \ d\tau =e^{-\kappa}\,. 
\end{equation}

\noindent Long before recombination $\kappa$ was very large, resulting in $\exp(-\kappa)\rightarrow 0$. After recombination, $n_e=0$ and hence $\kappa$ will become constant (without reionization, it would be $0$). Therefore the integral of $g$ during recombination is a fixed constant that is insensitive to events away from the LSS. Next, we would like to understand how a modified dark energy density changes the position of the peak and the width of $g$.

For the peak of $g$, we can use the Saha equation. For a gas composed of a single atomic species, the Saha equation is written as:

\begin{equation}
\label{saha}
\frac{n_{i+1} n_e}{n_i}=\frac{2}{\lambda^3} \frac{g_{i+1}}{g_i} \exp \left[-\frac{\left(\epsilon_{i+1}-\epsilon_i\right)}{k_B T}\right],   
\end{equation}

\noindent where $n_i$ is the density of atoms in the $i$-th state of ionization, that is with $i$ electrons removed, $g_i$ is the degeneracy of states for the $i$-ions, $\epsilon_i$ is the energy required to remove $i$ electrons from a neutral atom, creating an $i$-level ion, $n_e$ is the free electron density, and $\lambda$ is the thermal de Broglie wavelength of an electron
\begin{equation}
\lambda \stackrel{\text { def }}{=} \sqrt{\frac{2\pi}{ m_e k_B T}},
\end{equation}
where $m_e$ is the mass of an electron, and $T$ is the temperature of the gas. 
Hence from the Saha equation, we can see that $n_e$ depends only on the temperature, which is inversely proportional to the scale factor. If we change the dark energy fraction, the dependence of the ionization fraction $(x_e)$ on the scale factor will not change. Even from the Peebles or the Recfast recombination~\cite{Seager_1999}, the ionization fraction as a function of redshift will vary only slightly, which we have verified numerically. The position of the visibility peak, therefore, remains the same in $a$. In conformal time $\tau$, it shifts as $\tau$ changes as a function of $a$. An increase in dark energy density makes $H(z)$ larger, and hence the same scale factor corresponds to an earlier conformal time. A negative shift in the dark energy density moves the peak to a larger conformal time.

\begin{figure}
    \centering
    \includegraphics[width=.49\textwidth,trim = 150 420 200 420,clip]{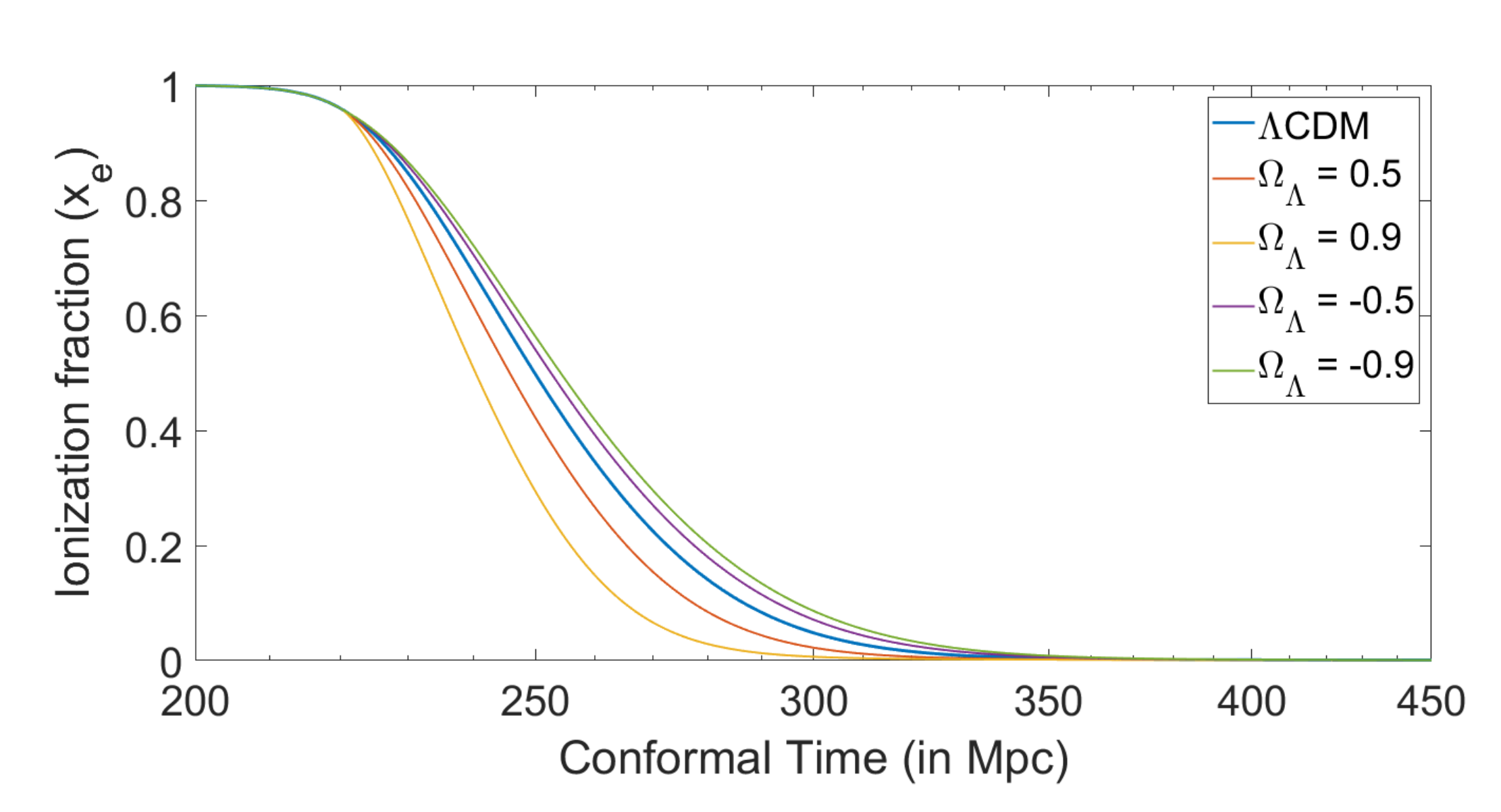}
    \includegraphics[width=.49\textwidth,trim={180 420 200 420},clip]{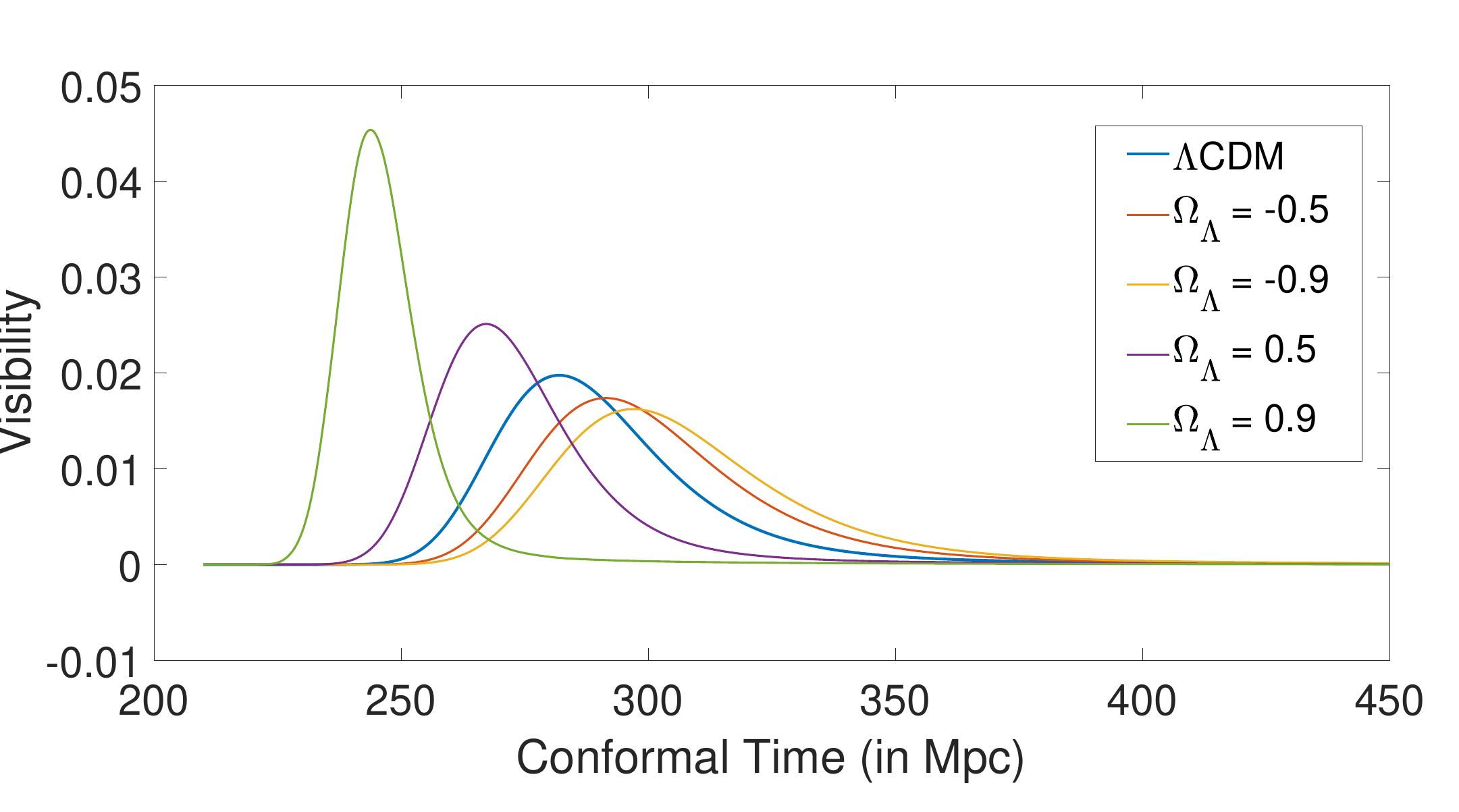}
    \includegraphics[width=.49\textwidth,trim = 150 420 200 420,clip]{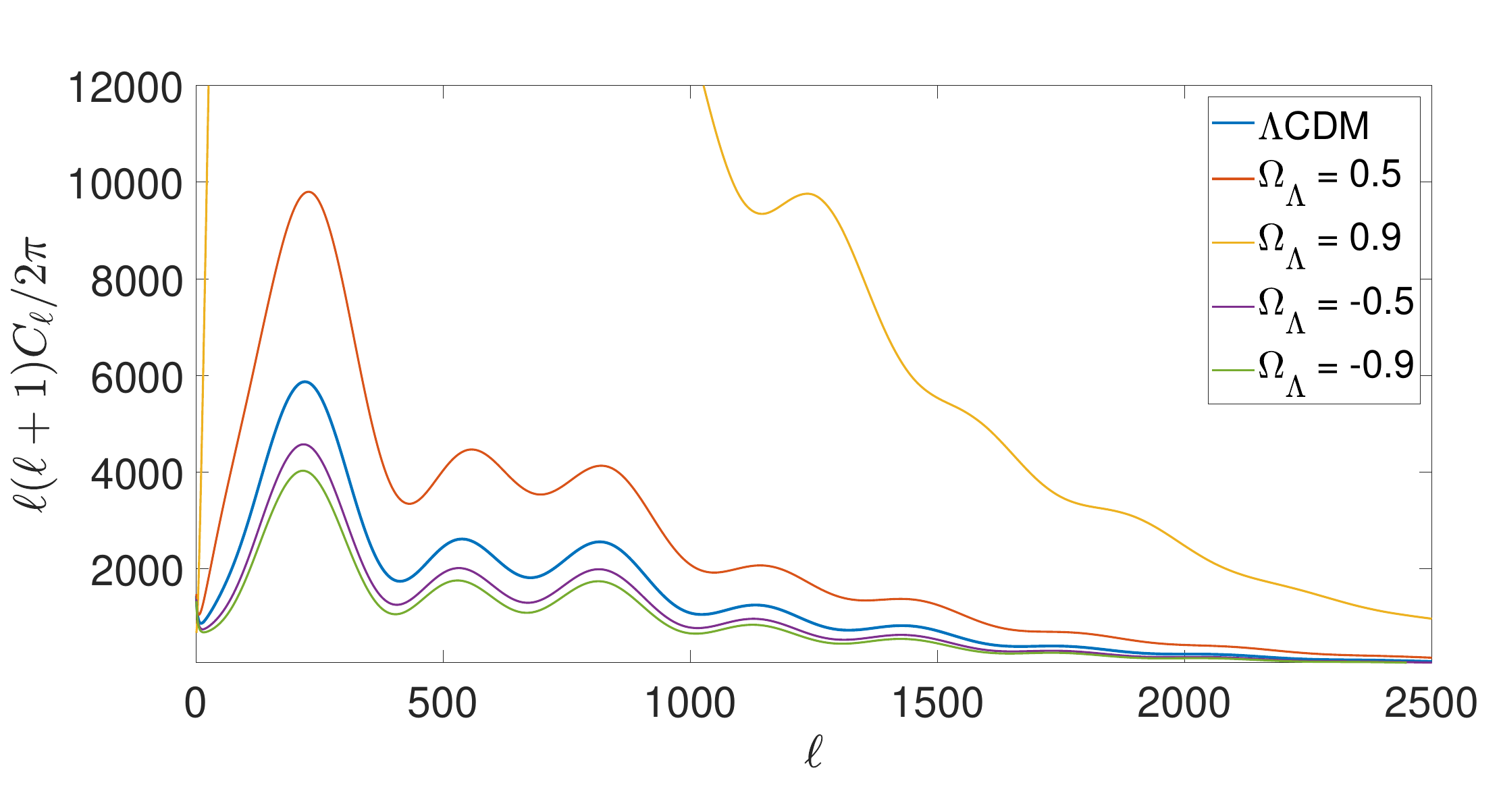}
    \includegraphics[width=.49\textwidth,trim={160 420 200 420},clip]{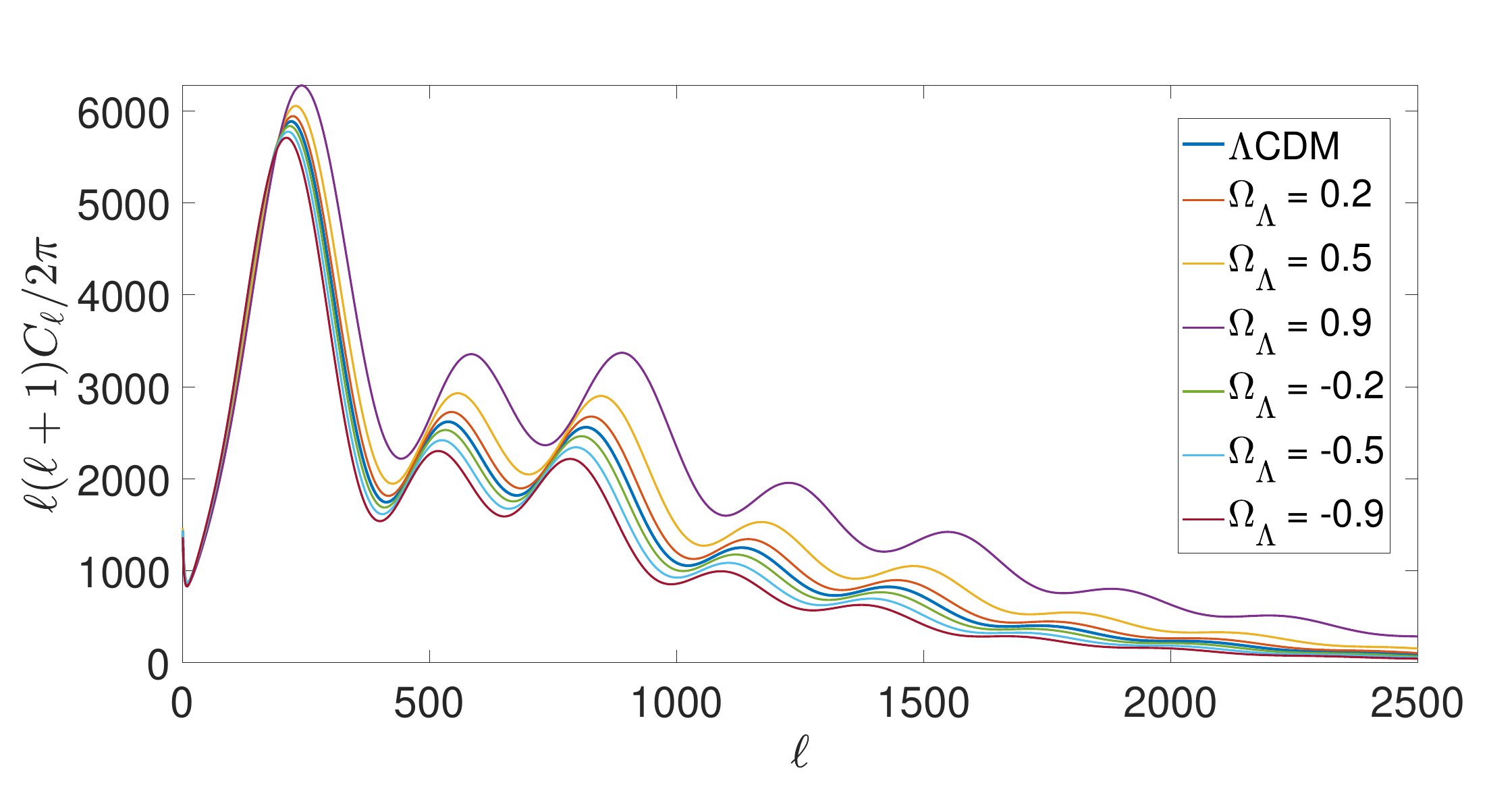}
    \caption{Top-left: The ionization fraction $(x_e)$ as a function of conformal time for different dark energy fractions. Top-right: The visibility function $(g)$ as a function of conformal time, Bottom-left: $C_\ell$ for different dark energy fractions between $z=900$ and $1300$. Bottom-right:  $C_\ell$ for different dark energy fractions between $z=900$ and $1300$ provided the dark energy only changes the ionization fraction. }
    \label{fig:recombination_4pot}
\end{figure}

In Appendix \ref{appendixB}, we show that with some approximations, the width of the visibility function in conformal time is given by

\begin{equation}
\frac{z_*^2}{10^4 H_*}.
\end{equation}
The starred quantities are evaluated at the peak of the visibility function. By adding a positive dark energy component during recombination, $H_*$ increases. As a result, the width of $g$ decreases, and $g$  becomes sharper, keeping the area under its curve the same. A negative shift in dark energy has the opposite effect.

Changing the height of the visibility function, $g$, modifies the power in $C_\ell$, which is eventually reflected in the $A_s$ parameter. A change in the conformal time of the peak of $g$ moves the CMB power spectrum to a higher or lower $\ell$. If the spread of $g$ changes, it modifies the ratio of the peaks as the powers of the SW term and the velocity term change differently.

These effects are shown in Fig.~\ref{fig:recombination_4pot}. For this analysis, we keep $\Omega_\Lambda(z)$ constant around recombination between redshifts $z=900$ and $1300$. For other redshifts, we keep $\Lambda$ constant such that $\Omega_\Lambda^0 = 0.7$. 
In the top-left plot of Fig.~\ref{fig:recombination_4pot}, we show how the ionization fraction $(x_e)$  as a function of conformal time changes if we modify the expansion history. As we add additional positive or negative dark energy, $\dot{a}$ increases or decreases; $x_e$ then gets stretched or squeezed as a function of the conformal time (as argued before, the dependence of $x_e$ on the scale factor will stay the same). The top-right plot shows how the visibility function changes as a function of the conformal time.  Note how the larger $\Omega_\Lambda$ is, the further the peak of $g$ moves to the left and increases in magnitude, such that the area under the curve is conserved. The bottom-left plot shows the effects on $C_\ell$ from changing the expansion history during  recombination (redshift 900 - 1300). 

Changing the expansion history during recombination changes its conformal time and also the perturbation equations to some extent. As there is a line-of-sight integration over conformal time, it affects the final power spectrum. 
Therefore, the changes we observe in $C_\ell$  are the combined effect of the shift in recombination time, and the SW, velocity, and the early ISW terms. 
 
To isolate the effect of recombination, we calculate the ionization fraction $(x_e)$, $\kappa$, $g$, $\dot{g}$ and $\ddot{g}$ for different expansion histories, then replace them in CAMB as a function of conformal time,  and calculate the CMB power spectrum. By doing this, the cosmological perturbations do not get affected by the expansion history modification and it helps us understand the sole effect of changing the visibility function and its derivatives during recombination. In the bottom-right plot, we plot the $C_\ell$ for this analysis. The size and positions of the peaks change, though the impact is less dramatic than in the previous case. The positions of the peaks change due to the change in the peak of the visibility function in conformal time. Also, with decreasing $\Omega_\Lambda$ and therefore increased spreading of $g$, the small fluctuations in the CMB get washed out due to diffusion damping. This decreases the power at higher multipoles as can be seen in the figure.

\subsection{After Recombination}

In this section we first review the ISW effect, which is the main component of the temperature source term that is sensitive to changes after recombination.  In the second subsection, we investigate the allowed variation of $\Lambda$ at late times. 

\subsubsection{Understanding the ISW Effect}\label{sec:isw}
The ISW effect is mainly sensitive to the late time expansion history of the Universe, especially the expansion history during recombination (early ISW) and after recombination (late ISW). It modifies the CMB temperature power spectrum at low multipoles. \eqref{Tsource} and \eqref{Esource} show that the ISW effect only impacts the temperature source term, and has no polarization component~\cite{Das_2014,Das:2013gta,Mukherjee:2014wva}. This property can distinguish the ISW effect from the effect of the inflationary power spectrum, which affects both the temperature and polarization source terms at low multipoles. Apart from that, the power at low multipoles can also be affected by observational artifacts such as beam functions and scan patterns which also need to be adequately accounted for in any analysis. 

As introduced in \eqref{Tsource}, the ISW source term is given by $e^{-\kappa}(\dot{\eta}+\ddot{\vartheta})$, where  $\vartheta$ is given by $\vartheta=(\dot{h}+6 \dot{\eta}) / 2 k^2$. While the metric perturbation variable $h$ is mainly related to the density perturbations of different matter and energy components, the variable $\eta$ is related to the velocity perturbations and is an order of magnitude smaller than $h$. Therefore, the major contribution to the ISW source term comes from $\ddot{\vartheta}$. $\vartheta$ satisfies the equation~\cite{ma1995cosmological,CMBAns1910}

\begin{equation}
\label{thetadouble}
\ddot{\vartheta}=-\frac{3 \dot{\sigma}}{2 k^2}+\dot{\eta}-2\left(\frac{\dot{a}}{a}\right) \dot{\vartheta}-2 \frac{d}{d \tau}\left(\frac{\dot{a}}{a}\right) \vartheta .
\end{equation}

\noindent Here, $\sigma$ is the shear term. Its contribution is small after LSS. Empirically $\dot{\eta}$ is also much smaller than $\vartheta$ and its derivatives as described before. Therefore, ignoring these two terms, the remaining part of the equation can be written as
\begin{equation}
\ddot{\vartheta}\simeq-2 \frac{d}{d \tau}\left(\frac{\dot{a}}{a} \vartheta\right) .
\end{equation}

Using separation of variables we can break $\vartheta$ into a $\tau$ dependent part and a $k$ dependent part as
\begin{equation}
\vartheta(k, \tau)=\vartheta_\tau(\tau) \vartheta_k(k) .
\end{equation}

\noindent $\vartheta_k(k)$ does not depend on the evolution history; hence, it remains almost unchanged throughout the expansion history of the Universe. A small contribution may come from the $\dot{\sigma}$ and $\dot{\eta}$ terms as their evolution depends on the wave number~\cite{CMBAns1910,ma1995cosmological}. However, these contributions are negligible. 

The evolution of $\vartheta_\tau$ in the conformal time domain can be written as 
\begin{equation}
\ddot{\vartheta}_\tau=-2 \frac{d}{d \tau}\left(\frac{\dot{a}}{a} \vartheta_\tau\right) \quad\;\implies \quad\;
\dot{\vartheta}_\tau+2 \frac{\dot{a}}{a} \vartheta_\tau = K\,.
\end{equation}

\noindent Here, $K$ is an integration constant. The equation shows that  $\vartheta_\tau$ is directly related to the expansion history of the Universe and has almost no dependence on the perturbations. Therefore, provided the initial conditions of $\vartheta_\tau$ and  $\vartheta_k(k)$ are known, the ISW source terms only depend on the expansion history. 

\begin{figure}
    \centering
    \includegraphics[width=.99\textwidth,trim = 10 20 30 20,clip]{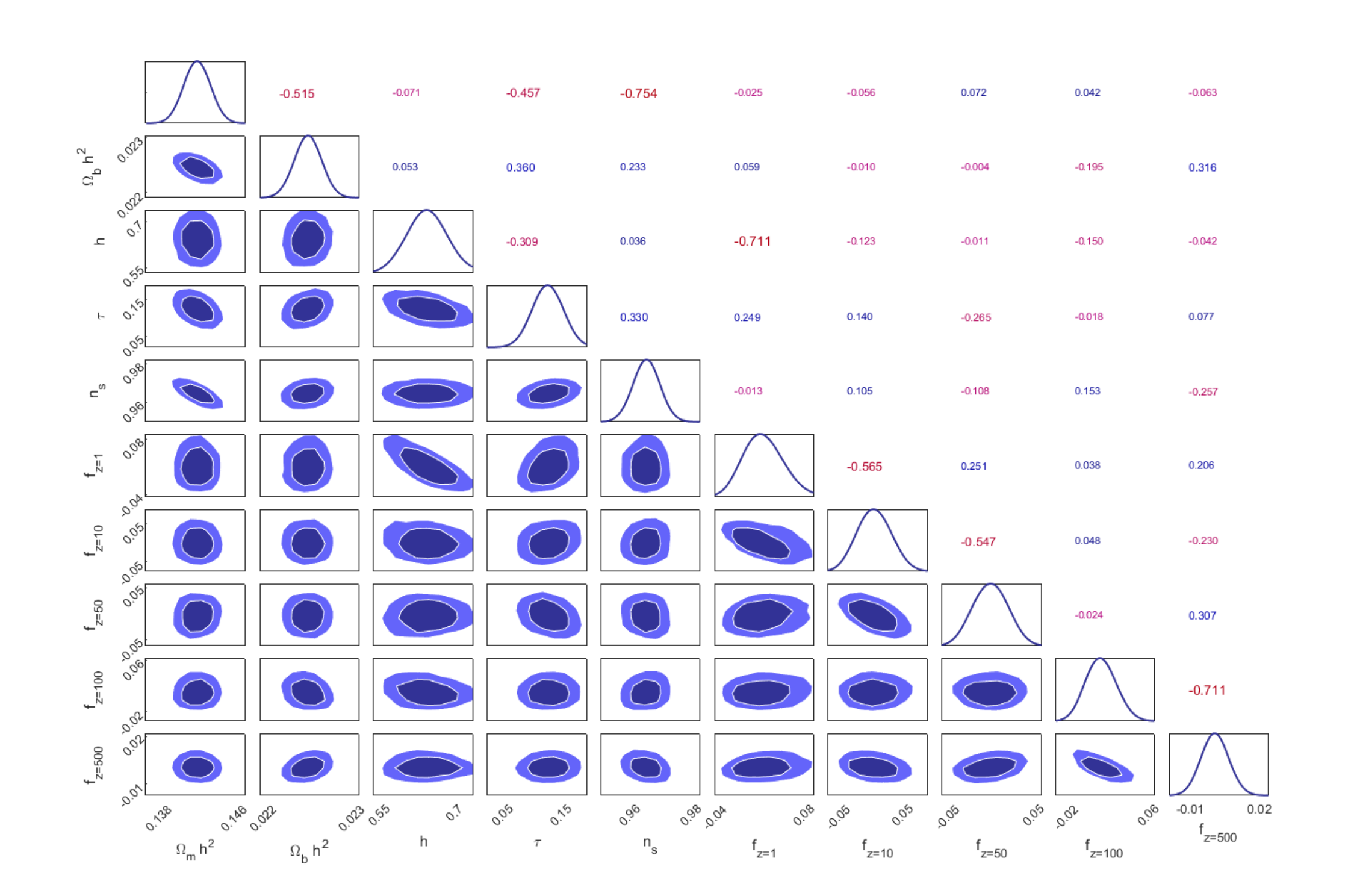}
    \caption{Results from an 11 parameter MCMC analysis constraining the expansion history after the last scattering surface, are shown. $A_s$, which just gives a scaling on the CMB power spectrum, is omitted to make the plot less cluttered. The lower triangle panels
show plots of the 68\% and 95\% confidence contours for pairs of parameters. The upper triangle shows the correlation-coefficient between the pair of parameters. The negative correlation-coefficient are in red and the positive correlation-coefficients are shown in blue. The diagonal plots are the 1-dimensional marginalized distribution of the parameters. The average, standard deviation and the best fit values are stated in Table~\ref{tab:my_label}.}
    \label{fig:ISW2D}
\end{figure}

By substituting the dominant part of the ISW source term into the equation \eqref{brightfunc} for the temperature brightness fluctuation function $\Delta_{T \ell}(k)$, and then substituting the brightness fluctuations into \eqref{scalarpower}, we obtain the power spectrum from the ISW term as
\begin{equation}
C_\ell^{T T, ISW}=(4 \pi)^2 \int \mathrm{~d} k\ k^2  P^s(k)\vartheta_k(k)^2\left[\int_0^{\tau_0} \mathrm{~d} \tau e^{-\kappa}\ddot{\vartheta}_\tau j_\ell(k(\tau_0-\tau))\right]^2 \,.
\end{equation}

\noindent For any $\ell$, the spherical Bessel function $j_\ell(x)$ has its first peak near $x = \frac{\pi}{2}\ell$. If we add positive energy density to the dark energy sector at time $\tau_1$ then it increases the amplitude of the ISW source term at $\tau_1$. This boosts the brightness function $\Delta_{T \ell}(k)$ at $k(\tau_1,\ell) \approx \frac{\pi \ell}{2 (\tau_0-\tau_1)}$, as the maximum contribution to the brightness fluctuation function comes from the first peak of $j_\ell(x)$.
In order for Everpresent $\Lambda$ to be a good cosmological model, it should not deform the power spectrum away from the the standard one. Therefore, if the expansion history at a particular redshift distorts a particular multipole, say $\ell$, of the power spectrum, then the expansion history should compensate for it at another redshift to obtain a power spectrum compatible with observation. However, this is not possible if the modification of the expansion history is significant. To see this, we can use the properties of the spherical Bessel functions.  As we want to keep the power at $C_\ell$ the same, we need to keep the area under the brightness function the same. For this, we must decrease the amplitude of the brightness fluctuation function at some other $k$, which must be done by adding a negative energy density in the dark energy at a $\tau \approx \tau_0 - \frac{\pi \ell}{2 k}$. This allows us to keep the power fixed at the particular $C_\ell$. However, while we are preserving the power at a particular $\ell$, these modifications most certainly change the power at other multipoles. This is because to keep the power unchanged at another multipole, say $\ell'$, we need to lower the power at $\tau \approx \tau_0 - \frac{\pi \ell'}{2 k}$. As we only have a single $\vartheta_T$, if we keep the power fixed at $\ell$ then the power at $\ell'$ is bound to change. Here we have assumed that to keep $C_\ell$ unchanged, the brightness fluctuation function has to be unchanged. While this assumption is true in general, this is not a mathematical proof of the fact that there exists no expansion history that can keep the $C_\ell$ unchanged. There may exist some inflationary power spectrum ($P^s(k)$) or $\vartheta_k(k)$, such that the integration over $k$ counteracts an altered brightness function to give the same $C_\ell$. This, however requires the primordial power spectrum and expansion history before recombination to be correlated with the expansion history after recombination, which is not realistic.
 Therefore, if we try to modify the expansion history, then that change will reflect on $C_\ell$. A random choice of expansion history proposed by Everpresent $\Lambda$ will certainly alter the $C_\ell$. 
However, if we make slight modifications in the expansion history, it can cause small deformations in the temperature power spectrum, and the change in $\chi^2$ would be small. It may give an acceptable $C_\ell$ or even a $C_\ell$ with better likelihood than the $\Lambda$CDM. 

\begin{figure}
    \centering
    \includegraphics[width=.49\textwidth,trim = 350 380 400 730,clip]{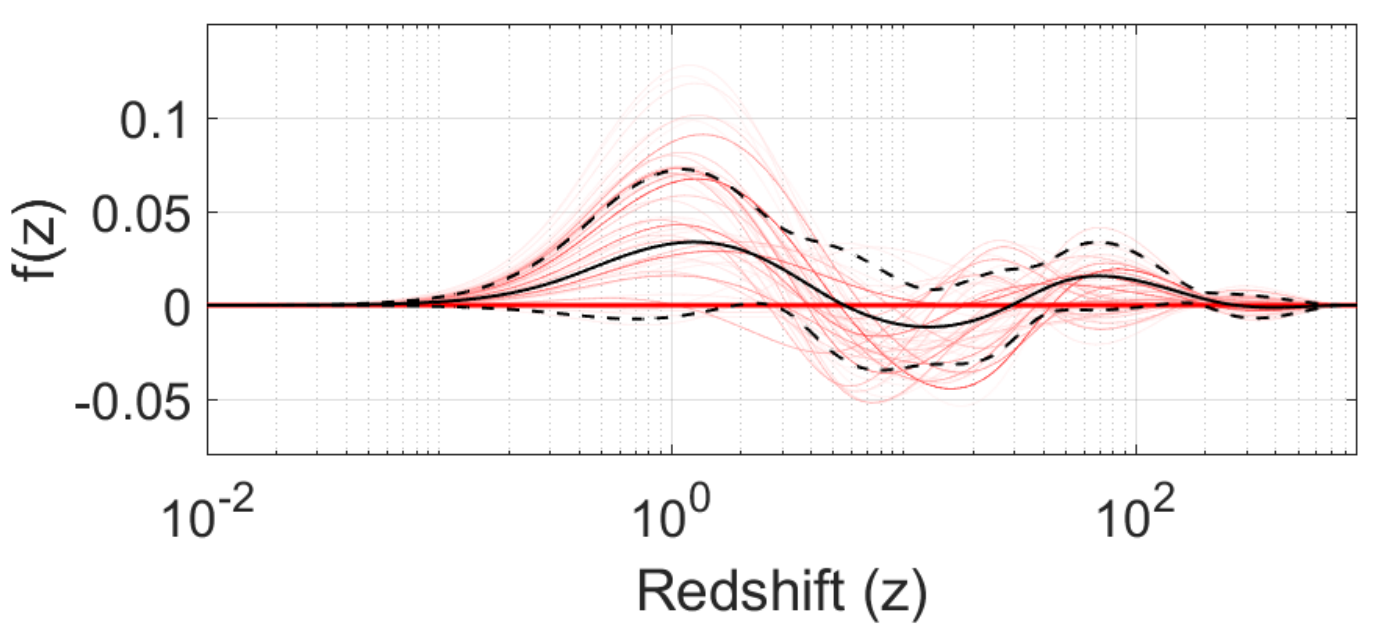}    
    \includegraphics[width=.49\textwidth,trim = 350 700 400 410,clip]{CMBFigs/OmegaLambda_new1.pdf}    
    \caption{A few random expansion histories from the $200$ best fit $\chi^2$ are shown. The opacity of each curve is proportional to its likelihood. The thick red line shows the standard $\Lambda$CDM values for reference. The mean and the standard deviation are shown in solid black and dashed black curves.  
    Left: plot of $f(z)$. We keep the $f(z)$ and its derivatives at redshifts $z=0$ and $z=1100$ to be $0$. A bumpy feature at low redshift and a dip at higher redshift give a better likelihood than standard $\Lambda$CDM. Right: $\Omega_\Lambda$ is shown for different $f(z)$ from the left plot. }
    \label{fig:ISW_f_omega}
\end{figure}

\begin{figure}
    \centering
    \includegraphics[width=.49\textwidth,trim = 360 365 400 710,clip]{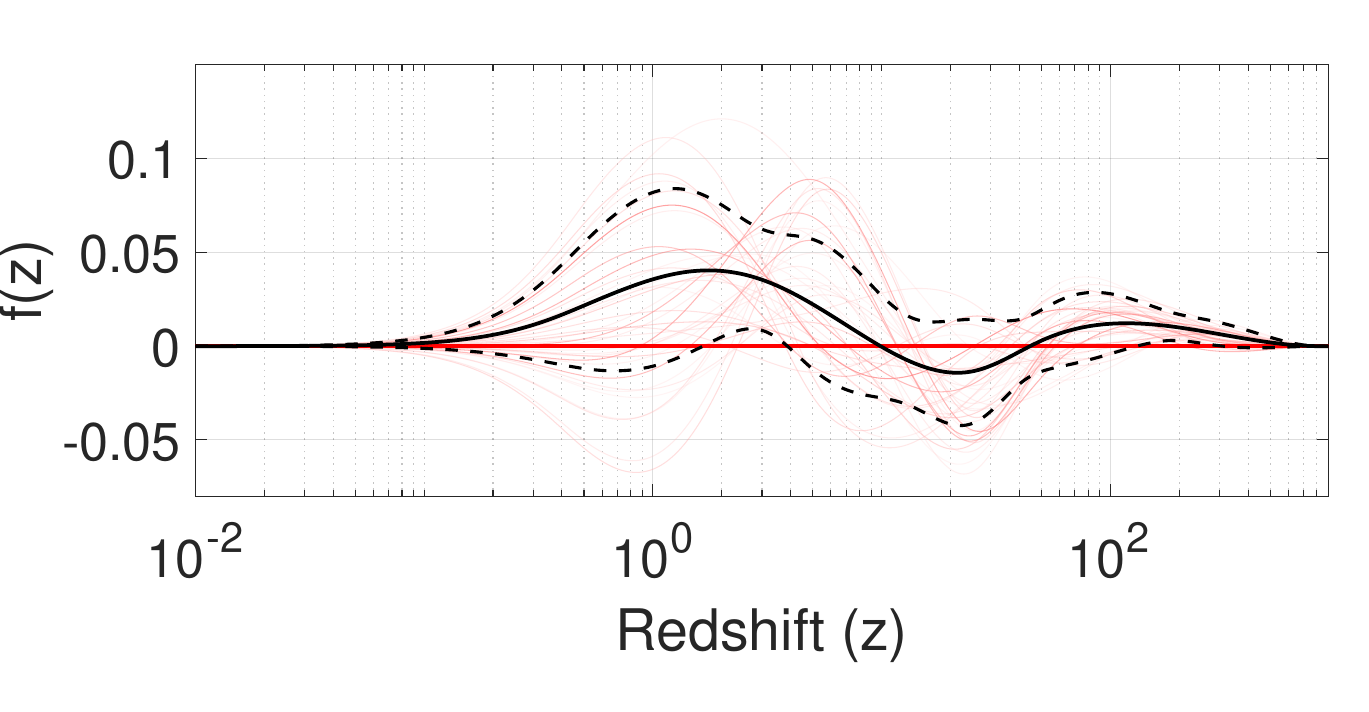}    
    \includegraphics[width=.49\textwidth,trim = 360 680 400 390,clip]{CMBFigs/OmegaLambda_new_anotherset.pdf}
    \caption{A few random expansion histories from the $200$ best fit $\chi^2$ for a different set of $z$ points (i.e. the points where $f(z)$ is a free parameter) are shown. The opacity of a curve is proportional to its likelihood. The thick red line shows the standard $\Lambda$CDM values for reference. The mean and the standard deviation are shown in solid black and dashed black curves.  
    Left: plot of $f(z)$. We keep the $f(z)$ and its derivatives at redshifts $z=0$ and $z=1100$ to be $0$. A bumpy feature at low redshift and a dip at higher redshift give a better likelihood than standard $\Lambda$CDM. Even though the $z$ points are different the space is similar to that in Fig.~\ref{fig:ISW_f_omega}. Right: $\Omega_\Lambda$ is shown for different $f(z)$ from the left plot.}
    \label{fig:ISW_f_omega_1}
\end{figure}

\subsubsection{Numerical Analysis} \label{sec:num_analysis}
We do an MCMC analysis to constrain the allowed variation in the dark energy density after recombination. We select $5$ redshifts $z_i$ between $0$ and $1100$: $1$, $10$, $50$, $100$ and $500$. We allow the dark energy to vary randomly at these five redshifts and then fit a spline, $f(z)$, through the 5 points. At redshifts $1$ and $1100$ we fix the spline and its derivatives to $0$. The dark energy density fraction is taken to be
\begin{equation}
    \Omega_\Lambda = \frac{\left[\Omega_{\Lambda_0} + (1+z)^3 f(z)\right]}{\left[\Omega_{\Lambda_0} + (1+z)^3 f(z)\right] + (1+z)^3 \Omega_m^0}\,.
    \label{omega_after_lss}
\end{equation}
Since we are only considering redshifts after the LSS, the radiation energy density can be ignored compared to matter. 

We use SCoPE for our MCMC analysis~\cite{das2014scope}. A total of $11$ parameters are considered, which include the standard 6 $\Lambda$CDM parameters - $\Omega^0_m h^2$, $\Omega^0_b h^2$, $h$, $A_s$, $n_s$, and $\kappa$; as well as the $5$ $f(z_i)$ values at the redshifts mentioned above. We consider a flat Universe such that $\Omega^0_{\Lambda} = 1- \Omega^0_{m}$. The prior on $f(z_i)$ allows the dark energy density to be of the order of the ambient matter density. 

The results from our analysis are shown in Fig.~\ref{fig:ISW2D}. The best fit and the average values of the parameters are listed in Table~\ref{tab:my_label}. The average values and the standard deviation of all the $f(z_i)$ are small, showing that a significant departure from the standard $\Lambda$CDM behaviour is not allowed by the data. It should also be noted that the rejection rate is very high in this analysis. This is because the likelihood contour for the $f(z_i)$ values does not have a single minimum. This shows that unless the expansion history is specially chosen, an arbitrarily generated expansion history from Everpresent $\Lambda$ is unlikely to provide a suitable power spectrum. In addition, the existence of several minima indicates that the mean value of dark energy plotted in solid lines in Fig.~\ref{fig:ISW_f_omega} and Fig.~\ref{fig:ISW_f_omega_1} does not necessarily give a good fit. 

In Fig.~\ref{fig:ISW_f_omega}, we plot $f(z)$ (left) and  $\Omega_\Lambda$ (right) for some random expansion histories from the $200$ best-fit expansion histories from our analysis. As mentioned, we set $f(z)$ and its derivatives at redshifts $z=0$ and $z=1100$ to $0$. The opacity of a plot is proportional to the likelihood of that particular expansion history. It is interesting to note that a positive bump-like feature at low redshift and a dip-like feature at high redshift ($z= 5 - 100$) in $f(z)$ provide a better fit than the $\Lambda$CDM model. This feature has been explored theoretically before in~\cite{Das_2014}, and the present numerical analysis indicates the same behaviour. $f(z)$  decreases slowly with increasing $z$ leading to a small window for modifying dark energy at higher redshifts. Finally, the  $\Omega_\Lambda$  plot in Fig.~\ref{fig:ISW_f_omega} shows that most dark energies that give realistic $\chi^2$ closely follow the standard $\Lambda$CDM, therefore disfavouring significant deviations from known expansion histories. These results are consistent with the results in~\cite {Wang2018, Zhao2017}.

 We also confirmed the results of this analysis by repeating it with a different set of $z_i$ and observing that it led to similar results. Here one should note that the position of the peaks in $f(z)$ depends on the choice of $z_i$ points. For a different $z_i$ point set, the position of these peaks may slightly change, but the decaying nature of $f(z)$ with $z$ remains the same, with a bump-like feature at low $z$ and a dip-like feature at a slightly higher redshift ($z= 5 - 100$). In Fig.~\ref{fig:ISW_f_omega_1}, we plot $f(z)$ (left),  $\Omega_\Lambda$ (right) for a few random expansion histories from the $200$ best fit models obtained by fixing $f(z)$ at another 5 redshift points, namely 0.5, 5, 25, 50, 250. We can see that the nature of the allowed variation is the same as shown in Fig.~\ref{fig:ISW_f_omega}. The peak positions at low redshifts remained almost the same, but at higher redshifts, they changed slightly due to the choice of the redshift points.

\begin{figure}
    \centering
    \includegraphics[width=.90\textwidth,trim = 20 20 20 50,clip]{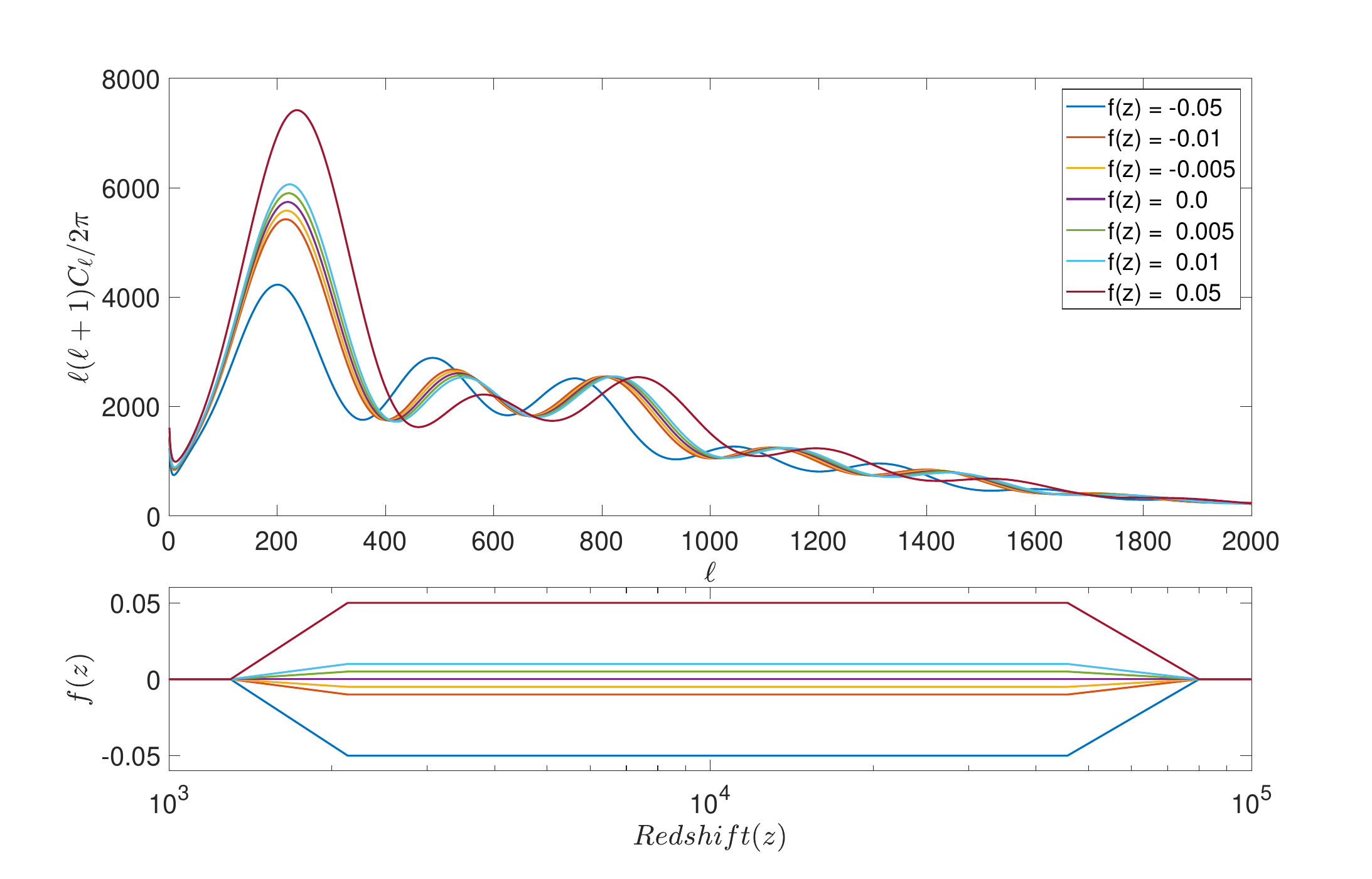}
    \caption{$C_\ell$'s for different fixed $f(z)$ before recombination are shown. As we are adding positive $f(z)$ the power spectrum is moving towards higher $\ell$, while negative $f(z)$ is pushing the spectra towards lower $\ell$. Also, as expected, changing $f(z)$ is changing the ratio of the even and the odd peaks. Positive $f(z)$ are giving higher power to the odd peaks while negative $f(z)$ are giving the higher power to the even peaks.}
    \label{fig:Cl_SWV}
\end{figure}

\subsection{Before Recombination} \label{sec:before_rec}

In this subsection we investigate how a fluctuating dark energy prior to the LSS affects the CMB power spectrum. 
Before recombination, the major contribution to the CMB power spectrum comes from the SW and velocity terms in \eqref{Tsource}. In this era, the photons and baryons are strongly coupled. Their perturbations also depend on the perturbations of the other components, such as dark matter and neutrinos. Therefore, it is difficult to study the behaviour analytically. However, we can expect two primary effects. 

First, from \eqref{Tsource}, we see that the SW term contains a $\dot{\vartheta}$, and the velocity term includes a $\vartheta$ term. As  $\vartheta$ depends mostly on the expansion history of the Universe (as discussed in Section \ref{sec:isw}), any change in the dark energy density (positive or negative), changes $\vartheta$ and $\dot{\vartheta}$. This modifies the base level of the SW and the velocity term fluctuations; it also adjusts the heights of both the even and the odd peaks in the CMB power spectrum. 

Secondly, adding additional positive or negative dark energy influences the time interval from inflation to recombination, which we call $\tau^*$. As the baryon and photon densities stay the same (since recombination occurs under specific conditions), the baryon to photon ratio, $R$ remains unchanged. The sound speed in the medium during the tight coupling era is given by $c_s^2=\frac{1}{3(1+R)}$, and therefore remains unaltered by the change in dark energy. As the fluctuation length scale during recombination is roughly given by $c_s \tau^*$, the angular length scale of the perturbation at the LSS is $\theta^* \approx \frac{c_s \tau^*}{\tau_0 -\tau^*}$. Here, $\tau_0 -\tau^*$ is the distance from the present time to the LSS, and remains unaltered as we are not considering a change in the expansion history after recombination in this section. Therefore, we have that $\theta^* \propto \tau^*$. Adding positive dark energy before the LSS decreases $\tau^*$ and subsequently lowers the value of $\theta^*$. This then moves the CMB power spectrum towards higher multipoles. Conversely, if we add negative dark energy, then $\theta^*$ increases, causing the power spectrum to shift towards lower multipoles. 

Similar to Section \ref{sec:num_analysis}, we want to study the amount of variation allowed in the dark energy.  This variation is represented by the function $f(z)$.  Before the LSS then, we can write the dark energy density fraction as 

\begin{equation}
    \Omega_\Lambda = \frac{\left[\Omega_{\Lambda_0} + \left((1+z)^3 + \frac{\Omega_r^0}{\Omega_m^0}(1+z)^4\right) f(z)\right]}{\left[\Omega_{\Lambda_0} + \left((1+z)^3 + \frac{\Omega_r^0}{\Omega_m^0}(1+z)^4\right) f(z)\right] + (1+z)^3 \Omega_m^0 + (1+z)^4 \Omega_r^0} \,.
    \label{omega_after_lss2}
\end{equation}

\begin{figure}
    \centering
    \includegraphics[width=.99\textwidth,trim = 272 370 230 400,clip]{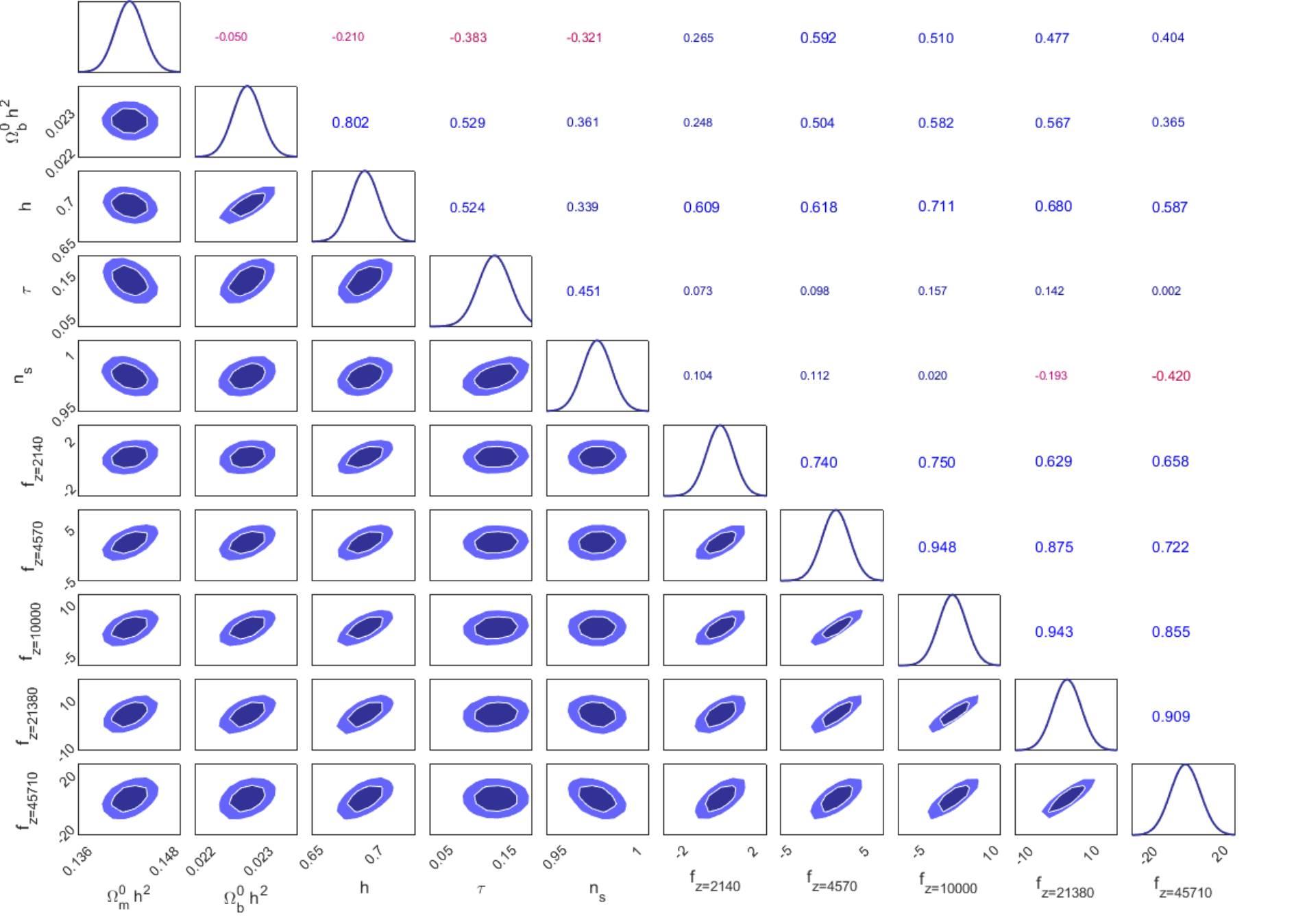}
    \caption{Results from an 11 parameter MCMC analysis constraining the expansion history before the last scattering surface, are shown. $A_s$, which just gives a scaling on the CMB power spectrum, is omitted to make the plot less cluttered. The lower triangle panels show plots of the 68\% and 95\% confidence contours for pairs of parameters. The upper triangle shows the correlation-coefficient between the pair of parameters. Negative correlation-coefficients are shown in red and positive correlation-coefficients are shown in blue. The diagonal plots are the 1-dimensional marginalized distribution of the parameters. The average standard deviation and the best fit values are stated in Table~\ref{tab:my_label}. In this plot the $f(z)$ values have been multiplied by 1000 to avoid the zeros after the decimal point.}
    \label{fig:SWV2D}
\end{figure}

\noindent In this pre-recombination history, we must consider the radiation density, as it is dominant for part of this period. According to the Everpresent $\Lambda$ model, the energy density of dark energy is of the order of the ambient density. If we take the dark energy density, as in the previous section, to be $f(z) (1+z)^3$, then it will be negligible during the radiation-dominated era. Therefore, we add an additional term behaving as $(1+z)^4$ that allows the dark energy density to be of the order of the ambient radiation density. 

In Fig.~\ref{fig:Cl_SWV} we show the effect of adding a positive or negative contribution to the dark energy sector before the LSS. We keep $f(z)$ nearly fixed before recombination - holding it constant, but allowing it to go continuously to $0$ near $z=10^3$ and $10^5$. As explained earlier, an increase in $f(z)$ gives a smaller $\tau^*$, decreases $\theta^*$, and pushes the power spectrum towards higher multipoles. Furthermore, the power from the SW effect increases, which gives high power to the odd peaks of the CMB power spectrum. The opposite effect happens for a negative $f(z)$.

\subsubsection{MCMC Analysis}

\begin{figure}
    \centering
    \includegraphics[width=.80\textwidth,trim = 10 30 10 30,clip]{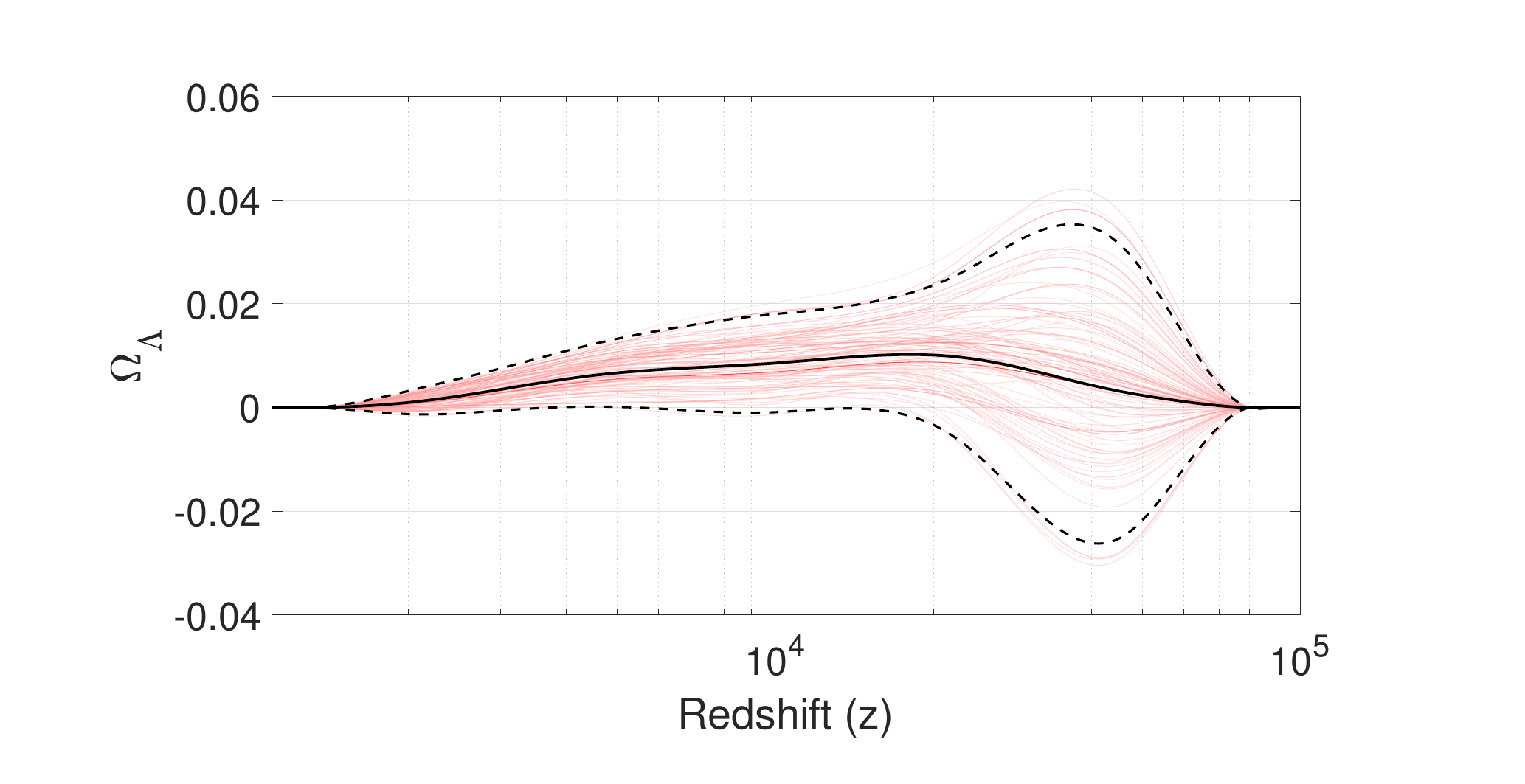}
    \caption{A few random $\Omega^0_\Lambda$ from the 200 best fit $\chi^2$ are shown. The solid black and the dashed black curves show the mean and the standard deviation of the distribution. The opacity of a curve is proportional to its likelihood. As $f(z)$ and its derivatives are taken to be $0$ at $z=1100$ and $z=10^5$, $\Omega^0_\Lambda$ is also $0$. Most of the best fit values have a positive $\Omega_\Lambda$ at high redshift. }
    \label{fig:SWV_OML}
\end{figure}

To constrain the possible variation in dark energy before recombination, as in the previous section, we perturb the dark energy at $5$ redshifts $z_j$ and then fit a spline  $f(z)$ through the $5$ points. The $5$ selected redshifts are logarithmically spaced between the recombination redshift ($z=1100$) and $z=10^5$: the precise redshifts are $2140$, $4570$, $10000$, $21380$, and $45710$. We allow our MCMC code to freely choose the values of  $f(z)$ at these $5$ redshifts along with the other $6$ standard model parameters. We fix $f(z)$ and its derivatives to be $0$ at $z=1100$ and $10^5$. 

Fig.~\ref{fig:SWV2D} summarizes the results from our analysis. The average and the best fit values of $f(z)$ are shown on the right hand side of Table~\ref{tab:my_label}. Interestingly, the average and best fit values for all the $f(z_i)$ are positive, showing that positive dark energy density before recombination is favoured by observations. The best fit $\chi^2$ is about $9.02$ lower than the best fit $\Lambda$CDM model. We can also relate the results to the neutrino mass and number of neutrino species. The CMB data predicts the effective number of neutrinos to be slightly larger than $3$, and also prefers massive neutrinos over massless neutrinos~\cite{aghanim2020planck,escudero2020cmb}. This adds additional energy to the background before the recombination. One interesting thing to note from the contour plots in Fig.~\ref{fig:SWV2D} is that all the $f(z_j)$'s are positively correlated to the other $f(z_j)$'s, especially at high redshifts. Therefore, $f(z)$ should not fluctuate much or vary randomly. Another interesting feature in the contour plots is that $f(z)$, in this case, is strongly correlated with the Hubble parameter, baryon density, and the matter density as well. 

In Fig.~\ref{fig:SWV_OML}, we show $\Omega_\Lambda$ for some random expansion histories chosen from the top $200$ best fit parameters. Unlike for the late time Universe, in this case, most of the best-fit expansion histories give a positive $\Omega_\Lambda$ at all redshifts, although a few do change sign. The allowed values of $\Omega_\Lambda$ are small, which is in conflict with the typically greater in magnitude $\Omega_\Lambda$ values from Everpresent $\Lambda$ realizations. We can also see that in the radiation-dominated era, the allowed variation of $\Omega_\Lambda$ is similar in magnitude to that of the matter dominated era in the right hand plot of Fig. \ref{fig:ISW_f_omega}.  This is again different from the typical behaviour in Everpresent $\Lambda$, in which $\Omega_\Lambda$ tends to have larger fluctuations at earlier times.

These results are in line with the results in \cite{poulin2019early}, where the authors used two phenomenological scalar field models to implement early dark energy. They found that an extra $5\%$ energy density fraction in the form of dark energy at $z\sim5000$ which later dilutes faster than radiation gives a better fit to CMB than $\Lambda$CDM, and furthermore, alleviates the Hubble tension. We have found that with the phenomenological model \eqref{omega_after_lss2}, we can have a dark energy fraction of $0.6\pm0.6\%$ at $z\sim5000$, and $1.0\pm1.5\%$ at $z\sim20000$ which is consistent with the ``$n=2$" model in \cite{poulin2019early}  within $2\sigma$ uncertainty. At the same time our result is consistent with zero dark energy within $1\sigma$. The reason for a less significant dark energy fraction in our case can be attributed to having set the dark energy equal to zero after LSS. Introducing early dark energy in the form of \eqref{omega_after_lss2} also increases the value of the Hubble parameter, decreasing the Hubble tension with the SH0ES measurement \cite{riess2022comprehensive} from $5\sigma$ to $3.5\sigma$.

\noindent 
\begin{table}
    \centering
\begin{tabular}{|c|c|c||c|c|c|c|}
\hline  \multicolumn{3}{|c||}{\rule{0pt}{0.4cm}Modified DE after LSS } & \multicolumn{3}{c|}{ Modified DE before LSS  } \\[0.1cm]
\hline
\hline\rule{0pt}{0.4cm} & Average & Best Fit & & Average & Best Fit \\[0.1cm]
\hline\rule{0pt}{0.4cm}$\Omega^0_m h^2$ & $0.1415 \pm 0.0015$ &$0.1415$& $\Omega^0_m h^2$  & $0.1426 \pm 0.0021$ & 0.1430\\[0.1cm]
\hline\rule{0pt}{0.4cm}$\Omega^0_b h^2$ & $0.0224 \pm 0.00016$ &$0.0224$& $\Omega^0_b h^2$  & $0.0227 \pm 0.00026$ & $0.0226$ \\[0.1cm]
\hline\rule{0pt}{0.4cm}$h$ & $0.6370 \pm 0.0384$ & $0.6275$ & $h$ &$0.6910 \pm 0.0121$ & $0.6866$ \\[0.1cm]
\hline\rule{0pt}{0.4cm}$\kappa$ & $0.1238 \pm 0.0295$ & $0.1231$ & $\kappa$ & $0.1303 \pm 0.0287$ & 0.1245 \\[0.1cm]
\hline\rule{0pt}{0.4cm}$n_s$ & $0.9650 \pm 0.0045$ & $0.9655$ & $n_s$  & $0.9758 \pm 0.01$ & 0.9713\\[0.1cm]
\hline\rule{0pt}{0.4cm} $\log \left(10^{10} A_s\right)$ & $3.17 \pm 0.0571$ & $3.1340$  & $\log \left(10^{10} A_s\right)$ & $3.1910 \pm 0.0550$ & $3.1811$ \\[0.1cm]
\hline\rule{0pt}{0.4cm}$f_{z=1}$ & $0.0248 \pm 0.028$ & $0.0467$ &$f_{z=2140}$ & $0.00037 \pm 0.00078$ & $0.00029$ \\[0.1cm]
\hline\rule{0pt}{0.4cm}$f_{z=10}$ & $0.0018 \pm 0.0289$ & $-0.0249$ &$f_{z=4570}$ & $0.0019 \pm 0.0018$ & $0.0016$ \\[0.1cm]
\hline\rule{0pt}{0.4cm}$f_{z=50}$ & $0.0034 \pm 0.0201$ & $-0.0015$ & $f_{z=10000}$ & $0.0026 \pm 0.0028$ & $0.0021$ \\[0.1cm]
\hline\rule{0pt}{0.4cm}$f_{z=100}$ & $0.0133 \pm 0.0160$ & $0.0224$ & $f_{z=21380}$ & $0.0030 \pm 0.0045$ & $0.0026$ \\[0.1cm]
\hline\rule{0pt}{0.4cm}$f_{z=500}$ & $0.00075 \pm 0.0057$ & $-0.0028$ & $f_{z=45710}$ & $0.0010 \pm 0.0084$ & $0.0009$ \\[0.1cm]
\hline\rule{0pt}{0.4cm}$-2\Delta \log \mathcal{L}$ & & $5.68$ & $-2\Delta\log \mathcal{L}$ &  & $9.02$\\[0.1cm]
\hline
\end{tabular}
    \caption{ Average, standard error, and the best fit values of the parameters from the 11
dimensional parameter estimation. Results for modified expansion histories after and before recombination are shown on the left and right sides of the table, respectively.}
    \label{tab:my_label}
\end{table}

\section{Discussion and Conclusion}\label{sec:discussion_conclusion}
In this paper, we tested a particular implementation of Everpresent $\Lambda$, namely Model 1, against SN Ia and CMB data. Our analysis of the supernova data indicates that a small fraction of seeds ($\sim0.14\%$)  can produce expansion histories with a similar $\chi^2$ value to the $\Lambda$CDM model, and a small fraction ($\sim0.015\%$) can do better, assuming all other parameters ($\alpha$ and $\Omega_m^0 h^2$) are chosen appropriately. To better understand these results, we produced sets of mock supernova data using realizations of Model 1 with realistic values of $H_0$. We carried out a similar analysis on these mock data sets, i.e. fitting Model 1 realizations with arbitrary seeds to the mock data. Our results indicate that almost $10\%$ of the seeds can provide better fits to the mock data set than $\Lambda$CDM, versus only $0.015\%$ for the real data. This suggests that the good realization are indeed distinguished. Our analysis also revealed that at low redshifts the atypical good realizations of Model 1 have smaller dark energy variation and magnitude than in the case of the bad realizations. Furthermore, at sufficiently low redshifts, the good realizations had a smaller dark energy density than matter density, whereas the bad ones had a dark energy density larger than matter density.

Given that SN Ia data only tests the late time behaviour of a cosmological model, we also considered CMB data. In this regard and in our studies, neither of the Everpresent $\Lambda$ models (Model 1 and Model 2) were able to improve on $\Lambda$CDM. For the CMB data, the $\chi^2$ values for the best fit Model 1 realizations in our investigations were more than $150.0$ greater than $\Lambda$CDM. Considering only those realizations with $\chi^2$ below a threshold of $\chi^2<3000.0$, fits to the CMB data led to unrealistically small best fit values of $H_0$ (up to around 50 km/s/MPc), and on average higher best fit values of $\Omega_m^0 h^2$ (averaging around 0.16, although values close to 0.14 were also achieved), thus favouring a matter dominated universe. This indicates that Model 1 as is has difficulty accounting for the present energy density of the Universe. Tests of Model 2  yielded similar behaviour.  However, as suggested in \cite{Zwane_2018}, we found that suppressing the dark energy around the last scattering surface leads to much better fits to the data (with $\chi^2$ about 20.0 greater than $\Lambda$CDM), achieves larger $h$ (up to $\sim0.6$), and has more instances of $\Omega_m^0 h^2$ near 0.14. This together with the supernova results suggests that a stochastic model may still yield a realistic history if it has the right character at the appropriate times during its evolution.

To understand the variation in the expansion history allowed by CMB data, we constrained the variation before and after recombination separately. Our results suggest that while some variation in dark energy density is possible, for the kinds of variation we considered it is significantly smaller than what is typical for Everpresent $\Lambda$. Interestingly, these smaller variations were able to yield better $\chi^2$ values than $\Lambda$CDM when fitting them to the CMB spectra and even alleviate the Hubble tension for the case of pre-recombination dark energy. This is consistent with some earlier results in the literature on early dark energy models. Incorporating some variation in the cosmological constant can therefore be a valuable avenue for model development. While this analysis is suggestive, it remains an open question whether larger fluctuations are possible if one considers a larger number of fluctuations. Namely, it may be in principle possible to have more fluctuations if they have the right character to lead to overall cancellations in some of the integrated effects. Meanwhile, as mentioned in the introduction, observational data continue to hint that $\Lambda$ is time dependent \cite{DESI:2024mwx}. As also noted in Section 4, the fluctuations of $\Lambda$ allow for good fits of the SN Ia data.  One can therefore view the highly fluctuating nature of Everpresent $\Lambda$, and Model 1 in particular as a promising modification to the late-time expansion history.

In conclusion, our analysis shows that the most basic models of Everpresent $\Lambda$ have difficulty describing both the SN Ia and CMB data. However, as there were some positive results for special realizations of Model 1 and the more generic variations we considered, it shows that we should further develop and consider Everpresent $\Lambda$ and stochastic models in general as potential models of the Universe. There are several avenues in which this can be pursued.  One would be to modify the existing models by applying further (theoretically or phenomenologically motivated) constraints on the amplitude of the fluctuations of $\Lambda$.  Another approach would be to more carefully consider the theoretical foundation of Everpresent $\Lambda$ and its relation to the existing models.  Model 1 was introduced under the assumption that the Friedmann equation remains valid in this stochastic framework.  If it does remain valid, in the case of the $H^2<0$ crashes, one may consider whether quantum effects prohibit this from happening and effectively dampen the size of the fluctuations, or if the evolution continues in a nonclassical manner. Alternatively, one could more carefully consider what new effective equations of motion should replace the current ones.

\section*{Acknowledgements} We thank Fay Dowker, Rafael Sorkin, and Nosiphiwo Zwane for  helpful discussions, and we especially thank Niayesh Afshordi for detailed discussions on the analysis. This work was partially funded by a Leverhulme Trust Research Project Grant. YY acknowledges financial support from Imperial College London through an Imperial College Research Fellowship grant. AN is funded by the President’s PhD Scholarships from Imperial College London.
\appendix
\section{MCMC Analysis with a Seed Number}\label{appendixA}

Everpresent $\Lambda$ is a stochastic model of the Universe. To calculate the expansion history from an Everpresent $\Lambda$ model, in addition to the continuous parameters $(\alpha, \Omega^0_m h^2)$, we need the discrete parameter $s$, the seed, to generate the series of random numbers. Unlike the other continuous parameters, $s$ can assume any integer value in $[0,\infty)$, and does not converge in a given range. Instead, we get an entirely different expansion history and a different CMB power spectrum for each value of $s$, most of which may not even resemble the CMB power spectrum. The $\chi^2$ for the power spectrum or distance modulus may differ by several 100s, for different seeds while keeping other parameters fixed. Therefore, we cannot use $s$ as an independent parameter in the MCMC process.

In this appendix we will summarize 3 methods for doing an MCMC analysis to determine the posterior distribution of model parameters.  The main difference between these 3 methods is in how the random seed that produces the unique histories in Models 1 and 2 is handled.  Method 1 attempts to find the distributions of model parameters by considering the distributions of those parameters with a large but finite selection of seeds.  Method 2, incorporated in \cite{Zwane_2018}, treats the seed itself as one of the model parameters that the MCMC iterates over.  Finally, Method 3 is how the MCMC is implemented in this paper, where the distribution is determined for each seed separately, before combining the results for the model as a whole.

\subsection*{Method 1}
For our model, there are continuous parameters, $\theta$, (e.g. the parameters relevant to the supernova analysis are $\theta = (\alpha, \Omega^0_m h^2)$), and one auxiliary parameter, the seed $s$ for which we want to calculate the likelihood. The seed $s$ needs to be marginalized to get the posterior of different parameters. We can write the likelihood function as

\begin{equation}
L\left( \theta \right) =P\left(  D| \theta \right) =\sum _{s}P\left(  D| \theta ,s\right) P\left( s\right) 
=\dfrac{1}{N}\sum _{s}P\left(  D| \theta ,s\right) =\dfrac{1}{N}\sum _{s}L\left( \theta ,s\right)\,.
\end{equation}

\noindent Here $D$ is the observational data and $N$ is the number of seeds in the sample space. We assume that all the seeds are equally likely, and thus $P(s) = \frac{1}{N}$. Hence the likelihood $L(\theta)$ is an average over $L(\theta,s)\propto \exp(-1/2\chi^2(\theta,s))$.

For sampling this likelihood using a Metropolis-Hastings MCMC, we can set the transition probability as

\begin{equation}
T\left( \theta _{i+1} | \theta_i\right)=
\text{min}\left[1,\,\, \dfrac{g(\theta_i | \theta_{i+1}) \sum _{s}\exp \left( -\dfrac{1}{2}\chi ^{2}\left( \theta _{i+1},s\right) \right)  }{ g(\theta_{i+1} | \theta_{i})\sum _{s'}\exp \left( -\dfrac{1}{2}\chi ^{2}\left( \theta _{i},s'\right) \right) } \right]\,,
\end{equation}

\noindent where $g$ is the proposal distribution and $g(\theta_{i+1} | \theta_{i})$  is the probability of choosing $\theta_{i+1}$ when the chain is in $\theta_i$. If $g(\theta_{i+1} | \theta_{i}) = g(\theta_{i} | \theta_{i+1})$, we can run the MCMC analysis using the following recipe. 
Assume that the unbiased sample space of seeds consists of $N$ seeds. At each step of the MCMC, for the given $\theta$, take all $N$ seeds, and compute the $\chi^2$ for each seed. Then sum these likelihoods to find the transition probabilities. If we run the MCMC chain long enough, we can sample the posterior distribution, $P(D|\theta)$.

However, the problem with this method is that it is extremely slow as we are calculating the likelihood for all of the seeds at each step. Also, the process is not easily parallelizable. 

\subsection*{Method 2}
A second option that is proposed in~\cite{Zwane_2018}, is to use the seed $s$ as an independent parameter and keep changing the seed value at each step. 
In such a scenario, the transition probability is given by 

\begin{equation}
  T\left(  \theta _{i+1},s_{i+1}| \theta _{i},s_{i}\right)
  =\text{min}\left[1,\,\,\dfrac{g(\theta_i | \theta_{i+1})g(s_i | s_{i+1})\exp \left( -\dfrac{1}{2}\chi^{2}\left( \theta _{i+1},s_{i+1}\right) \right) }{g(\theta_{i+1} | \theta_{i})g(s_{i+1} | s_{i})\exp \left( -\dfrac{1}{2}\chi^{2}\left( \theta _{i},s_{i}\right) \right)}\right]  \,.
\end{equation}

\noindent Unless we take a finite subset of the seed numbers, $g(s_{i+1} | s_{i})$ and $g(s_{i} | s_{i+1})$ are both $0$ (assuming a flat distribution), and hence the MCMC algorithm can not sample the posterior distribution properly.

To understand it intuitively, let us imagine the scenario that only two kinds of seeds exist that give two entirely different types of expansion histories. As there are an infinite number of seeds in both categories, the MCMC run may keep choosing seeds from a particular kind of distribution. This may happen if the one set has in general better $\chi^2$ values, or simply because it is easier to optimize the $\theta$ parameters for distributions that are similar. It will be impossible to marginalize over the seeds. 

Therefore, to sample the posterior, we need to choose the seeds from a finite unbiased sample of seeds and not from a pool of an infinite number of seeds. In such a case, assuming all the seeds are equally likely, $g(s_{i+1} | s_{i})=g(s_{i} | s_{i+1})$. Assuming  $g(\theta_{i+1} | \theta_{i})=g(\theta_{i} | \theta_{i+1})$, the transition probability can be given by 

\begin{equation}
  T\left(  \theta _{i+1},s_{i+1}| \theta _{i},s_{i}\right)
  =\text{min}\left[1,\,\,\dfrac{\exp \left( -\dfrac{1}{2}\chi^{2}\left( \theta _{i+1},s_{i+1}\right) \right) }{\exp \left( -\dfrac{1}{2}\chi^{2}\left( \theta _{i},s_{i}\right) \right)}\right]  .
\end{equation}

\noindent However, even in this case, the problem is that in general, if we change the seed, the change in $\chi^2$ is very high, and hence the transition probability is negligibly small, and thus the chains do not move. Even though mathematically it may be able to sample the posterior distribution, provided we run the chains for an infinite amount of time, in practice the chains cannot sample the distribution in a reasonable time, if the sample space is too large.

\subsection*{Method 3}

A more efficient option is to run the MCMC for each of the seeds independently, i.e. to keep $s$ constant and only update $\theta$. The transition probability in that case is given by

\begin{equation}
 T\left(  \theta _{i+1},s| \theta _{i},s\right) 
 =\text{min}\left[1,\,\,\dfrac{\exp \left( -\dfrac{1}{2}\chi^{2}\left( \theta _{i+1},s\right) \right) }{\exp \left( -\dfrac{1}{2}\chi ^{2}\left( \theta _{i},s\right) \right) } \right]\,.
\end{equation}

Therefore, if we run this chain while keeping the seed $s$ constant, we can get the posterior distribution of $\theta$ for that particular seed, $s$.  After one run of the MCMC, we  sample the posterior 
$P\left( D | \theta, s\right) =\dfrac{L\left( \theta ,s\right) }{\int d\theta 'L\left( \theta ',s\right) }$. We then calculate the  posterior over all seeds as
\begin{equation}
P(D | \theta) = \sum_s P(D, |\theta , s)P(s), 
\end{equation}
where $P(s)$ will be constant if we take an equal number of sample points for each seed. Therefore we take the average over the  $P(\theta|s,D)$'s for all seeds to get $P(\theta|D)$. This method of calculating the probability distribution is faster than Methods 1 and 2 because we can calculate the $P(\theta|s,D)$ for different seeds independently in parallel and then take their average. Therefore, we use this method in this paper.  

\section{On the Width of the Visibility Function}\label{appendixB}
Taking $X_e=n_e/n_b$ as the free-electron fraction with $n_b=\eta_b n_\gamma\approx 10^{-9}(k_BT)^3$ being the baryon number density, $\eta_b$ being the baryon to photon fraction, and $\Delta\epsilon=13.6$eV,  \eqref{saha} can be rewritten as \cite{piattella2018lecture}
\begin{equation}
    \frac{X_e^2}{1-X_e}=10^9\left(\frac{m_e}{2\pi k_BT}\right)^{3/2}e^{-\Delta\epsilon/(k_BT)}\,,
\end{equation}
where $\mathcal{H}=aH$. Multiplying both sides by $1-X_e$ and differentiating with respect to the conformal time gives 
\begin{equation}
    \dot{X}_e=\frac{1-X_e}{2-X_e}\left(\frac{3}{2}-\frac{\Delta\epsilon}{k_BT}\right)\mathcal{H}X_e\,.
\end{equation}
During recombination, $k_BT$ is at least an order of magnitude smaller than $\Delta\epsilon$, and $(1-X_e)/(2-X_e)$ is an $\mathcal{O}(1)$ number. Also noting that 
\begin{equation}
    \frac{\Delta\epsilon}{k_BT}=\frac{\Delta\epsilon}{k_BT_0}\frac{1}{1+z}\approx\frac{\Delta\epsilon}{k_BT_0}\frac{1}{z},
\end{equation}
and substituting in for the known constants, one finds
\begin{equation}
\label{Inonm}
    \dot{X}_e\simeq-10^4\frac{1}{z}\mathcal{H}X_e.
\end{equation}
On the other hand, the peak of the visibility function (at the recombination conformal time $\tau_*$) is characterized by
\begin{equation}
    \dot{g}|_{\tau_*}=0.
\end{equation}
Using the definition of $g$ ($g=-\dot{\kappa} \exp (-\kappa)$; $\kappa=\int_\tau^{\tau_0} a\, n_e \,\sigma_T \,\mathrm{~d}\tau$), this yields
\begin{equation}
\label{gdoteq}
    \frac{d}{d\tau}\left(a n_e \sigma_T\right)_*=-\left(a n_e \sigma_T\right)_*^2.
\end{equation}
Now the relations $\dot{n}_b=-3\mathcal{H}n_b$, and $\dot{a}=a\mathcal{H}$, simplify the expression to
\begin{equation}
\label{nonam}
    \dot{X}_e(\tau_*)=2\mathcal{H_*}X_e(\tau_*)-(an_b\sigma_T)_*X_e(\tau_*)^2.
\end{equation}
The collision rate of photons at $\tau_*$ is still an order of magnitude larger than the Hubble rate, as we verify below. Therefore, we ignore the first term on the right-hand side of \eqref{nonam} compared to the second term. Comparing the result with \eqref{Inonm}, we find
\begin{equation}
    X_e(\tau_*)=\frac{10^4\mathcal{H}_*}{z_*(an_b\sigma_T)_*}\,.
\end{equation}
Therefore, since $X_e=n_e/n_b$, we have
\begin{equation} \label{ans}
    (an_e\sigma_T)_*=\frac{10^4\mathcal{H}_*}{z_*}.
\end{equation}
This equation implies that $(an_e\sigma_T)_*\approx10\mathcal{H}_*$ and justifies the above claim that $\mathcal{H}_*$ can be ignored compared to $(an_e\sigma_T)_*$. Now, for finding the width of the visibility function, note that $g=\exp(-\kappa+\ln(-\dot{\kappa}))$. Taylor expanding around $\tau_*$ inside the exponent, we see that the coefficient of $\tau-\tau_*$ vanishes due to $\dot{g}|_{\tau_*}=0$. To leading order $g$ is then a Gaussian function in $\tau$, and the coefficient of $-(\tau-\tau_*)^2/2$ is thus the inverse of the variance of $g$. Writing $g$ as a Gaussian, 
\begin{equation}
    g(\tau)=\frac{1}{\sqrt{2\pi\sigma_g^2}}e^{-(\tau-\tau_*)^2/(2\sigma_g^2) + \mathcal{O}(\tau^3)},
\end{equation}
we can solve for its variance as
\begin{equation}
    \frac{1}{\sigma_g^2}=\frac{d^2}{d\tau^2}(\kappa-\ln(-\dot{\kappa}))_*=
    \frac{d}{d\tau}\left( -a n_e \sigma_T -\frac{1}{a n_e \sigma_T} \frac{d}{d\tau}(a X_e n_b \sigma_T)\right)_*.
\end{equation}
We can argue that the second term on the right-hand side is negligible. To this end, note that during recombination $\dot{X}_e=-10^4\mathcal{H}X_e/z\approx -10\mathcal{H}X_e$. Therefore
\begin{equation}
\label{sg2}
    \frac{1}{\sigma_g^2}\approx\frac{d}{d\tau}\left(-a n_e \sigma_T+12\mathcal{H}\right)_*.
\end{equation}
In a matter dominated universe, $\dot{\mathcal{H}}=-\mathcal{H}^2/2$. On the other hand, from \eqref{gdoteq} and \eqref{ans}, $\frac{d}{d\tau}\left(-a n_e \sigma_T\right)_*=\left(-a n_e \sigma_T\right)_*^2=(10^4\mathcal{H}_*/z_*)^2\approx 100\mathcal{H}_*^2$. Hence, the first term in \eqref{sg2} dominates the second one, and we have found
\begin{equation}
    \frac{1}{\sigma_g^2}\simeq\frac{d}{d\tau}\left(-a n_e \sigma_T\right)_*=\left(\frac{10^4\mathcal{H}_*}{z_*}\right)^2,
\end{equation}
as claimed in the text.

\bibliographystyle{jhep}
\bibliography{EverLam2.bib}

\providecommand{\href}[2]{#2}\begingroup\raggedright\begin{thebibliography}{100}

\bibitem{Zwane_2018}
N.~Zwane, N.~Afshordi, and R.~D. Sorkin, {\it {Cosmological tests of
  Everpresent $\Lambda$}},  {\em Classical and Quantum Gravity} {\bf 35} (sep,
  2018) 194002, [\href{http://arxiv.org/abs/1703.06265}{{\tt
  arXiv:1703.06265}}].

\bibitem{SupernovaCosmologyProject:1998vns}
{\bf Supernova Cosmology Project} Collaboration, S.~Perlmutter et~al., {\it
  {Measurements of $\Omega$ and $\Lambda$ from 42 high redshift supernovae}},
  {\em Astrophys. J.} {\bf 517} (1999) 565--586,
  [\href{http://arxiv.org/abs/astro-ph/9812133}{{\tt astro-ph/9812133}}].

\bibitem{SupernovaSearchTeam:1998fmf}
{\bf Supernova Search Team} Collaboration, A.~G. Riess et~al., {\it
  {Observational evidence from supernovae for an accelerating universe and a
  cosmological constant}},  {\em Astron. J.} {\bf 116} (1998) 1009--1038,
  [\href{http://arxiv.org/abs/astro-ph/9805201}{{\tt astro-ph/9805201}}].

\bibitem{Di_Valentino_2021}
E.~D. Valentino, O.~Mena, S.~Pan, L.~Visinelli, W.~Yang, A.~Melchiorri, D.~F.
  Mota, A.~G. Riess, and J.~Silk, {\it In the realm of the hubble tension - a
  review of solutions},  {\em Classical and Quantum Gravity} {\bf 38} (jul,
  2021) 153001.

\bibitem{riess2018new}
A.~G. Riess, S.~Casertano, W.~Yuan, L.~Macri, J.~Anderson, J.~W. MacKenty,
  J.~B. Bowers, K.~I. Clubb, A.~V. Filippenko, D.~O. Jones, et~al., {\it New
  parallaxes of galactic cepheids from spatially scanning the hubble space
  telescope: Implications for the hubble constant},  {\em The Astrophysical
  Journal} {\bf 855} (2018), no.~2 136.

\bibitem{Riess_2019}
A.~G. Riess, S.~Casertano, W.~Yuan, L.~M. Macri, and D.~Scolnic, {\it Large
  magellanic cloud cepheid standards provide a 1{\%} foundation for the
  determination of the hubble constant and stronger evidence for physics beyond
  $\lambda${CDM}},  {\em The Astrophysical Journal} {\bf 876} (may, 2019) 85.

\bibitem{riess2020expansion}
A.~G. Riess, {\it The expansion of the universe is faster than expected},  {\em
  Nature Reviews Physics} {\bf 2} (2020), no.~1 10--12.

\bibitem{wong2020h0licow}
K.~C. Wong, S.~H. Suyu, G.~C. Chen, C.~E. Rusu, M.~Millon, D.~Sluse, V.~Bonvin,
  C.~D. Fassnacht, S.~Taubenberger, M.~W. Auger, et~al., {\it H0licow--xiii. a
  2.4 per cent measurement of h 0 from lensed quasars: 5.3 $\sigma$ tension
  between early-and late-universe probes},  {\em Monthly Notices of the Royal
  Astronomical Society} {\bf 498} (2020), no.~1 1420--1439.

\bibitem{di2021combined}
E.~Di~Valentino, {\it A combined analysis of the h0 late time direct
  measurements and the impact on the dark energy sector},  {\em Monthly Notices
  of the Royal Astronomical Society} {\bf 502} (2021), no.~2 2065--2073.

\bibitem{riess2021cosmic}
A.~G. Riess, S.~Casertano, W.~Yuan, J.~B. Bowers, L.~Macri, J.~C. Zinn, and
  D.~Scolnic, {\it Cosmic distances calibrated to 1\% precision with gaia edr3
  parallaxes and hubble space telescope photometry of 75 milky way cepheids
  confirm tension with $\lambda$cdm},  {\em The Astrophysical Journal Letters}
  {\bf 908} (2021), no.~1 L6.

\bibitem{riess2022comprehensive}
A.~G. Riess, W.~Yuan, L.~M. Macri, D.~Scolnic, D.~Brout, S.~Casertano, D.~O.
  Jones, Y.~Murakami, G.~S. Anand, L.~Breuval, et~al., {\it A comprehensive
  measurement of the local value of the hubble constant with 1 km s- 1 mpc- 1
  uncertainty from the hubble space telescope and the sh0es team},  {\em The
  Astrophysical Journal Letters} {\bf 934} (2022), no.~1 L7.

\bibitem{secco2022dark}
L.~F. Secco, S.~Samuroff, E.~Krause, B.~Jain, J.~Blazek, M.~Raveri, A.~Campos,
  A.~Amon, A.~Chen, C.~Doux, et~al., {\it Dark energy survey year 3 results:
  Cosmology from cosmic shear and robustness to modeling uncertainty},  {\em
  Physical Review D} {\bf 105} (2022), no.~2 023515.

\bibitem{li2023kids}
S.-S. Li, H.~Hoekstra, K.~Kuijken, M.~Asgari, M.~Bilicki, B.~Giblin,
  C.~Heymans, H.~Hildebrandt, B.~Joachimi, L.~Miller, et~al., {\it Kids-1000:
  Cosmology with improved cosmic shear measurements},  {\em Astronomy \&
  Astrophysics} {\bf 679} (2023) A133.

\bibitem{aghanim2020planck}
N.~Aghanim, Y.~Akrami, M.~Ashdown, J.~Aumont, C.~Baccigalupi, M.~Ballardini,
  A.~Banday, R.~Barreiro, N.~Bartolo, S.~Basak, et~al., {\it Planck 2018
  results-vi. cosmological parameters},  {\em Astronomy \& Astrophysics} {\bf
  641} (2020) A6.

\bibitem{troxel2018cosmological}
M.~Troxel, D.~Collaboration, et~al., {\it Cosmological constraints from galaxy
  clustering and weak lensing},  in {\em American Astronomical Society Meeting
  Abstracts\# 231}, vol.~231, pp.~219--02, 2018.

\bibitem{asgari2021kids}
M.~Asgari, C.-A. Lin, B.~Joachimi, B.~Giblin, C.~Heymans, H.~Hildebrandt,
  A.~Kannawadi, B.~St{\"o}lzner, T.~Tr{\"o}ster, J.~L. van~den Busch, et~al.,
  {\it Kids-1000 cosmology: Cosmic shear constraints and comparison between two
  point statistics},  {\em Astronomy \& Astrophysics} {\bf 645} (2021) A104.

\bibitem{Addison_2016}
G.~E. Addison, Y.~Huang, D.~J. Watts, C.~L. Bennett, M.~Halpern, G.~Hinshaw,
  and J.~L. Weiland, {\it {QUANTIFYING} {DISCORDANCE} {IN} {THE} 2015 {PLANCK}
  {CMB} {SPECTRUM}},  {\em The Astrophysical Journal} {\bf 818} (feb, 2016)
  132.

\bibitem{Kitching_2016}
T.~D. Kitching, L.~Verde, A.~F. Heavens, and R.~Jimenez, {\it Discrepancies
  between {CFHTLenS} cosmic shear and planck: new physics or systematic
  effects?},  {\em Monthly Notices of the Royal Astronomical Society} {\bf 459}
  (mar, 2016) 971--981.

\bibitem{Couchot_2017}
F.~Couchot, S.~Henrot-Versill{\'{e} }, O.~Perdereau, S.~Plaszczynski, B.~R.
  d'Orfeuil, M.~Spinelli, and M.~Tristram, {\it Relieving tensions related to
  the lensing of the cosmic microwave background temperature power spectra},
  {\em Astronomy \& Astrophysics} {\bf 597} (jan, 2017) A126.

\bibitem{lusso2019tension}
E.~Lusso, E.~Piedipalumbo, G.~Risaliti, M.~Paolillo, S.~Bisogni, E.~Nardini,
  and L.~Amati, {\it Tension with the flat $\lambda$cdm model from a
  high-redshift hubble diagram of supernovae, quasars, and gamma-ray bursts},
  {\em Astronomy \& Astrophysics} {\bf 628} (2019) L4.

\bibitem{verde2013planck}
L.~Verde, P.~Protopapas, and R.~Jimenez, {\it Planck and the local universe:
  Quantifying the tension},  {\em Physics of the Dark Universe} {\bf 2} (2013),
  no.~3 166--175.

\bibitem{addison2018elucidating}
G.~Addison, D.~Watts, C.~Bennett, M.~Halpern, G.~Hinshaw, and J.~Weiland, {\it
  Elucidating $\lambda$cdm: impact of baryon acoustic oscillation measurements
  on the hubble constant discrepancy},  {\em The Astrophysical Journal} {\bf
  853} (2018), no.~2 119.

\bibitem{evslin2017isolating}
J.~Evslin, {\it Isolating the lyman alpha forest bao anomaly},  {\em Journal of
  Cosmology and Astroparticle Physics} {\bf 2017} (2017), no.~04 024.

\bibitem{guandalin2023theoretical}
C.~Guandalin, J.~Piat, C.~Clarkson, and R.~Maartens, {\it Theoretical
  systematics in testing the cosmological principle with the kinematic quasar
  dipole},  {\em The Astrophysical Journal} {\bf 953} (2023), no.~2 144.

\bibitem{luongo2022larger}
O.~Luongo, M.~Muccino, E.~{\'O}. Colg{\'a}in, M.~Sheikh-Jabbari, and L.~Yin,
  {\it Larger h 0 values in the cmb dipole direction},  {\em Physical Review D}
  {\bf 105} (2022), no.~10 103510.

\bibitem{horstmann2022inference}
N.~Horstmann, Y.~Pietschke, and D.~J. Schwarz, {\it Inference of the cosmic
  rest-frame from supernovae ia},  {\em Astronomy \& Astrophysics} {\bf 668}
  (2022) A34.

\bibitem{zhao2021tomographic}
D.~Zhao and J.-Q. Xia, {\it A tomographic test of cosmic anisotropy with the
  recently-released quasar sample},  {\em The European Physical Journal C} {\bf
  81} (2021), no.~10 948.

\bibitem{bengaly2018probing}
C.~A. Bengaly, R.~Maartens, and M.~G. Santos, {\it Probing the cosmological
  principle in the counts of radio galaxies at different frequencies},  {\em
  Journal of Cosmology and Astroparticle Physics} {\bf 2018} (2018), no.~04
  031.

\bibitem{BOSS:2013igd}
{\bf BOSS} Collaboration, A.~Font-Ribera et~al., {\it {Quasar-Lyman $\alpha$
  Forest Cross-Correlation from BOSS DR11 : Baryon Acoustic Oscillations}},
  {\em JCAP} {\bf 05} (2014) 027, [\href{http://arxiv.org/abs/1311.1767}{{\tt
  arXiv:1311.1767}}].

\bibitem{BOSS:2014hwf}
{\bf BOSS} Collaboration, T.~Delubac et~al., {\it {Baryon acoustic oscillations
  in the Ly\ensuremath{\alpha} forest of BOSS DR11 quasars}},  {\em Astron.
  Astrophys.} {\bf 574} (2015) A59, [\href{http://arxiv.org/abs/1404.1801}{{\tt
  arXiv:1404.1801}}].

\bibitem{Wang:2018fng}
Y.~Wang, L.~Pogosian, G.-B. Zhao, and A.~Zucca, {\it {Evolution of dark energy
  reconstructed from the latest observations}},  {\em Astrophys. J. Lett.} {\bf
  869} (2018) L8, [\href{http://arxiv.org/abs/1807.03772}{{\tt
  arXiv:1807.03772}}].

\bibitem{Aubourg:2014yra}
E.~Aubourg et~al., {\it {Cosmological implications of baryon acoustic
  oscillation measurements}},  {\em Phys. Rev. D} {\bf 92} (2015), no.~12
  123516, [\href{http://arxiv.org/abs/1411.1074}{{\tt arXiv:1411.1074}}].

\bibitem{colgain2023putting}
E.~O. Colgáin, M.~M. Sheikh-Jabbari, R.~Solomon, M.~G. Dainotti, and
  D.~Stojkovic, {\it Putting flat $\lambda$cdm in the (redshift) bin},  2023.

\bibitem{malekjani2023negative}
M.~Malekjani, R.~M. Conville, E.~O. Colgáin, S.~Pourojaghi, and M.~M.
  Sheikh-Jabbari, {\it Negative dark energy density from high redshift
  pantheon+ supernovae},  2023.

\bibitem{colgain2023mcmc}
E.~O. Colgáin, S.~Pourojaghi, M.~M. Sheikh-Jabbari, and D.~Sherwin, {\it Mcmc
  marginalisation bias and $\lambda$cdm tensions},  2023.

\bibitem{DESI:2024mwx}
{\bf DESI} Collaboration, A.~G. Adame et~al., {\it {DESI 2024 VI: Cosmological
  Constraints from the Measurements of Baryon Acoustic Oscillations}},
  \href{http://arxiv.org/abs/2404.03002}{{\tt arXiv:2404.03002}}.

\bibitem{Sorkin:2007bd}
R.~D. Sorkin, {\it {Is the cosmological 'constant' a nonlocal quantum residue
  of discreteness of the causal set type?}},  {\em AIP Conf. Proc.} {\bf 957}
  (2007), no.~1 142--153, [\href{http://arxiv.org/abs/0710.1675}{{\tt
  arXiv:0710.1675}}].

\bibitem{deCesare:2016dnp}
M.~de~Cesare, F.~Lizzi, and M.~Sakellariadou, {\it {Effective cosmological
  constant induced by stochastic fluctuations of Newton's constant}},  {\em
  Phys. Lett. B} {\bf 760} (2016) 498--501,
  [\href{http://arxiv.org/abs/1603.04170}{{\tt arXiv:1603.04170}}].

\bibitem{Cree:2018mcx}
S.~S. Cree, T.~M. Davis, T.~C. Ralph, Q.~Wang, Z.~Zhu, and W.~G. Unruh, {\it
  {Can the fluctuations of the quantum vacuum solve the cosmological constant
  problem?}},  {\em Phys. Rev. D} {\bf 98} (2018), no.~6 063506,
  [\href{http://arxiv.org/abs/1805.12293}{{\tt arXiv:1805.12293}}].

\bibitem{originallambda}
R.~D. Sorkin, {\it {A Modified Sum-Over-Histories for Gravity reported in the
  article by D. Brill and L. Smolin: “Workshop on quantum gravity and new
  directions”}},  in {\em Highlights in gravitation and cosmology:
  Proceedings of the International Conference on Gravitation and Cosmology,
  Goa, India, 14–19 December 1987} (B.~R. Iyer, A.~Kembhavi, J.~V. Narlikar,
  and C.~V. Vishveshwara, eds.), pp.~184--186, 1988.

\bibitem{sorkin1991spacetime}
R.~D. Sorkin, {\it Spacetime and causal sets.},  {\em Relativity and
  Gravitation} (1991) 150.

\bibitem{sorkin1994role}
R.~D. Sorkin, {\it Role of time in the sum-over-histories framework for
  gravity},  {\em International journal of theoretical physics} {\bf 33}
  (1994), no.~3 523--534.

\bibitem{Ahmed_2004}
M.~Ahmed, S.~Dodelson, P.~B. Greene, and R.~Sorkin, {\it {Everpresent
  $\Lambda$}},  {\em Physical Review D} {\bf 69} (may, 2004)
  [\href{http://arxiv.org/abs/astro-ph/0209274}{{\tt astro-ph/0209274}}].

\bibitem{Ahmed_2013}
M.~Ahmed and R.~D. Sorkin, {\it {Everpresent $\Lambda$ {II}: Structural
  stability}},  {\em Physical Review D} {\bf 87} (mar, 2013).

\bibitem{das2023aspects1}
S.~Das, A.~Nasiri, and Y.~K. Yazdi, {\it Aspects of everpresent $\lambda$. part
  i. a fluctuating cosmological constant from spacetime discreteness},  {\em
  Journal of Cosmology and Astroparticle Physics} {\bf 2023} (2023), no.~10
  047.

\bibitem{lehners2023review}
J.-L. Lehners, {\it Review of the no-boundary wave function},  {\em Physics
  Reports} {\bf 1022} (2023) 1--82.

\bibitem{scolnic2018complete}
D.~M. Scolnic, D.~Jones, A.~Rest, Y.~Pan, R.~Chornock, R.~Foley, M.~Huber,
  R.~Kessler, G.~Narayan, A.~Riess, et~al., {\it The complete light-curve
  sample of spectroscopically confirmed sne ia from pan-starrs1 and
  cosmological constraints from the combined pantheon sample},  {\em The
  Astrophysical Journal} {\bf 859} (2018), no.~2 101.

\bibitem{scolnic2022pantheon+}
D.~Scolnic, D.~Brout, A.~Carr, A.~G. Riess, T.~M. Davis, A.~Dwomoh, D.~O.
  Jones, N.~Ali, P.~Charvu, R.~Chen, et~al., {\it The pantheon+ analysis: the
  full data set and light-curve release},  {\em The Astrophysical Journal} {\bf
  938} (2022), no.~2 113.

\bibitem{Jimenez_2002}
R.~Jimenez and A.~Loeb, {\it Constraining cosmological parameters based on
  relative galaxy ages},  {\em The Astrophysical Journal} {\bf 573} (jul, 2002)
  37--42, [\href{http://arxiv.org/abs/astro-ph/0106145v}{{\tt
  astro-ph/0106145v}}].

\bibitem{M_ller_2018}
O.~Müller, M.~Rejkuba, and H.~Jerjen, {\it Distances from the tip of the red
  giant branch to the dwarf galaxies dw1335-29 and dw1340-30 in the centaurus
  group},  {\em Astronomy and Astrophysics} {\bf 615} (jul, 2018) A96.

\bibitem{freedman2023progress}
W.~L. Freedman and B.~F. Madore, {\it Progress in direct measurements of the
  hubble constant},  {\em Journal of Cosmology and Astroparticle Physics} {\bf
  2023} (2023), no.~11 050.

\bibitem{anand2022comparing}
G.~S. Anand, R.~B. Tully, L.~Rizzi, A.~G. Riess, and W.~Yuan, {\it Comparing
  tip of the red giant branch distance scales: an independent reduction of the
  carnegie-chicago hubble program and the value of the hubble constant},  {\em
  The Astrophysical Journal} {\bf 932} (2022), no.~1 15.

\bibitem{Riess_2020}
A.~G. Riess, W.~Yuan, S.~Casertano, L.~M. Macri, and D.~Scolnic, {\it The
  accuracy of the hubble constant measurement verified through cepheid
  amplitudes},  {\em The Astrophysical Journal} {\bf 896} (jun, 2020) L43,
  [\href{http://arxiv.org/abs/2005.02445}{{\tt arXiv:2005.02445}}].

\bibitem{Sabour2014}
M.~Abdel-Sabour, M.~E. Nouh, I.~A. Issa, M.~S. El-Nawawy, A.~Kordi,
  Z.~Almostafa, A.~E. El-Said, and G.~B. Ali, {\it Determination of the hubble
  constant using cepheids},  \href{http://arxiv.org/abs/1409.4168 }{{\tt
  arXiv:1409.4168 }}.

\bibitem{rosenberg2022cmb}
E.~Rosenberg, S.~Gratton, and G.~Efstathiou, {\it Cmb power spectra and
  cosmological parameters from planck pr4 with camspec},  {\em Monthly Notices
  of the Royal Astronomical Society} {\bf 517} (2022), no.~3 4620--4636.

\bibitem{seraille2024constraining}
E.~Seraille, J.~Noller, and B.~D. Sherwin, {\it Constraining dark energy with
  the integrated sachs-wolfe effect},  {\em arXiv preprint arXiv:2401.06221}
  (2024).

\bibitem{Nunes_2020}
R.~C. Nunes, S.~K. Yadav, J.~F. Jesus, and A.~Bernui, {\it Cosmological
  parameter analyses using transversal {BAO} data},  {\em Monthly Notices of
  the Royal Astronomical Society} {\bf 497} (jul, 2020) 2133--2141.

\bibitem{PhysRevD.93.023530}
G.~C. Carvalho, A.~Bernui, M.~Benetti, J.~C. Carvalho, and J.~S. Alcaniz, {\it
  Baryon acoustic oscillations from the sdss dr10 galaxies angular correlation
  function},  {\em Phys. Rev. D} {\bf 93} (Jan, 2016) 023530,
  [\href{http://arxiv.org/abs/1507.08972}{{\tt arXiv:1507.08972}}].

\bibitem{Jailson2016}
J.~S. Alcaniz, G.~C. Carvalho, A.~Bernui, J.~C. Carvalho, and M.~Benetti, {\it
  Measuring baryon acoustic oscillations with angular two-point correlation
  function},  2016.

\bibitem{de_Carvalho_2018}
E.~de~Carvalho, A.~Bernui, G.~Carvalho, C.~Novaes, and H.~Xavier, {\it Angular
  baryon acoustic oscillation measure at z=2.225 from the sdss quasar survey},
  {\em Journal of Cosmology and Astroparticle Physics} {\bf 2018} (apr, 2018)
  064--064.

\bibitem{DeBoer_2017}
D.~R. DeBoer et~al., {\it Hydrogen epoch of reionization array ({HERA})},  {\em
  Publications of the Astronomical Society of the Pacific} {\bf 129} (mar,
  2017) 045001.

\bibitem{Carilli_2020}
C.~L. Carilli et~al., {\it Imaging and modeling data from the hydrogen epoch of
  reionization array},  {\em The Astrophysical Journal Supplement Series} {\bf
  247} (apr, 2020) 67.

\bibitem{Amiri_2022}
M.~Amiri et~al., {\it An overview of {CHIME}, the canadian hydrogen intensity
  mapping experiment},  {\em The Astrophysical Journal Supplement Series} {\bf
  261} (jul, 2022) 29.

\bibitem{Mandana2022}
{CHIME Collaboration}, M.~Amiri, et~al., {\it Detection of cosmological 21 cm
  emission with the canadian hydrogen intensity mapping experiment},  2022.

\bibitem{barry2022ska}
N.~Barry, G.~Bernardi, B.~Greig, N.~Kern, and F.~Mertens, {\it Ska-low
  intensity mapping pathfinder updates: deeper 21 cm power spectrum limits from
  improved analysis frameworks},  {\em Journal of Astronomical Telescopes,
  Instruments, and Systems} {\bf 8} (2022), no.~1 011007--011007.

\bibitem{Cunnington2022}
S.~Cunnington et~al., {\it Hi intensity mapping with meerkat: power spectrum
  detection in cross-correlation with wigglez galaxies},  2022.

\bibitem{van_Haarlem_2013}
M.~P. van Haarlem et~al., {\it {LOFAR}: The {LOw}-frequency {ARray}},  {\em
  Astronomy \& Astrophysics} {\bf 556} (jul, 2013) A2.

\bibitem{Newburgh_2016}
L.~B. Newburgh et~al., {\it {HIRAX}: a probe of dark energy and radio
  transients},  in {\em {SPIE} Proceedings} (H.~J. Hall, R.~Gilmozzi, and H.~K.
  Marshall, eds.), {SPIE}, aug, 2016.

\bibitem{li2023fast}
Y.~Li, Y.~Wang, F.~Deng, W.~Yang, W.~Hu, D.~Liu, X.~Zhao, S.~Zuo, S.~Shu,
  J.~Li, et~al., {\it Fast drift scan survey for hi intensity mapping: I.
  preliminary data analysis},  {\em arXiv preprint arXiv:2305.06405} (2023).

\bibitem{grasha2020evolution}
K.~Grasha, J.~Darling, A.~K. Leroy, and A.~D. Bolatto, {\it The evolution of
  neutral hydrogen over the past 11 gyr via h i 21 cm absorption},  {\em
  Monthly Notices of the Royal Astronomical Society} {\bf 498} (2020), no.~1
  883--898.

\bibitem{wolz2022h}
L.~Wolz, A.~Pourtsidou, K.~W. Masui, T.-C. Chang, J.~E. Bautista, E.-M.
  M{\"u}ller, S.~Avila, D.~Bacon, W.~J. Percival, S.~Cunnington, et~al., {\it H
  i constraints from the cross-correlation of eboss galaxies and green bank
  telescope intensity maps},  {\em Monthly Notices of the Royal Astronomical
  Society} {\bf 510} (2022), no.~3 3495--3511.

\bibitem{ansari2020design}
R.~Ansari, J.~Campagne, D.~Charlet, M.~Moniez, C.~Pailler, O.~Perdereau,
  M.~Taurigna, J.~Martin, F.~Rigaud, P.~Colom, et~al., {\it Design, operation
  and performance of the paon4 prototype transit interferometer},  {\em Monthly
  Notices of the Royal Astronomical Society} {\bf 493} (2020), no.~2
  2965--2980.

\bibitem{Zhang_2016}
J.~Zhang, R.~Ansari, X.~Chen, J.-E. Campagne, C.~Magneville, and F.~Wu, {\it
  Sky reconstruction from transit visibilities: {PAON}-4 and tianlai dish
  array},  {\em Monthly Notices of the Royal Astronomical Society} {\bf 461}
  (jun, 2016) 1950--1966.

\bibitem{Chae_2002}
K.-H. Chae, A.~D. Biggs, and et.al., {\it Constraints on cosmological
  parameters from the analysis of the cosmic lens all sky survey radio-selected
  gravitational lens statistics},  {\em Physical Review Letters} {\bf 89} (sep,
  2002) [\href{http://arxiv.org/abs/astro-ph/0209602}{{\tt astro-ph/0209602}}].

\bibitem{Biesiada_2010}
M.~Biesiada, A.~Pi{\'{o}}rkowska, and B.~Malec, {\it Cosmic equation of state
  from strong gravitational lensing systems},  {\em Monthly Notices of the
  Royal Astronomical Society} (may, 2010) no--no.

\bibitem{Cao_2012}
S.~Cao, Y.~Pan, M.~Biesiada, W.~Godlowski, and Z.-H. Zhu, {\it Constraints on
  cosmological models from strong gravitational lensing systems},  {\em Journal
  of Cosmology and Astroparticle Physics} {\bf 2012} (mar, 2012) 016--016,
  [\href{http://arxiv.org/abs/1105.6226 }{{\tt arXiv:1105.6226 }}].

\bibitem{Cao_2015}
S.~Cao, M.~Biesiada, R.~Gavazzi, A.~Pi{\'{o} }rkowska, and Z.-H. Zhu, {\it
  Cosmology with strong-lensing systems},  {\em The Astrophysical Journal} {\bf
  806} (jun, 2015) 185, [\href{http://arxiv.org/abs/1509.07649}{{\tt
  arXiv:1509.07649}}].

\bibitem{Waizmann_2008}
J.-C. Waizmann and M.~Bartelmann, {\it Impact of early dark energy on the
  planck {SZ} cluster sample},  {\em Astronomy \& Astrophysics} {\bf 493} (dec,
  2008) 859--870.

\bibitem{DeDeo2005}
S.~DeDeo, D.~N. Spergel, and H.~Trac, {\it The kinetic sunyaev-zel'dovitch
  effect as a dark energy probe}, .

\bibitem{abbott2018dark}
T.~Abbott, F.~Abdalla, J.~Annis, K.~Bechtol, J.~Blazek, B.~Benson,
  R.~Bernstein, G.~Bernstein, E.~Bertin, D.~Brooks, et~al., {\it Dark energy
  survey year 1 results: a precise h0 estimate from des y1, bao, and d/h data},
   {\em Monthly Notices of the Royal Astronomical Society} {\bf 480} (2018),
  no.~3 3879--3888.

\bibitem{haridasu2022scrutinizing}
B.~S. Haridasu, H.~Khoraminezhad, and M.~Viel, {\it Scrutinizing early dark
  energy models through cmb lensing},  {\em arXiv preprint arXiv:2212.09136}
  (2022).

\bibitem{ye2023shape}
G.~Ye, J.-Q. Jiang, and Y.-S. Piao, {\it Shape of cmb lensing in the early dark
  energy cosmology},  {\em arXiv preprint arXiv:2305.18873} (2023).

\bibitem{brout2022pantheon+}
D.~Brout, D.~Scolnic, B.~Popovic, A.~G. Riess, A.~Carr, J.~Zuntz, R.~Kessler,
  T.~M. Davis, S.~Hinton, D.~Jones, et~al., {\it The pantheon+ analysis:
  cosmological constraints},  {\em The Astrophysical Journal} {\bf 938} (2022),
  no.~2 110.

\bibitem{hoeting1999bayesian}
J.~A. Hoeting, D.~Madigan, A.~E. Raftery, and C.~T. Volinsky, {\it Bayesian
  model averaging: a tutorial (with comments by m. clyde, david draper and ei
  george, and a rejoinder by the authors},  {\em Statistical science} {\bf 14}
  (1999), no.~4 382--417.

\bibitem{paradiso2024convenient}
S.~Paradiso, M.~DiMarco, M.~Chen, G.~McGee, and W.~Percival, {\it A convenient
  approach to characterizing model uncertainty with application to early dark
  energy solutions of the hubble tension},  {\em Monthly Notices of the Royal
  Astronomical Society} {\bf 528} (2024), no.~2 1531--1540.

\bibitem{paradiso2024evaluating}
S.~Paradiso, G.~McGee, and W.~Percival, {\it Evaluating extensions to lcdm: an
  application of bayesian model averaging},  {\em arXiv preprint
  arXiv:2403.02120} (2024).

\bibitem{Bahcall1992}
N.~A. {Bahcall} and R.~{Cen}, {\it {Galaxy Clusters and Cold Dark Matter: A
  Low-Density Unbiased Universe?}},  {\em The Astrophysical Journal Letters}
  {\bf 398} (Oct., 1992) L81.

\bibitem{Costanzi2019}
M.~Costanzi, E.~Rozo, and et.al., {\it {Methods for cluster cosmology and
  application to the SDSS in preparation for DES Year 1 release}},  {\em
  Monthly Notices of the Royal Astronomical Society} {\bf 488} (07, 2019)
  4779--4800,
  [\href{http://arxiv.org/abs/https://academic.oup.com/mnras/article-pdf/488/4/4779/29155032/stz1949.pdf}{{\tt
  https://academic.oup.com/mnras/article-pdf/488/4/4779/29155032/stz1949.pdf}}].

\bibitem{2020Abdullah}
M.~H. {Abdullah}, A.~{Klypin}, and G.~{Wilson}, {\it {Cosmological Constraints
  on {\ensuremath{\Omega}}$_{m}$ and {\ensuremath{\sigma}}$_{8}$ from Cluster
  Abundances Using the GalWCat19 Optical-spectroscopic SDSS Catalog}},  {\em
  The Astrophysical Journal} {\bf 901} (Oct., 2020) 90,
  [\href{http://arxiv.org/abs/2002.11907}{{\tt arXiv:2002.11907}}].

\bibitem{sahni2008two}
V.~Sahni, A.~Shafieloo, and A.~A. Starobinsky, {\it Two new diagnostics of dark
  energy},  {\em Physical Review D} {\bf 78} (2008), no.~10 103502.

\bibitem{CMBAns1910}
S.~Das and A.~Phan, {\it {Cosmic Microwave Background Anisotropy numerical
  solution (CMBAns). Part I. An introduction to $C_l$ calculation}},  {\em
  JCAP} {\bf 05} (2020) 006, [\href{http://arxiv.org/abs/1910.00725}{{\tt
  arXiv:1910.00725}}].

\bibitem{Das:2013gta}
S.~Das, A.~Shafieloo, and T.~Souradeep, {\it {ISW effect as probe of features
  in the expansion history of the Universe}},  {\em JCAP} {\bf 1310} (2013)
  016, [\href{http://arxiv.org/abs/1305.4530}{{\tt arXiv:1305.4530}}].

\bibitem{Das_2014}
S.~Das and T.~Souradeep, {\it Suppressing {CMB} low multipoles with {ISW}
  effect},  {\em Journal of Cosmology and Astroparticle Physics} {\bf 2014}
  (feb, 2014) 002--002, [\href{http://arxiv.org/abs/1312.0025 }{{\tt
  arXiv:1312.0025 }}].

\bibitem{lewis2011camb}
A.~Lewis and A.~Challinor, {\it Camb: Code for anisotropies in the microwave
  background},  {\em Astrophysics source code library} (2011) ascl--1102.

\bibitem{das2014scope}
S.~Das and T.~Souradeep, {\it Scope: an efficient method of cosmological
  parameter estimation},  {\em Journal of Cosmology and Astroparticle Physics}
  {\bf 2014} (2014), no.~07 018.

\bibitem{privatecomm}
N.~Zwane. private communication.

\bibitem{Seljak_1996}
U.~Seljak and M.~Zaldarriaga, {\it A line-of-sight integration approach to
  cosmic microwave background anisotropies},  {\em The Astrophysical Journal}
  {\bf 469} (oct, 1996) 437.

\bibitem{Bernal2016}
J.~Bernal, L.~Verde, and A.~Riess, {\it The trouble with $h_0$}, .

\bibitem{Seager_1999}
S.~Seager, D.~D. Sasselov, and D.~Scott, {\it A new calculation of the
  recombination epoch},  {\em The Astrophysical Journal} {\bf 523} (sep, 1999)
  L1--L5.

\bibitem{Mukherjee:2014wva}
S.~Mukherjee, S.~Das, M.~Joy, and T.~Souradeep, {\it {Estimation of Inflation
  parameters for Perturbed Power Law model using recent CMB measurements}},
  {\em JCAP} {\bf 1501} (2015) 043, [\href{http://arxiv.org/abs/1410.8835}{{\tt
  arXiv:1410.8835}}].

\bibitem{ma1995cosmological}
C.-P. Ma and E.~Bertschinger, {\it Cosmological perturbation theory in the
  synchronous and conformal newtonian gauges},  {\em arXiv preprint
  astro-ph/9506072} (1995).

\bibitem{Wang2018}
Y.~{Wang}, L.~{Pogosian}, G.-B. {Zhao}, and A.~{Zucca}, {\it {Evolution of Dark
  Energy Reconstructed from the Latest Observations}},  {\em The Astrophysical
  Journal Letters} {\bf 869} (Dec., 2018) L8,
  [\href{http://arxiv.org/abs/1807.03772}{{\tt arXiv:1807.03772}}].

\bibitem{Zhao2017}
G.-B. {Zhao}, M.~{Raveri}, L.~{Pogosian}, Y.~{Wang}, R.~G. {Crittenden}, W.~J.
  {Handley}, W.~J. {Percival}, F.~{Beutler}, J.~{Brinkmann}, C.-H. {Chuang},
  A.~J. {Cuesta}, D.~J. {Eisenstein}, F.-S. {Kitaura}, K.~{Koyama},
  B.~{L'Huillier}, R.~C. {Nichol}, M.~M. {Pieri}, S.~{Rodriguez-Torres}, A.~J.
  {Ross}, G.~{Rossi}, A.~G. {S{\'a}nchez}, A.~{Shafieloo}, J.~L. {Tinker},
  R.~{Tojeiro}, J.~A. {Vazquez}, and H.~{Zhang}, {\it {Dynamical dark energy in
  light of the latest observations}},  {\em Nature Astronomy} {\bf 1} (Aug.,
  2017) 627--632, [\href{http://arxiv.org/abs/1701.08165}{{\tt
  arXiv:1701.08165}}].

\bibitem{escudero2020cmb}
M.~Escudero and S.~J. Witte, {\it A cmb search for the neutrino mass mechanism
  and its relation to the hubble tension},  {\em The European Physical Journal
  C} {\bf 80} (2020), no.~4 294, [\href{http://arxiv.org/abs/1909.04044}{{\tt
  arXiv:1909.04044}}].

\bibitem{poulin2019early}
V.~Poulin, T.~L. Smith, T.~Karwal, and M.~Kamionkowski, {\it Early dark energy
  can resolve the hubble tension},  {\em Physical review letters} {\bf 122}
  (2019), no.~22 221301.

\bibitem{piattella2018lecture}
O.~Piattella, {\em Lecture notes in cosmology}.
\newblock Springer, 2018.

\end{thebibliography}\endgroup
\end{document}